\documentclass[useAMS,usenatbib]{mn2e}
\usepackage{savesym}
\usepackage{txfonts}
\savesymbol{iint}
\savesymbol{iiint}
\savesymbol{iiiint}
\savesymbol{idotsint}
\usepackage{graphicx}
\usepackage{amssymb}
\usepackage[below]{placeins}
\usepackage{txfonts}
\usepackage{natbib}
\usepackage{amsmath}
\usepackage{float}
\restoresymbol{TXF}{iint}
\restoresymbol{TXF}{iiint}
\restoresymbol{TXF}{iiiint}
\restoresymbol{TXF}{idotsint}
\usepackage{multirow}
\usepackage{array}
\bibpunct{(}{)}{;}{a}{}{,}

\def \xmm{{\emph{XMM-Newton}}}
\def \rosat{{\emph{ROSAT}}}
\def \chandra{{\emph{Chandra}}}

\def \suzaku{{\emph{Suzaku}}}

\newcommand*{\mysub}[2]{\ensuremath{#1_{\mathrm{#2}}}}

\def\spose#1{\hbox to 0pt{#1\hss}}
\def\approxlt{\mathrel{\spose{\lower 3pt\hbox{$\sim$}}
        \raise 2.0pt\hbox{$<$}}}
\def\approxgt{\mathrel{\spose{\lower 3pt\hbox{$\sim$}}
        \raise 2.0pt\hbox{$>$}}}
\def\approxpropto{\mathrel{\spose{\lower 3pt\hbox{$\sim$}}
        \raise 2.0pt\hbox{$\propto$}}}
\mathchardef\twiddle="2218

\def\multleft#1{\hbox to size{\vbox {\halign {\lft{##}\cr #1}}\hfill}\par}
\def\multright#1{\hbox to size{\vbox {\halign {\rt{##}\cr #1}}\hfill}\par}

\def\today{\ifcase\month\or January\or February\or March\or April\or May\or
      June\or July\or August\or September\or October\or November\or December\fi
      \space\number\day, \number\year}
\def\<{\thinspace}

\def\cm{{\rm\thinspace cm}}
\def\erg{{\rm\thinspace erg}}

\def\ga{{\rm\thinspace gauss}}

\def\keV{{\rm\thinspace keV}}

\def\s{{\rm\thinspace s}}

\def\ergpcmsqps{\hbox{$\erg\cm^{-2}\s^{-1}\,$}}

\def\cha{{\it Chandra}}
\def\ae{{\small ACIS-EXTRACT}}

\def\rosat{{\it ROSAT}}

\def\xmm{{\it XMM-Newton}}

\def\wav{{\small WAVDETECT}}

\newcommand*{\thresh}{\mysub{P}{b} \ensuremath{< 10^{-3}}}

\makeatletter
\newcommand{\thickhline}{%
    \noalign {\ifnum 0=`}\fi \hrule height 1.2pt
    \futurelet \reserved@a \@xhline
}
\newcolumntype{"}{@{\hskip\tabcolsep\vrule width 1pt\hskip\tabcolsep}}
\makeatother

\def\cdfs{{\it CDFS}}

\bibpunct{(}{)}{;}{a}{}{,}

\title[To the Edge of the Perseus Cluster]
{Azimuthally Resolved X-Ray Spectroscopy to the Edge of the Perseus Cluster}

\author[O. Urban et al.]{O. Urban$^{1,2,3}$\thanks{E-mail: ondrej@stanford.edu}, 
A. Simionescu$^{1,2,4}$, N. Werner$^{1,2}$, S. W. Allen$^{1,2,3}$, S. Ehlert$^{1,2,3}$,\newauthor
I. Zhuravleva$^{1,2}$, R. G. Morris$^{1,2}$, A. C. Fabian$^{5}$, A. Mantz$^{6,7}$, P. E. J. Nulsen$^{8}$,\newauthor
J. S. Sanders$^{9}$, Y. Takei$^{4}$\\
$^{1}$Kavli Institute for Particle Astrophysics and Cosmology, Stanford University, 452 Lomita Mall, Stanford, CA 94305-4085, USA\\
$^{2}$Department of Physics, Stanford University, 382 Via Pueblo Mall, Stanford, CA 94305-4060, USA\\
$^{3}$SLAC National Accelerator Laboratory, 2575 Sand Hill Road, Menlo Park, CA 94025, USA\\
$^{4}$Institute of Space and Astronautical Science (ISAS), JAXA, 3-1-1 Yoshinodai, Chuo-ku, Sagamihara, Kanagawa 252-5210, Japan\\
$^{5}$Institute of Astronomy, Madingley Road, Cambridge CB3 0HA, UK\\
$^{6}$Kavli Institute for Cosmological Physics, University of Chicago, 5640 South Ellis Avenue, Chicago, IL 60637, USA\\
$^{7}$Department of Astronomy and Astrophysics, University of Chicago, 5640 South Ellis Avenue, Chicago, IL 60637-1433, USA\\
$^{8}$Harvard-Smithsonian Center for Astrophysics, 60 Garden Street, Cambridge, MA 02138, USA\\
$^{9}$Max-Planck-Institut f\"ur extraterrestrische Physik, Giessenbachstrasse 1, 85748 Garching, Germany\\
}

\begin{document}
\maketitle
\begin{abstract}
We present the results from extensive, new observations of the Perseus Cluster of galaxies, obtained as a \suzaku\ Key Project. The 85~pointings analyzed span eight azimuthal directions out to $2^\circ=2.6\,\text{Mpc}$, to
and beyond the virial radius $r_{200}\sim1.8\,\text{Mpc}$, offering the most detailed X-ray observation of the intracluster medium (ICM) at large radii in any cluster to date.
The azimuthally averaged density profile for $r>0.4r_{200}$ is relatively flat, with a best-fit power-law index $\delta=1.69\pm0.13$ significantly smaller than expected from numerical simulations. The entropy profile in the outskirts lies systematically below the power-law behavior expected 
from large-scale structure formation models which include only the heating associated with gravitational collapse. The pressure profile beyond $\sim0.6r_{200}$ shows an excess with respect to the best-fit model describing the SZ measurements for a sample of clusters observed with Planck. The inconsistency between the expected and measured density, entropy, and pressure profiles can be explained primarily by an overestimation of the density due to inhomogeneous gas distribution in the outskirts; there is no evidence for a bias in the temperature measurements within the virial radius. We find significant differences in thermodynamic properties of the ICM at large radii along the different arms. Along the cluster minor axis, we find a flattening of the entropy profiles outside $\sim0.6r_{200}$, while along the major axis, the entropy rises all the way to the outskirts. Correspondingly, the inferred gas clumping factor is typically larger along the minor than along the major axis.

\end{abstract}


\section{Introduction}

Approximately 90~per~cent of the baryonic mass in massive clusters of galaxies lies in the intracluster medium (ICM), a hot diffuse plasma that fills the clusters' gravitational potentials. The physical properties of the ICM,
including the temperature, density, and chemical composition, can be inferred from X-ray spectra. Until recently, however, the thermodynamic properties of the majority of the ICM ($\sim80$~per cent by volume, spanning radii
beyond 60\% of the virial radius\footnote{Here we define the virial radius $r_{\rm vir}=r_{200}$ within which the mean enclosed density is 200-times the critical density of the Universe at the redshift of the cluster.}) could
not be measured directly because of its low surface brightness and the high particle backgrounds of orbiting X-ray satellites.

Due to its relatively low particle background compared to flagship X-ray satellites like the \chandra\ X-ray observatory and \xmm, \suzaku\ has become the instrument of choice for studying cluster outskirts in
X-rays. The growing list of clusters studied out to their virial radii includes PKS0745-191 \citep{george2009,walker2012a}, A2204 \citep{reiprich2009}, A1795 \citep{bautz2009}, A1413 \citep{hoshino2010}, A1689
\citep{kawaharada2010}, A2142 \citep{akamatsu2011}, Hydra~A \citep{sato2012}, the Coma Cluster \citep{simionescu2013}, A2029 \citep{walker2012b}, the Centaurus Cluster \citep{walker2013}, and the Perseus Cluster \citep{simionescu2011}. The fossil group RX~J1159+5531 has 
been studied by \citet{humphrey2012}. The Virgo Cluster has been robustly studied to $r_{200}$ by \xmm\ \citep{urban2011}. For a comprehensive review, see \citet{reiprich2013}. 

A common finding of these studies is a deviation of the observed entropy profiles in the cluster outskirts from naive expectations. In a cluster formed by gravitational collapse, and in which no additional heating or cooling
occurs, the entropy is expected to follow a power-law with $K\sim r^{1.1}$ \citep{voit2005b}. However, the observed entropy profiles in clusters tend to flatten at large radii, differing from the expected value by a factor
of $2-2.5$ in the outskirts \citep{hoshino2010,simionescu2011,urban2011}. This phenomenon can in principle be attributed to several possibilities, including increased gas clumping at  large radii
\citep{simionescu2011,urban2011}, a hypothesis supported to some extent by simulations, e.g. \citet{nagai2011}. \citet{hoshino2010} and \citet{akamatsu2011} proposed that differences in the electron and ion temperatures
in the cluster outskirts, resulting from the relatively long  equilibration time at low densities \citep{ettori1998b}, can result in the temperatures being underestimated. At some level, both effects will be at play
\citep{akamatsu2011}.

\citet{lapi2010} and \citet{cavaliere2011} proposed the weakening of accretion shocks as the clusters get older to be the reason for the entropy profile flattening. As the accretion shock propagates outward, the gas
falls through a progressively smaller potential drop, thus reducing the entropy gain at the shock.

Radially and azimuthally resolved spectroscopy of galaxy cluster outskirts is challenging, and most measurements to date have been limited in spatial resolution to one data point between $r_{500}<r<r_{200}$.
Nonetheless, some \suzaku\ studies have been able to azimuthally resolve the spectroscopic properties of several nearby massive clusters. \citet{george2009} and \citet{walker2012a} analyzed four directions in PKS0745-191
($z=0.103$), finding no statistically significant azimuthal variation. \citet{bautz2009} observed Abell~1795 ($z=0.062$) along two directions, finding significant differences in the surface brightness in the
outskirts. \citet{kawaharada2010} defined four directions (offsets) in their observation of Abell~1689 ($z=0.183$). They measured an isotropic distribution of the physical properties, with the exception of the northeast
direction in the $r_{500}<r<r_{200}$ range, where the temperature and the entropy are significantly higher than in the other three directions. \citet{eckert2012} studied a sample of 31~clusters observed with \rosat\ to
infer that, beyond $\sim r_{500}$, galaxy clusters deviate significantly from spherical symmetry.

The nearby Perseus Cluster \citep[A~426, $z=0.0179$,][]{struble1999} is the brightest X-ray cluster in the sky. Because of this, it was one of the first and remains one of the best studied objects in X-rays
\citep[e.g.][]{gursky1971,forman1972,allen1992,ettori1998a,fabian2003,simionescu2011,fabian2011}. The cluster's X-ray emission is peaked on the optically brightest cluster galaxy (BCG) NGC~1275 (RA~3h19m48.1s,
DEC~$+41^\circ30'42''$). \citet{simionescu2011} used $260\,\text{ks}$ of \suzaku\ data to study the ICM of the Perseus Cluster out to the virial radius $r_{200}\sim1.8\,\text{Mpc}$ ($82'$, as inferred from these data)
along two directions towards the northwest and the east. With these data, they were able to resolve the radial interval between $r_{500}$ and $r_{200}$, for the first time, into independent radial bins. They found these two
arms of Perseus to be broadly consistent in the outskirts, but inside $\sim0.7\,\text{Mpc}$ the temperature profile along the eastern arm is systematically lower due to the presence of a cold front. Modeling the total mass,
they reported, based on the current \suzaku\ calibration, that the inferred gas mass fraction exceeds the cosmic mean in the outskirts. Assuming this excess to be due to the presence of gas density clumps, and correcting for
this, they obtained density, pressure, and entropy profiles consistent with the theoretical predictions for gravitational collapse. Recently, \citet{simionescu2012} used a combination of \rosat, \xmm\ and \suzaku\ observations
to find an evidence for large-scale sloshing motions of the ICM in the Perseus Cluster, the first time such motions have been observed on large scales. 

In this work we expand the effort of \citet{simionescu2011} adding six new arms of \suzaku\ data to the study of the Perseus Cluster, for a total of eight azimuthal directions covering the cluster out to beyond the virial radius.
The total \suzaku\ exposure time is $1.1\,\text{Ms}$. Results on the distribution of metals in the cluster derived from these data are discussed by~Werner et al. (submitted). The results on all other thermodynamic quantities
are reported here. In Sect.~\ref{sect:obsanal} we describe the reduction of the data and the extraction of science products. Sect.~\ref{sect:results} describes the main results of the analysis. In Sect.~\ref{sect:discus}, we
discuss the implications of our findings. Sect.~\ref{sect:concl} summarizes our results and conclusions. A discussion of the impact of systematic errors, including the influence of background fluctuations and stray light on the results, can be found in
the appendix.

Throughout the paper we adopt $\Lambda$CDM cosmology with $\Omega_{\rm m}=0.27$, $\Omega_\Lambda=0.73$ and $H_0=70\,\text{km}\,\text{s}^{-1}\,\text{Mpc}^{-1}$. At the redshift of the Perseus Cluster one arcminute corresponds to a physical scale of $21.8\,\text{kpc}$. Our 
reference value for the virial radius is $r_{200}=82'$ \citep{simionescu2011}. All errors quoted are 68~per~cent confidence unless otherwise stated. 

\section{Observations and Data Analysis}
\label{sect:obsanal}

\subsection{Reduction and Analysis of \suzaku\ Data}

A total of 85 \suzaku\ pointings targeting the Perseus Cluster were obtained as a Suzaku Key Project during AO~4--6. These pointings cover eight azimuthal arms from the center out to a radius of $\sim2^\circ$, with a nominal
exposure time of $130\,\text{ks}$ per arm. The data from the X-ray Imaging Spectrometers (XIS) 0, 1 and 3 were analyzed and are reported here.

The central pointing (obtained separately from the Key Project) and the eastern (E) and the northwestern (NW) arms, which each consist of 7~pointings, were observed in July and August~2009 (AO~4). These have been previously
discussed by \citet{simionescu2011}. We have reprocessed the data for these arms using new calibration products and an updated analysis of point sources and image artefacts, aiming for consistency in the analysis across all
8 arms. The western (W) and southern (S) arms consist of 11~pointings each, observed in August~2010 (AO~5). A change of observational strategy is responsible for the increased number of pointings. This was done in order to
expose the same area of the sky at large radii with two different parts of the detectors, and thus be able to better address systematic issues such as possible stray light contamination. The four remaining arms -- north (N),
northeast (NE), southeast (SE) and southwest (SW), consisting of 12 pointings each -- were observed in August and September~2011 (AO~6). The observations were carried out in ``diamond configuration'' with one of the detector
corners pointing towards the bright cluster center in order to minimize the stray light contamination.

\begin{figure*}
\centering
\includegraphics[width=.95\textwidth]{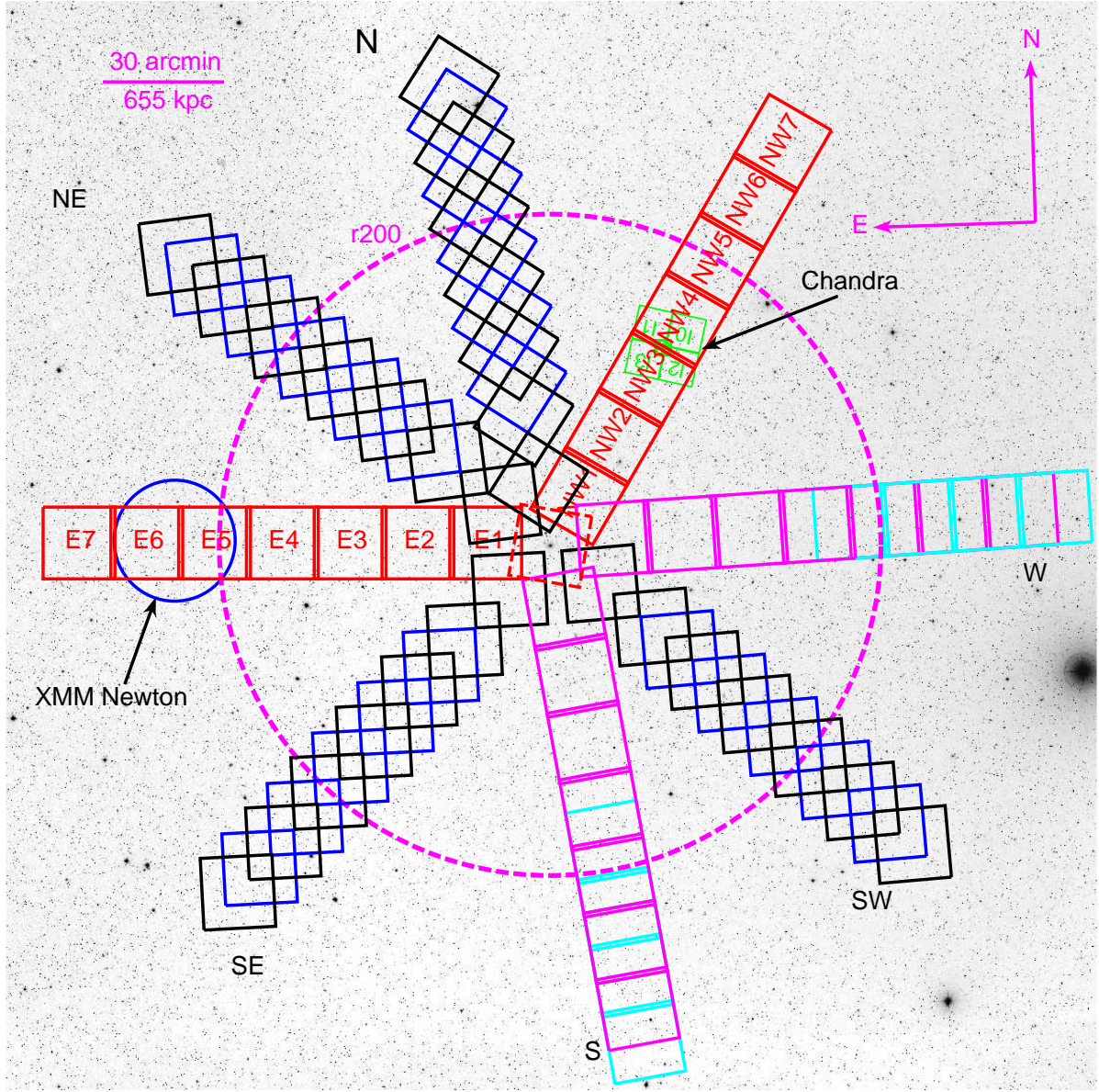}
\caption{Digitized Sky Survey (DSS) image of the Perseus Cluster region. Overplotted are the positions of the \suzaku\ pointings from AO~4 (\emph{red}), AO~5 (\emph{magenta} for integer and \emph{cyan} for half-integer
pointings, respectively) and AO~6 (\emph{black} and \emph{blue}). For clarity we individualy label only the AO~4 offset pointings. Observations with
corresponding indices from different AOs lie at the same distances from the cluster centre. In green we show the position of the \chandra\ pointings at large radius reported here (Sect.~\ref{subs:chandra}). The new
\xmm\ pointing at large radius (Sect.~\ref{subs:xmm}) is marked with a blue circle. The dashed circle shows the cluster's virial radius at $82'$. The brightest feature at the western edge of the DSS image is the star Algol
($\beta$ Per).}
\label{fig:pointingmap}
\end{figure*}

We label the pointings with integers starting at the offset pointing closest to the central one for each arm. In the case of overlapping pointings we use half-integer indexing (so that, for example, pointings W5 and NE5 have
the same distance from the cluster center and pointing N3.5 lies between and partly overlaps with N3 and N4). Fig.~\ref{fig:pointingmap} shows the layout of the pointings on the sky.

\subsubsection{Data Cleaning}
\label{subs:analysis}

Initial cleaned event lists were obtained using the standard screening criteria proposed by the XIS~team\footnote{Arida,~M., XIS Data Analysis, \\http://heasarc.gsfc.nasa.gov/docs/suzaku/analysis/abc/node9.html}. We also
checked for likely solar wind charge-exchange (SWCX) emission contamination using the WIND SWE (Solar Wind Experiment) data, following the analysis of \citet{fujimoto2007} and, in the case of affected pointings (N7, W7 and
W7.5), have used only data above $1.5\,\text{keV}$, where no SWCX lines are expected. We filtered out times of low geomagnetic cut-off rigidity (COR$\leq6\,\text{GV}$). Ray-tracing simulations of spatially uniform extended
emission were used to perform vignetting corrections \citep{ishisaki2007}. For the XIS~1 data obtained after the reported charge injection level increase on June~1st 2011, we have excluded two adjacent columns on either side
of the charge-injected columns (the standard is to exclude one column on either side).

\subsubsection{Imaging Analysis}
\label{subs:imaging_analysis}

\begin{figure*}
\centering
\includegraphics[width=0.95\textwidth]{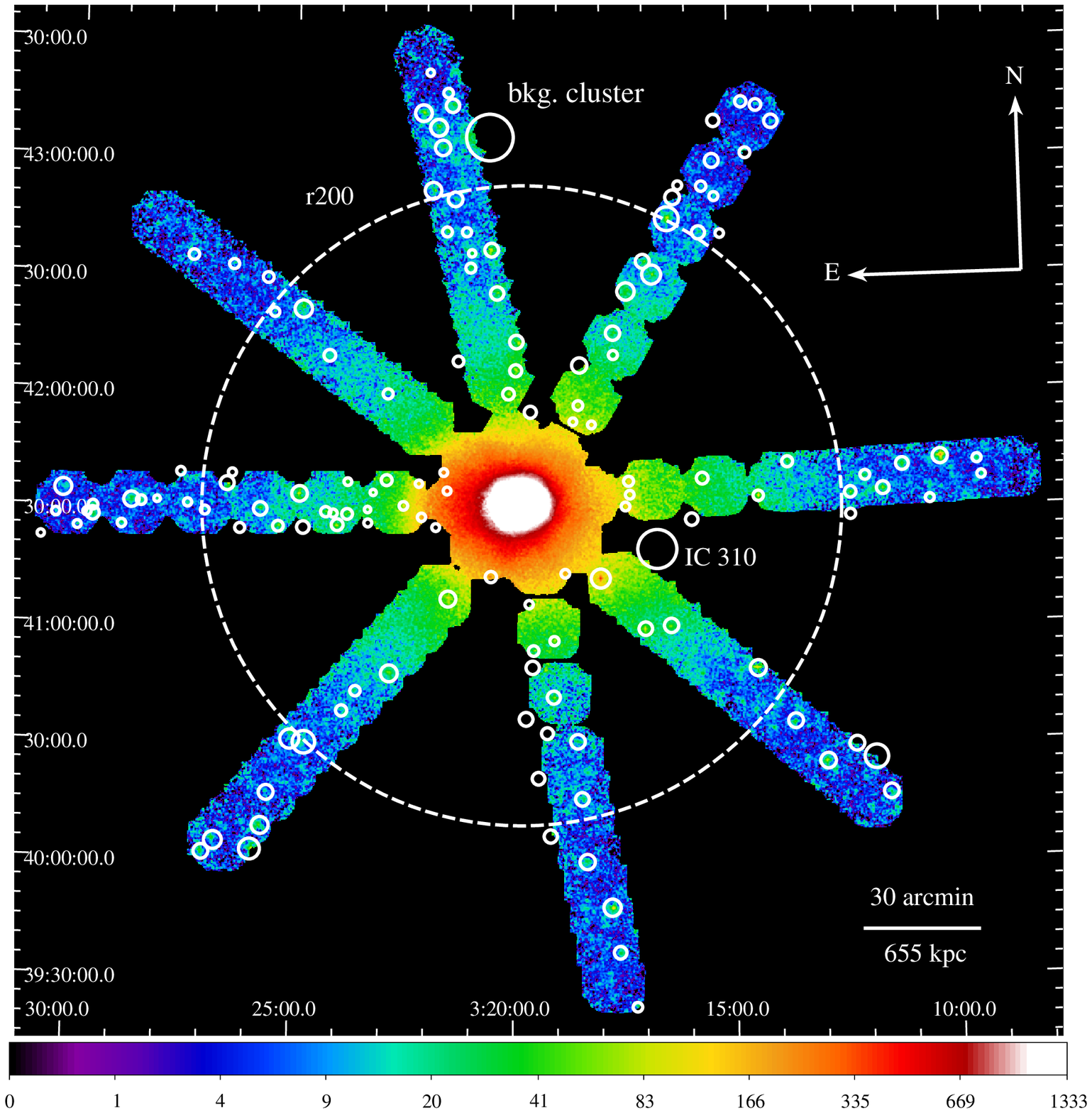}
\caption{Exposure- and vignetting-corrected mosaic of the Perseus observations in the $0.7-7.0\,\text{keV}$ range with the detector edge regions removed as described in the text. The white dashed circle centered on the BCG
NGC~1275 marks the nominal virial radius at $82'$. We highlight the locations of candidate point sources with white circles. The image has been smoothed with a Gaussian with width of $25''$, the color bar shows the surface
brightness in units of $10^{-6}$~counts/s/arcmin$^2$.}
\label{fig:octopus}
\end{figure*}

We extracted and co-added images from all three XIS~detectors in the $0.7-7.0\,\text{keV}$ energy band. Regions of $30''$ around the detector edges were removed, as were all pixels with an effective area less than 50~per cent
of the on-axis value. Instrumental background images in the same energy range were extracted from night Earth observations. These were subtracted from the cluster images before the vignetting correction was applied. The
resulting background- and vignetting-corrected mosaic image is shown in Fig.~\ref{fig:octopus}.

We have identified (and excluded from the subsequent analysis) a list of candidate point sources using the CIAO tool \texttt{wavdetect}. For this search we used a single wavelet radius of 1~arcmin (7.2~pixels),
which is approximately matched to the \suzaku\ half-power radius ($\sim1'$). In addition to the 140 candidate point sources identified in this way, we removed a $6'$~circle in an area of enhanced surface brightness in the N
arm \citep[associated with a background cluster; ][]{brunzendorf1999} and a $5'$~circle around the galaxy IC~310.

The flux of the dimmest identified candidate point source is $3\times10^{-14}\,\text{ergs}/\text{s}/\text{cm}^2$. Assuming this flux to be the detector sensitivity across the entire exposure in all the pointings, the number
of detected candidate point sources is consistent with the expected value obtained from the $\log{N}-\log{S}$ curve of the \chandra\ Deep Field South.

\subsubsection{Spectral Analysis}
\label{subs:spectral_analysis}

For each arm, we extracted a set of spectra from 17~annular regions centered on NGC~1275. These spectra exclude regions associated with the detector edges and candidate point sources. The spectra were rebinned to have at
least one count per channel. Night Earth observations were used to obtain the instrumental backgrounds. Appropriate response matrices were constructed for each region. Ancillary response files were calculated with
ray-tracing simulations using \texttt{xissimarfgen}, employing the detector contamination files from July~2012.

Spectral modelling was carried out using XSPEC \cite[version 12.7,][]{arnaud1996}, and the extended C-statistic estimator. We used the $0.7-7.0\,\text{keV}$ energy range for XIS0 and XIS3 and $0.6-7.0\,\text{keV}$ for XIS1.
The Galactic absorption was set individually for each pointing to the average column of the Leiden/Argentine/Bonn Survey calculated at the position of the given pointing \citep{kalberla2005}. We modeled the cluster emission
in each spectral region as single temperature plasma in collisional ionization equilibrium using the \texttt{apec} plasma code \citep{smith2001}. We adopt the solar abundance table from \citet{feldman1992}.

\subsubsection{Modeling the X-ray Foreground and Background}
\label{subs:cxfb}

Assuming, based on the results described below, an absence of significant cluster emission beyond a radius of $95'$ (or $2.1\,\text{Mpc}$), we have used the pointings indexed 6 and above (21~pointings in total) to determine
the X-ray foreground/background (CXFB). The N~arm was not used for this purpose due to the presence of a background group. In addition to \suzaku\ data, we also used spectra from the \rosat\ All Sky Survey
extracted from circular regions with a radius of 1~degree in the outer part of each arm to help constrain the low-energy spectral components.

Our background spectral model includes four components -- a power-law to model the unresolved point sources \citep[e.g.][]{deluca2004} and three thermal components modeling the local hot bubble
\citep[LHB,][]{1996A&A...305..308S}, the Galactic halo \citep[GH,][]{kuntz2000}, and a potential 0.6~keV foreground component, respectively. The power-law and the LHB components (which we term \emph{isotropic background
components}), as well as the temperatures of the GH and the 0.6~keV component, were assumed to be constant across the mosaic. The normalizations of the GH and 0.6~keV components were allowed to vary from arm to arm. Due to
the presence of the background group, model components for the N~arm were obtained by averaging the results for the neighboring NE and NW arms, weighted by their statistical errors. Due to their origin within the
Galaxy, all thermal background components had their redshifts set to zero and their metallicity set to Solar values. All CXFB components are absorbed by the Galactic column density, except for the LHB due to the proximity of
its origin to the Solar system. The fitted CXFB model components are listed in Tab.~\ref{tab:cxfb}.

\begin{table}
\setlength{\extrarowheight}{4pt}
\begin{center}
\caption{Summary of the CXFB spectral model components. The temperatures are given in~keV and fluxes in ergs/s/cm$^2$/arcmin$^2$ in the $0.7-7.0\,\text{keV}$ range.}
\begin{tabular}{l|cc}
\hline\hline
\multicolumn{3}{c}{Isotropic background components}\\
\hline
power-law index& \multicolumn{2}{c}{$1.52\pm0.02$}\\
power-law flux& \multicolumn{2}{c}{$(5.76\pm0.10)\times10^{-15}$}\\
$kT_{\text{LHB}}$&\multicolumn{2}{c}{$9.25_{-0.31}^{+0.30}\times10^{-2}$}\\
flux$_{\text{LHB}}$&\multicolumn{2}{c}{$(2.61\pm0.07)\times10^{-18}$}\\
$kT_{\text{GH}}$&\multicolumn{2}{c}{$0.138_{-0.013}^{+0.004}$}\\
$kT_{\text{0.6 keV}}$&\multicolumn{2}{c}{$0.632_{-0.020}^{+0.021}$}\\
\hline\hline
\multicolumn{3}{c}{Anisotropic background components}\\
\hline
   &flux$_{\text{GH}}$&flux$_{\text{0.6 keV}}$\\
\hline
E &$1.09_{-0.24}^{+0.56}\times10^{-16}$&$8.66_{-0.78}^{+0.78}\times10^{-16}$\\
NE&$1.72_{-0.34}^{+0.83}\times10^{-16}$&$6.46_{-1.02}^{+1.01}\times10^{-16}$\\
N &$1.69\times10^{-16}$&$5.01\times10^{-16}$\\
NW&$1.67_{-0.26}^{+0.74}\times10^{-16}$&$4.20_{-0.78}^{+0.74}\times10^{-16}$\\
W &$1.00_{-0.21}^{+0.46}\times10^{-16}$&$1.19_{-0.10}^{+0.10}\times10^{-15}$\\
SW&$1.08_{-0.21}^{+0.50}\times10^{-16}$&$7.59_{-0.88}^{+0.88}\times10^{-16}$\\
S &$9.86_{-1.75}^{+4.43}\times10^{-17}$&$6.12_{-0.69}^{+0.68}\times10^{-16}$\\
SE&$1.08_{-0.20}^{+0.47}\times10^{-16}$&$1.29_{-0.11}^{+0.11}\times10^{-15}$\\
\hline\hline
\end{tabular}
\label{tab:cxfb}
\end{center}
\end{table}

\subsection{Reduction and Analysis of \chandra\ Data}
\label{subs:chandra}

\chandra\ ACIS-I observations targeting the northwestern arm of the Perseus Cluster at a radius $r=57'$ ($0.7r_{200}$) were taken on 2011~November 5-11. The observations were performed in the VFAINT mode, which
improves the rejection of cosmic ray events. The total exposure time of the four overlapping pointings after cleaning is 144~ks. The standard level-1 event lists were reprocessed using the CIAO (version 4.4) software package. 
The images were produced following the procedures described in \citet{ehlert2013}. The analysis of the point source population detected by \chandra\ is described in detail in Sect.~\ref{chandraappendix}.

\section{Results}
\label{sect:results}

\subsection{\suzaku\ Surface Brightness Profiles}

\begin{figure*}
\centering
\includegraphics[width=0.95\textwidth]{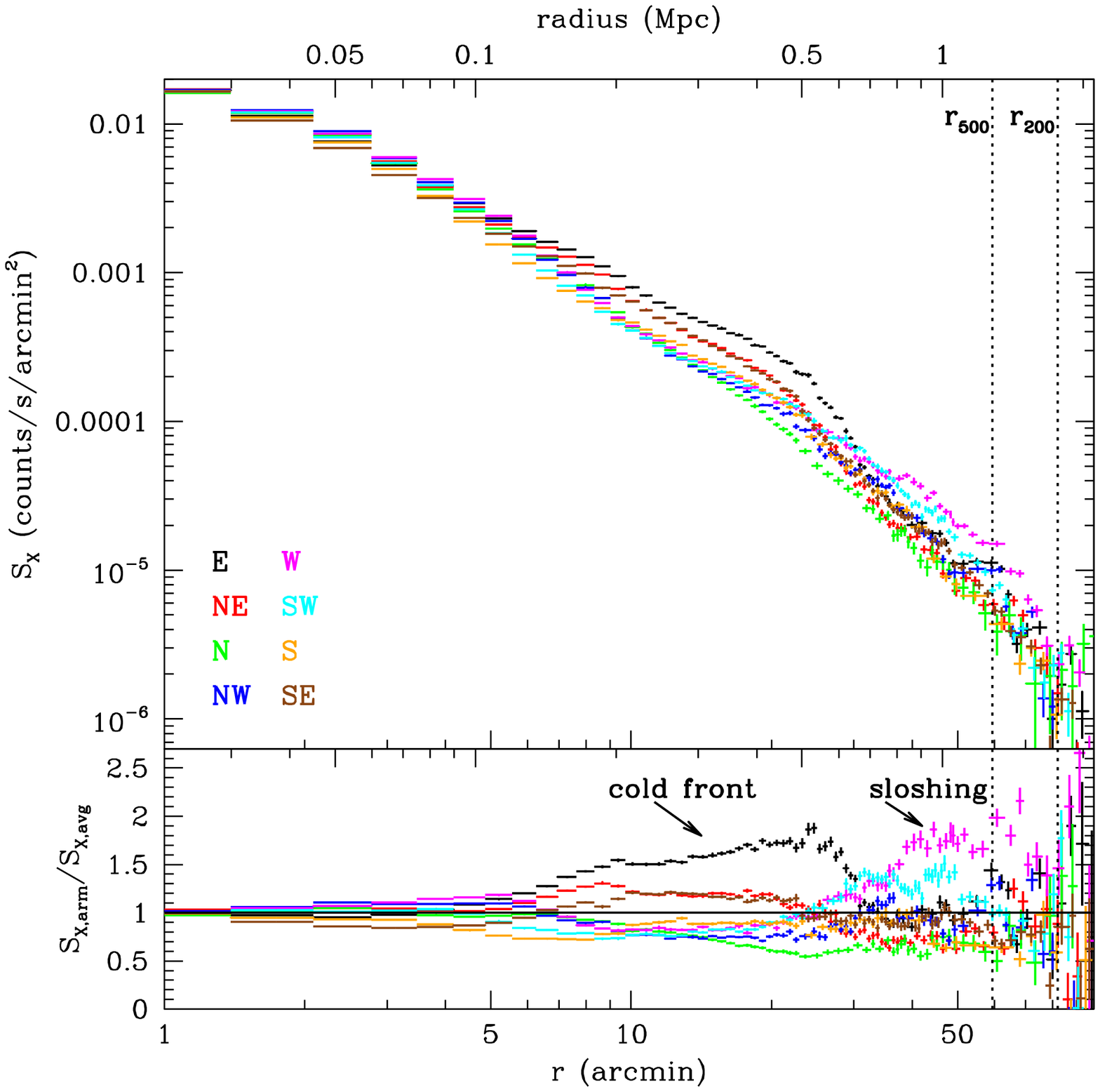}
\caption{The surface brightness profiles along the different arms corrected for CXFB  in the $0.7-7.0\,\text{keV}$ energy range (\emph{top}) and the residuals after dividing by the azimuthally averaged profile
(\emph{bottom}). The most prominent feature is the cold front along the E (black) arm inside $0.6\,\text{Mpc}$ that reaches into NE (red) and SE (brown) arms. Outside $\sim0.6\,\text{Mpc}$ ($27'$) the W and SW arms are the
brightest, revealing the presence of the large-scale spiral discussed in \citet{simionescu2012}. Positions of $r_{500}$ and $r_{200}$ are indicated with dotted lines.}
\label{fig:sbprof}
\end{figure*}

Fig.~\ref{fig:sbprof} shows the surface brightness profiles in the $0.7-7.0\,\text{keV}$ band along the eight Perseus arms after removing candidate point sources. In constructing the profiles, the CXFB surface brightness was
determined separately for each arm from the weighted average brightness measured beyond 2~Mpc. We subtracted this, assuming it to be constant across the arm, and added its error in quadrature to the statistical error in each
bin. Tab.~\ref{tab:sberr} lists the measured CXFB surface brightness for each arm. 

In order to better show the surface brightness features in the individual arms, we plot the ratio of the surface brightness in a given arm to the average surface brightness at a given radius in the bottom panel of
Fig.~\ref{fig:sbprof}.

\begin{table}
\begin{center}
\caption{\emph{Left column:} CXFB surface brightness for each of the arms including the statistical error obtained by weighed fitting of the image surface brightness outside 2~Mpc by a constant. \emph{Right column:} Power
law index describing the surface brightness profiles in the $0.75-2.0\,\text{Mpc}$ range.}
\begin{tabular}{l|cc}
\hline\hline
\multirow{2}{*}{arm}&$S_{X,\text{CXFB}}$&\multirow{2}{*}{$\alpha$}\\
&$\left(10^{-6}\,\text{cnts}/\text{s}/\text{arcmin}^2\right)$&\\
\hline
E &$5.14\pm0.34$&$2.87\pm0.17$\\
NE&$5.74\pm0.36$&$3.06\pm0.17$\\
N &$6.52\pm0.81$&$2.88\pm0.14$\\
NW&$7.37\pm0.25$&$2.74\pm0.14$\\
W &$6.51\pm0.33$&$2.88\pm0.12$\\
SW&$6.00\pm0.31$&$3.49\pm0.11$\\
S &$5.93\pm0.25$&$3.61\pm0.17$\\
SE&$6.18\pm0.18$&$3.43\pm0.13$\\
\hline\hline
\end{tabular}
\label{tab:sberr}
\end{center}
\end{table}

The cold front along the E arm between radii of $7'-30'$ ($0.15-0.7\,\text{Mpc}$), first discussed by \citet{simionescu2011}, is clearly evident in the profile. The effects of this cold front extend to the neighboring arms,
NE and SE, also enhancing their surface brightness at those radii. From $30'-60'$ ($0.6<r<1.2\,\text{Mpc}$) the SW and W arms display a brightness excess, showing evidence for the large-scale sloshing/swirling motion of the
ICM discussed by \citet{simionescu2012}.

The results of fitting the surface brightness data in the $34'-92'$ ($0.75<r<2.0\,\text{Mpc}$) range with a power-law model, excluding those regions influenced by the presence of the eastern cold front, are shown in
Tab.~\ref{tab:sberr}. The X-ray surface brightness profiles to the N, NW, E and W are significantly flatter than those toward the south. 

\subsection{Projected Spectral Results}

\begin{figure*}
\centering
\includegraphics[width=.95\textwidth]{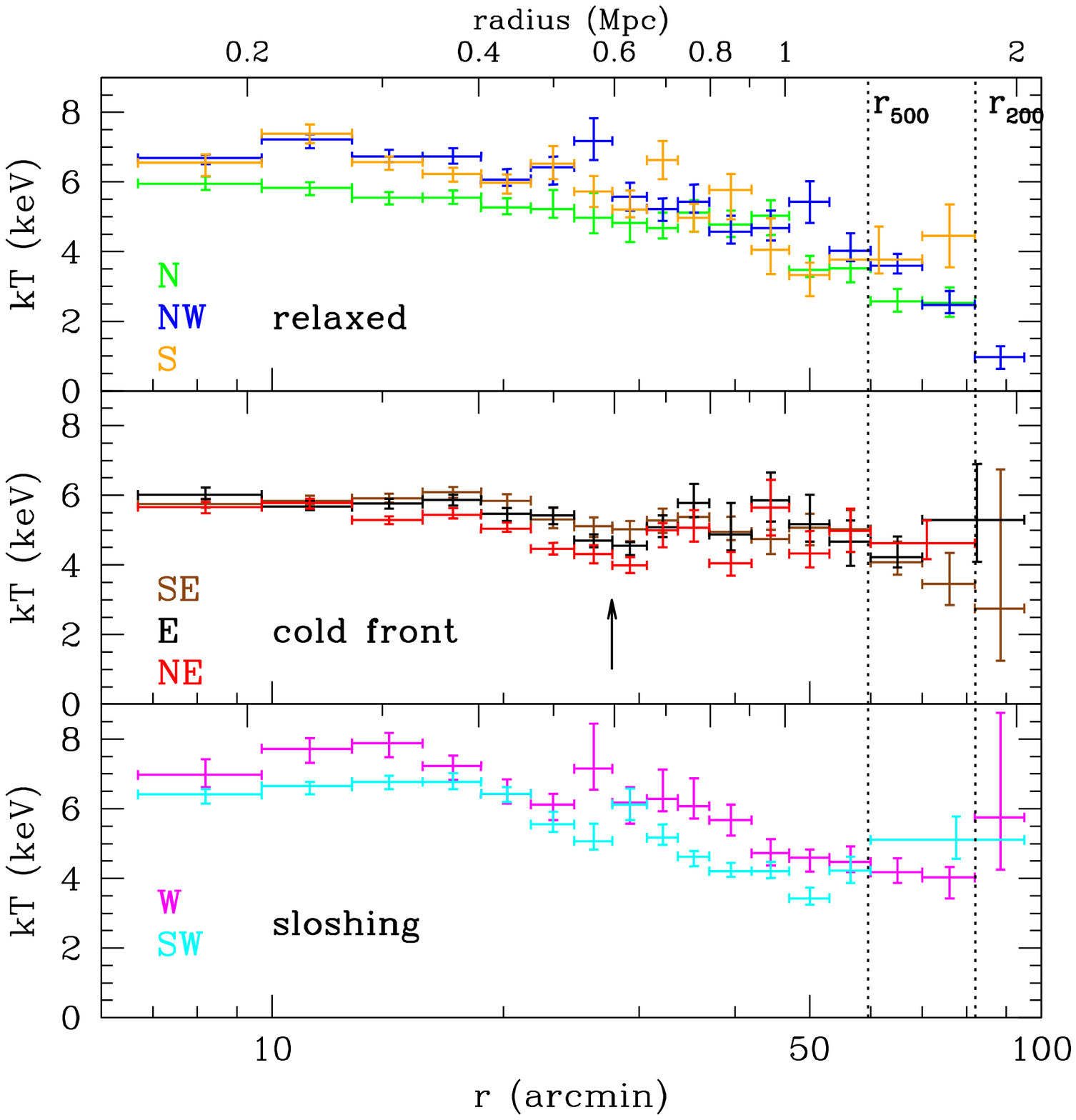}
\caption{Projected temperature profiles along the individual arms. Dotted lines mark the positions of $r_{500}$ and the virial radius $r_{200}$ at $82'=1.79\,\text{Mpc}$ \citep{simionescu2011} corresponding to the outer bound
of the last-but-one annulus. The last data point lies completely outside the virial radius. \emph{Top:} arms that appear the most relaxed, \emph{center:} arms influenced by the presence of the eastern cold front (indicated
by the black arrow), \emph{bottom:} arms exhibiting increased surface brightness and decreased temperature at large radii due to gas sloshing \citep{simionescu2013}.}
\label{fig:temp}
\end{figure*}

Temperature profiles measured from the projected spectra are shown in Fig.~\ref{fig:temp}. For clarity, all of the profiles are shown divided into three groups, according to the morphological similarities among the arms.

No major substructure, i.e. cold front or evidence for large scale sloshing \citep{simionescu2012}, is observed along the N, NW and S arms, and we will refer to these as the relaxed directions. Residual contamination from
the extended emission of a background group gives rise to a region of enhanced surface brightness beyond $r_{200}$ towards N, which influences the spectral results. Towards the NW the temperature quickly drops, reaching
values where emission lines appear - this allows us to tightly constrain the temperature in the last data point.

The SE, E and NE arms are all influenced by the presence of the cold front at $\sim30'$ (0.7~Mpc), as discussed for the E arm by \citet{simionescu2011}, evidenced by a dip in the respective temperature profiles at the same
radii where their surface brightness is boosted.

The W and SW arms show an increase of the surface brightness beyond the radius of the eastern cold front. \citet{simionescu2012} associated this brightness excess with large scale sloshing/swirling;
therefore, we group these two arms and refer to them as the ``sloshing'' arms.

\begin{figure*}
\centering
\includegraphics[width=.95\textwidth]{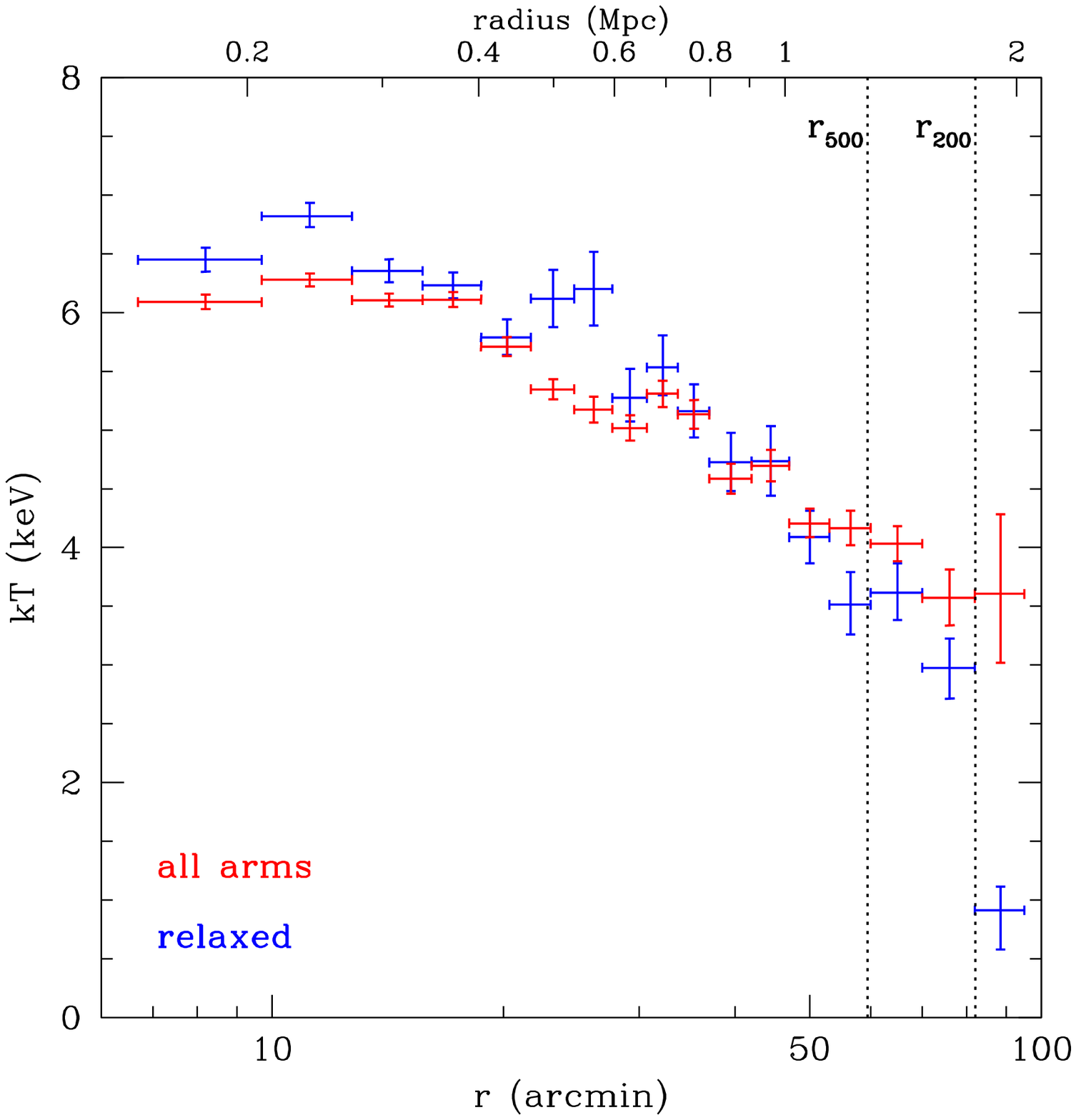}
\caption{Average profiles of projected temperature for the simultaneous fitting of all (\emph{blue}) and relaxed (\emph{red}) arms. In both cases, data for the N~arm beyond $r_{200}$ were excluded due to contaminating emission
from a background galaxy group.}
\label{fig:relaxed}
\end{figure*}

In order to reduce the impact of statistical uncertainties, we have also examined the results of simultaneously fitting spectra from multiple arms. These are shown in~Fig.~\ref{fig:relaxed}, in red the results using the three
relaxed (N+NW+S) and in blue using all eight arms. In both cases, we exclude N~spectra from the outermost annulus due to the contamination of the ICM emission by a background group. 

There are differences among the individual arms (and subsets of arms, as shown by Fig.~\ref{fig:relaxed}). Most notably, along the cluster's minor (north--south) axis, which roughly corresponds to the relaxed arms, the
temperature tends to drop towards the outskirts faster than along the major (east--west) axis.

In the appendix, we present a thorough analysis of the systematic uncertainties related to background fluctuations, point source exclusion, stray light, etc. We find our results to be generally robust to all the systematic tests that have been 
conducted.

Results from the observed metallicity distribution are reported elsewhere (Werner et al. 2013, submitted).

\subsection{Deprojected Spectral Results}

\subsubsection{Method for Obtaining the Profiles}

\begin{figure*}
\begin{minipage}{0.92\columnwidth}
\includegraphics[width=\textwidth]{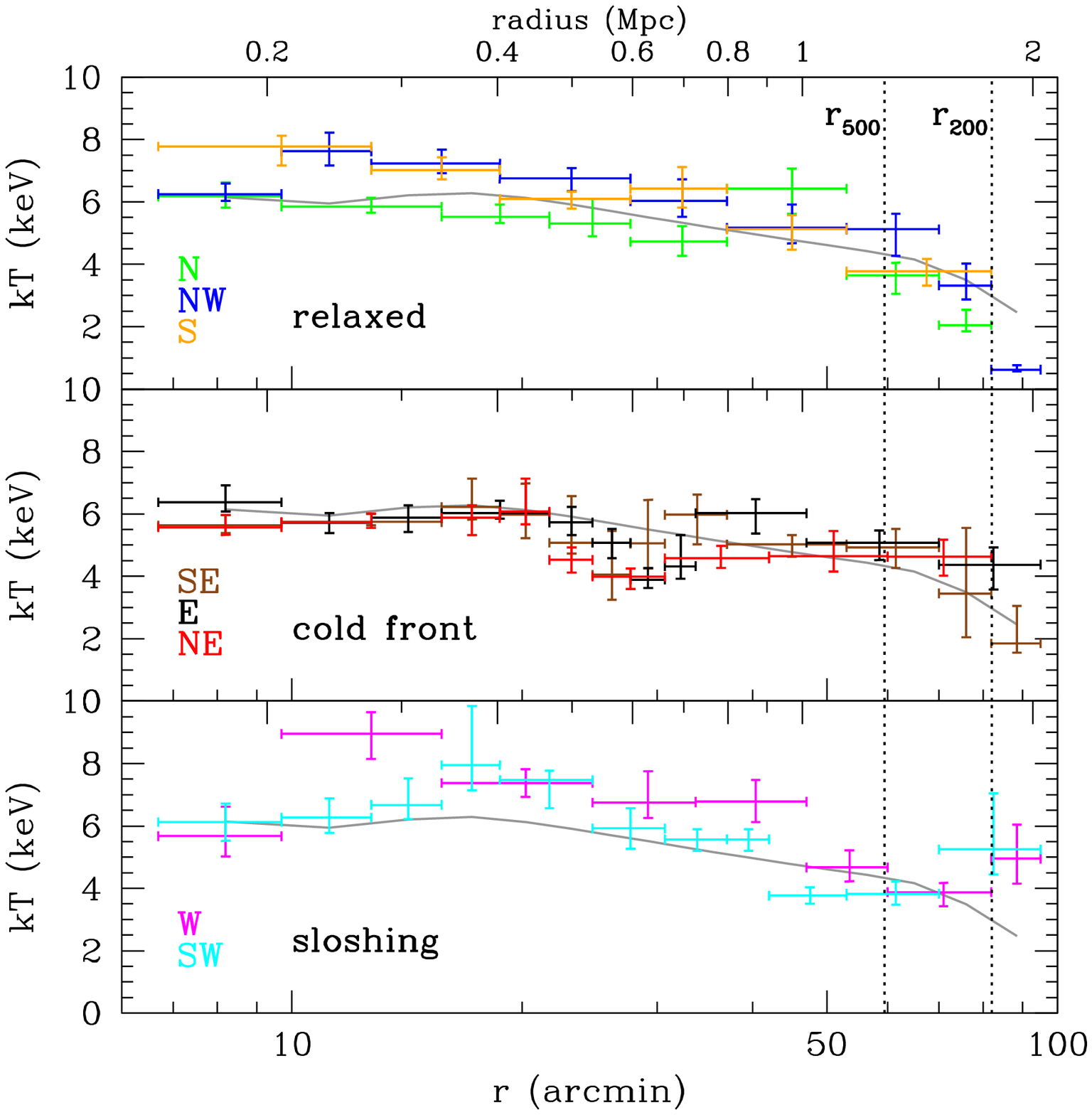}
\end{minipage}
\begin{minipage}{0.92\columnwidth}
\includegraphics[width=\textwidth]{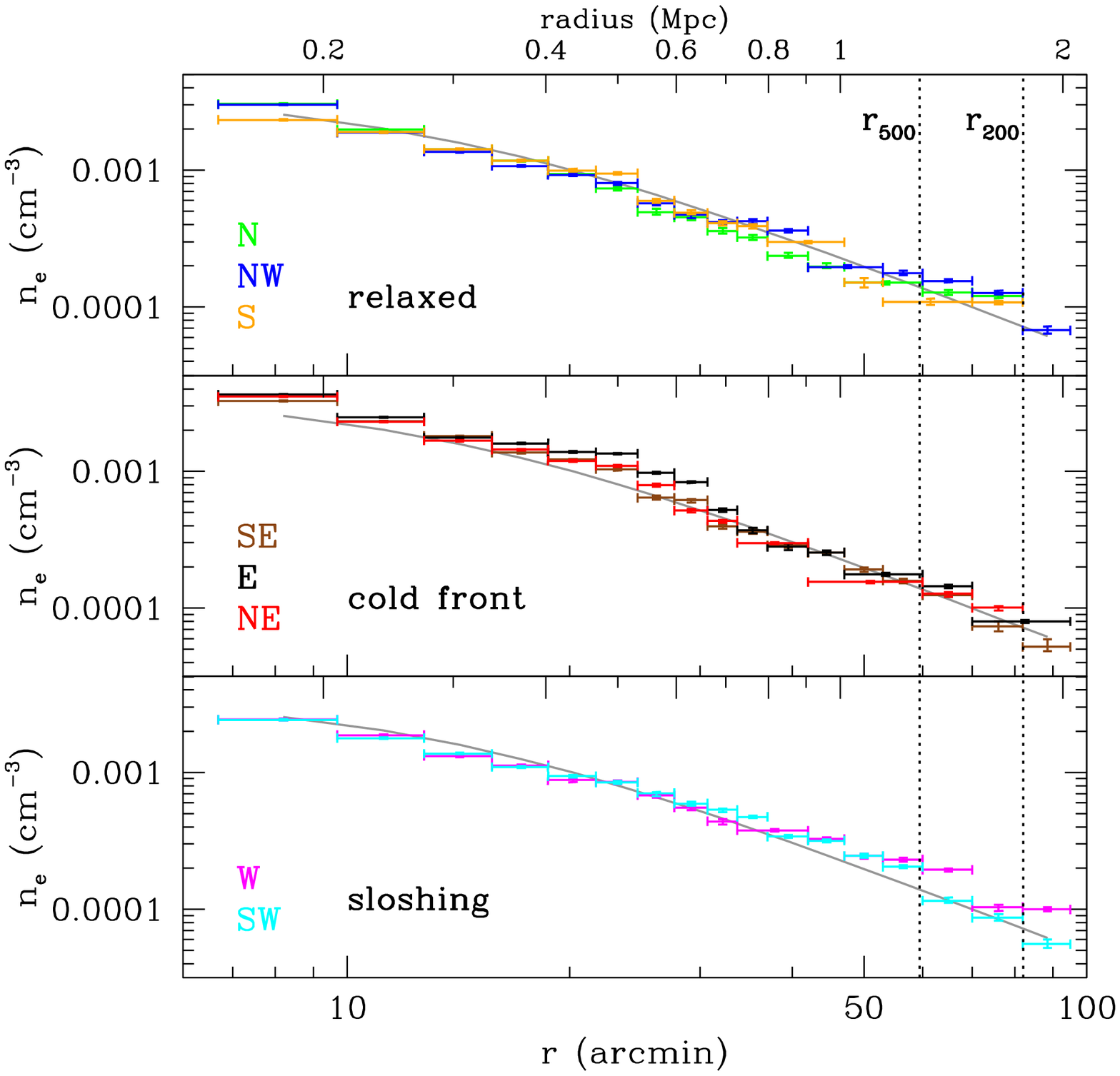}
\end{minipage}
\begin{minipage}{0.92\columnwidth}
\includegraphics[width=\textwidth]{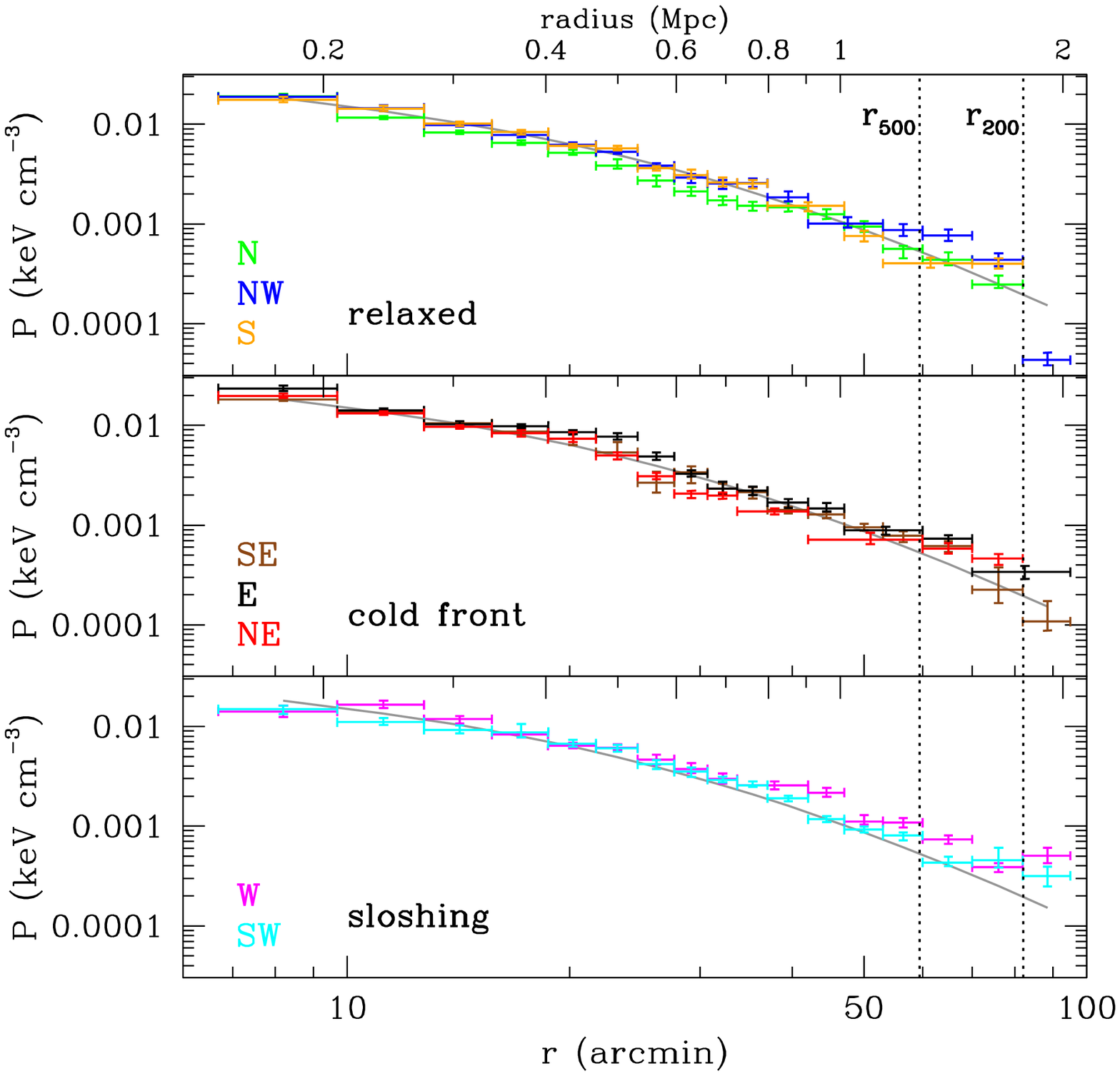}
\end{minipage}
\begin{minipage}{0.92\columnwidth}
\includegraphics[width=\textwidth]{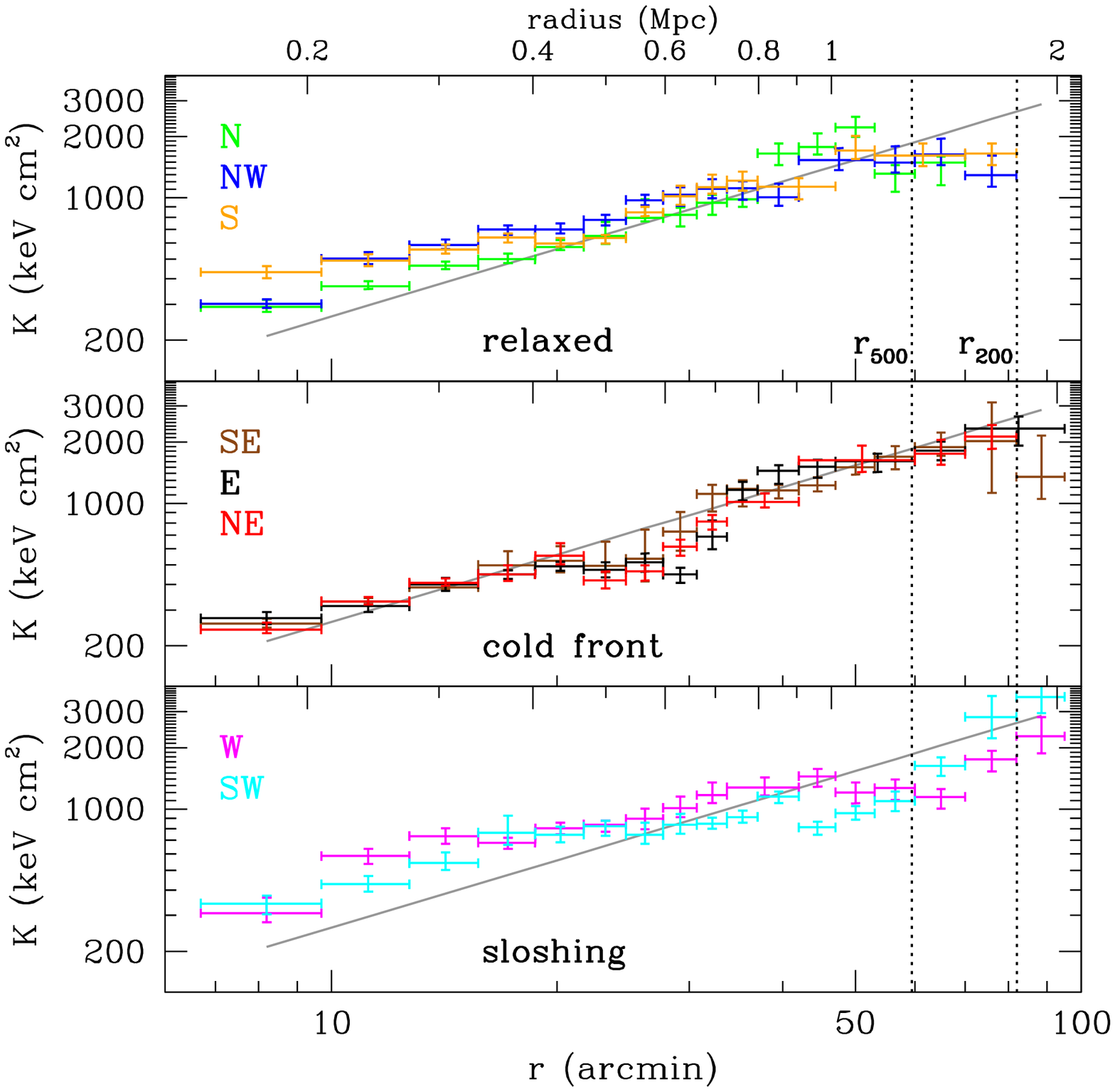}
\end{minipage}
\caption{Deprojected profiles. The arms are divided into the same groups as in Fig.~\ref{fig:temp}. Neighboring annuli in a given arm may be tied together to reduce `ringing' artifacts in the
deprojection. \emph{Top left panel:} the deprojected temperature profiles. The best fit of the temperature model of \citet{vikhlinin2006} to the average temperature profile is shown in grey.
\emph{Top right panel:} electron density profiles. The $\beta$-model fit to the azimuthaly averaged density profile is shown in grey. \emph{Bottom left panel:} pressure profiles. In grey we
overplot the \citet{planck2013} model. \emph{Bottom right panel:} entropy profiles. In grey we plot the baseline power-law relation $K\propto r^{1.1}$ of \citet{voit2005b} with the
normalization fixed to the expected value calculated following \citet[][see main text]{pratt2010}.}
\label{fig:deprojection}
\end{figure*}

Assuming spherical symmetry in the individual arms, we have carried out a deprojection analysis using the XSPEC model \texttt{projct}. For this analysis, we fixed the ICM metallicity to $Z=0.3Z_{\odot}$ everywhere
(this is in agreement with the measured projected $Z$ profile; Werner et al. 2013). We determined uncertainties in the derived properties using $\sim10^5$ Markov chain Monte Carlo simulation steps.
Fig.~\ref{fig:deprojection} shows the resulting profiles of deprojected temperature ($kT$), electron density ($n_{\rm e}$), pressure ($P=n_{\rm e}kT$), and entropy ($K=kT/n_{\rm e}^{2/3}$).

\begin{figure*}
\begin{minipage}{0.92\columnwidth}
\includegraphics[width=\textwidth]{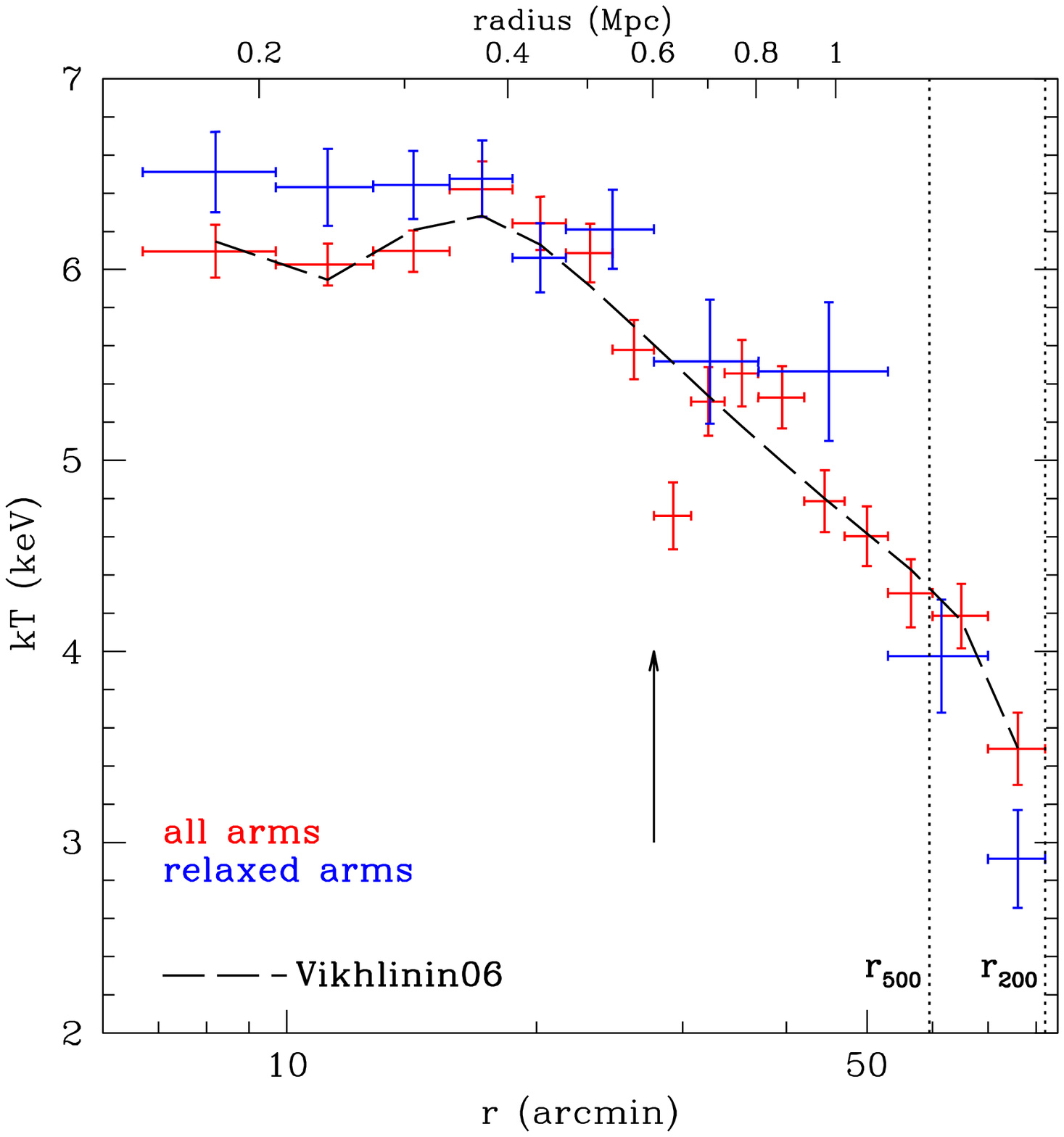}
\end{minipage}
\begin{minipage}{0.92\columnwidth}
\includegraphics[width=\textwidth]{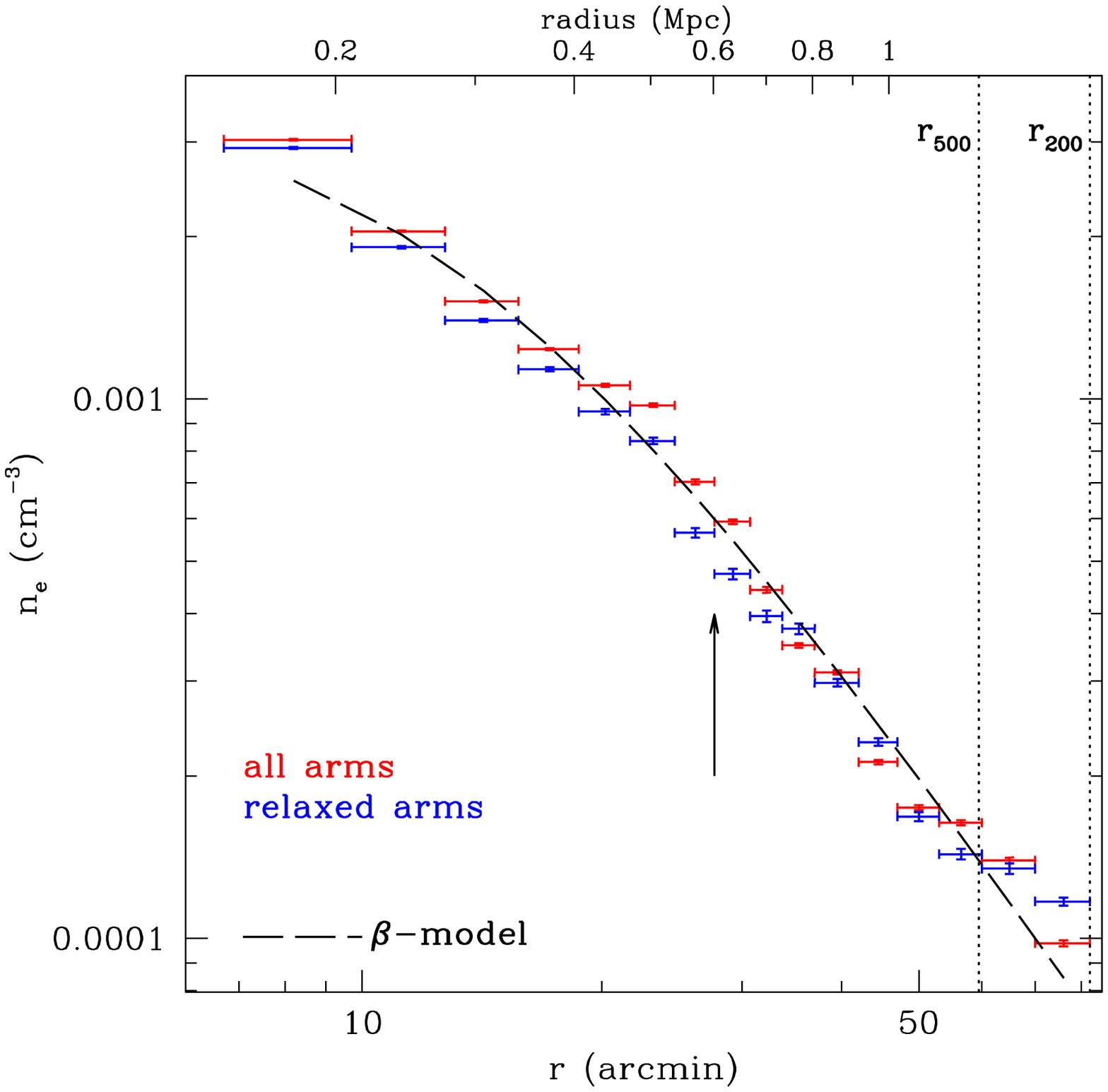}
\end{minipage}
\begin{minipage}{0.92\columnwidth}
\includegraphics[width=\textwidth]{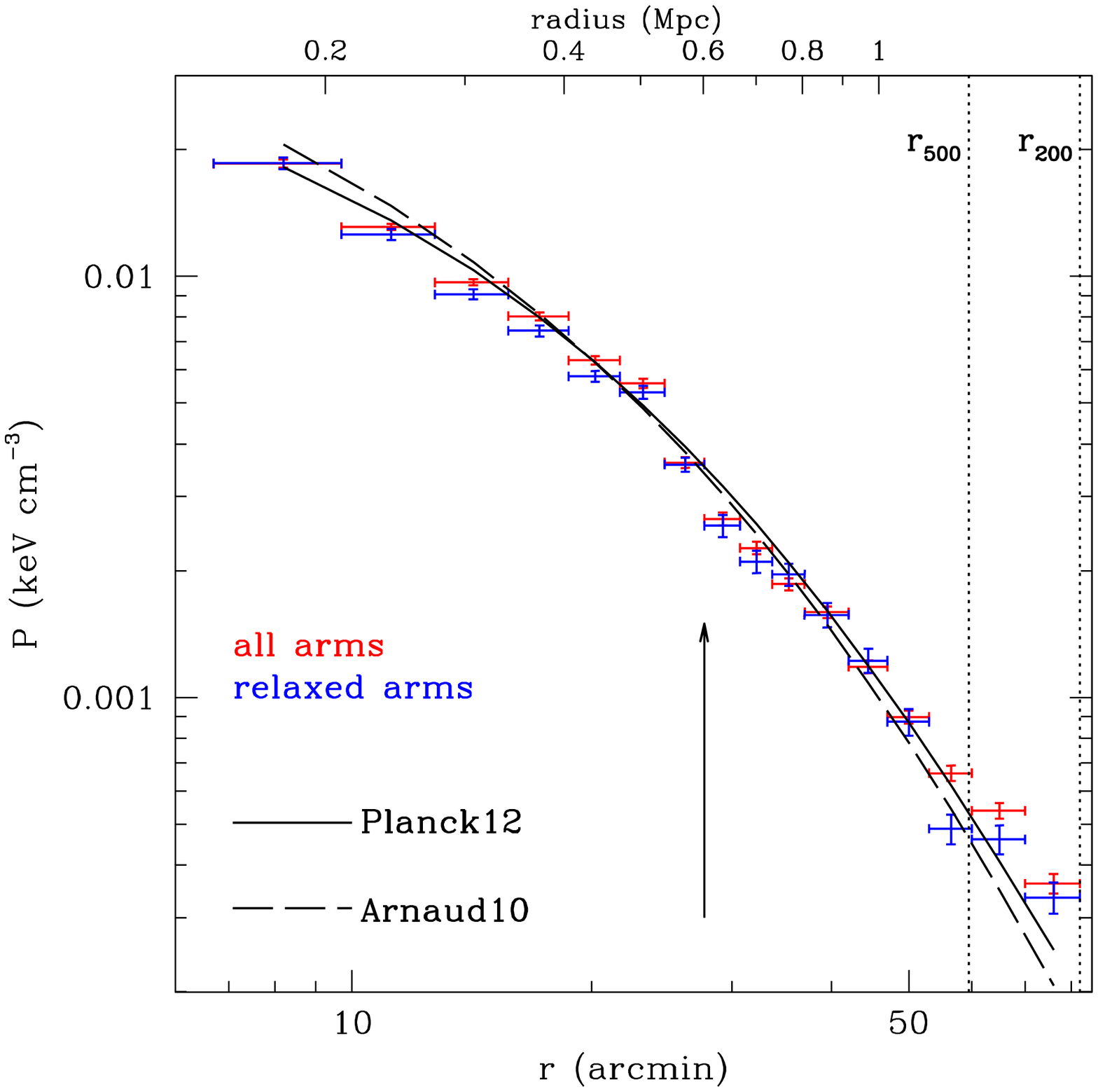}
\end{minipage}
\begin{minipage}{0.92\columnwidth}
\includegraphics[width=\textwidth]{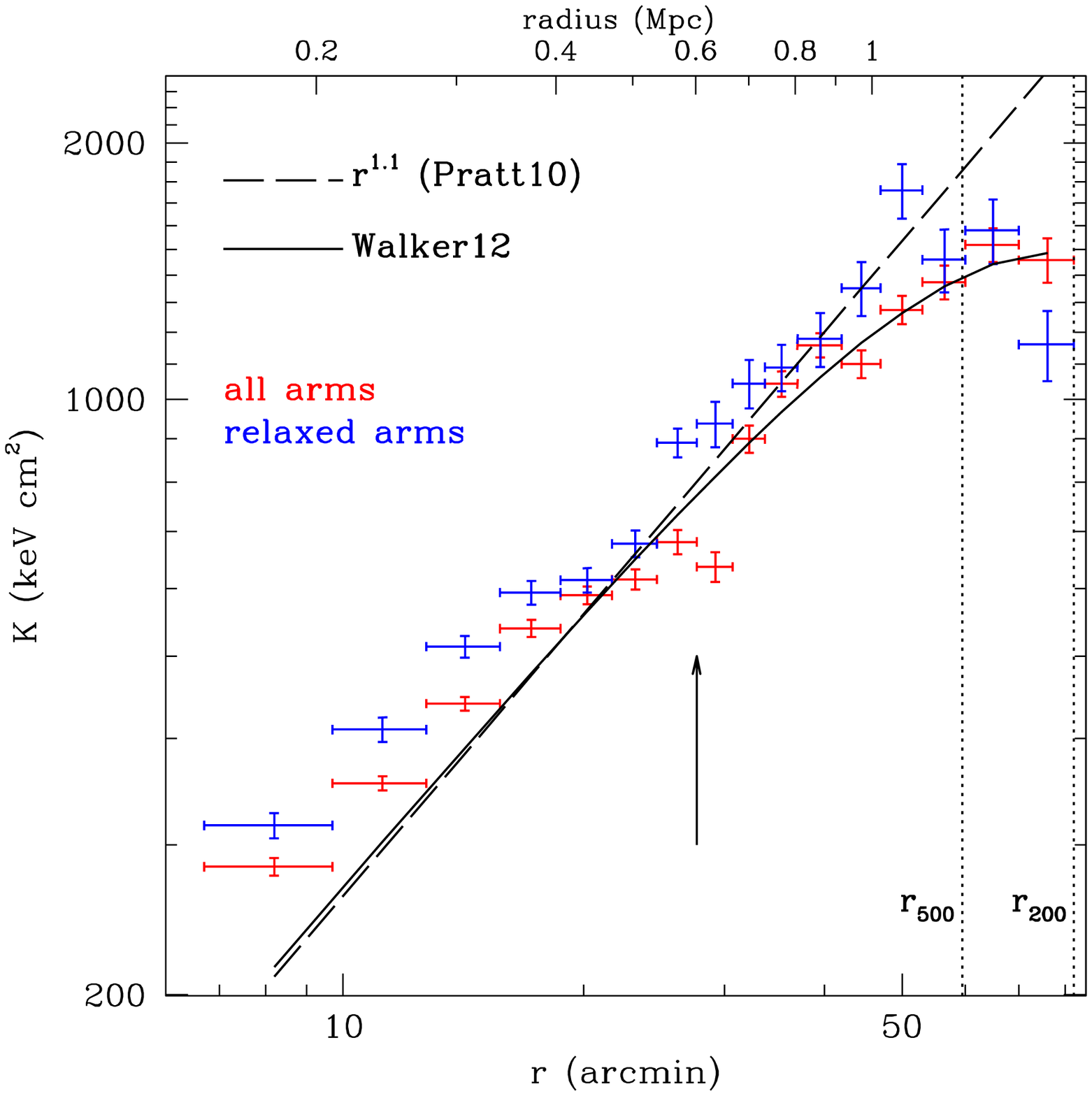}
\end{minipage}
\caption{Azimuthally averaged profiles of the ICM properties for all arms (\emph{red}) and the subset of relaxed arms (\emph{blue}). Black arrow indicates the position of the cold front to the east of the cluster center.
\emph{Top left:} Temperature profile with its best fit model from \citet{vikhlinin2006} shown as a dashed line. \emph{Top right:} Density profile with its best fit $\beta$-model (dashed line).  \emph{Bottom left:} Pressure
profile with overplotted best-fit theoretical models by \citet{arnaud2010} (dashed line) and \citet{planck2013} (solid line). The only free parameter in the fits was $r_{500}$. \emph{Bottom right:} Entropy profile.
Dashed line shows the $r^{1.1}$ power-law with normalization fixed to the value calculated according to \citet{pratt2010}. In solid black line we plot the best fit entropy profile by
\citet{walker2012c}.}
\label{fig:average}
\end{figure*}

Taking the weighed average of the deprojected results from all arms and for the subset of three relaxed arms (N, NW, S) we create the azimuthally-averaged profiles of temperature, density, pressure, and entropy shown
in~Fig.~\ref{fig:average}.

\subsubsection{Reference Model for the Temperature}

\citet[][ see also \citealt{allen2001}]{vikhlinin2006} introduce an analytical three-dimensional temperature model that extends to large radii in clusters

\begin{equation}
T_{\rm3D}(r)=T_0\frac{\left(\frac{r}{r_{\rm cool}}\right)^{a_{\rm c}}+\frac{T_{\rm min}}{T_0}}{1+\left(\frac{r}{r_{\rm cool}}\right)^{a_{\rm c}}}\frac{\left(\frac{r}{r_{\rm t}}\right)^{-a}}{\left[1+\left(\frac{r}{r_{\rm t}}\right)^b\right]^{c/b}}.
\label{eqn:viktemp}
\end{equation}
The model has eight free parameters that allow it to closely describe many features of a smooth temperature distribution. The best-fit parameters for the average profile of all arms and for the average of the subset of
relaxed arms (N, NW, S) are shown in Tab.~\ref{tab:temp}. We overplot the former profile in the top left panel of Fig.~\ref{fig:average}.

\begin{table}
\centering
\caption{Best fit parameters of the analytical three-dimensional temperature model from Eqn.~\ref{eqn:viktemp} for the average profile of two sets of arms.}
\begin{tabular}{l|cc}
\hline\hline
                           &all arms&relaxed arms\\
\hline
$T_0\,({\rm keV})$         &4.06    &5.30        \\
$T_{\rm min}/T_0$          &0.72    &0.88        \\
$r_{\rm cool}\,({\rm kpc})$&294     &261         \\
$a_{\rm c}$                      &6.72    &6.98        \\
$r_{\rm t}\,({\rm Mpc})$         &1.6     &1.1         \\
$a$                        &0.33    &0.18        \\
$b$                        &16.24   &271         \\
$c$                        &2.36    &1.29        \\
\hline\hline
\end{tabular}
\label{tab:temp}
\end{table}

\subsubsection{Reference Models for the Density}

We have fitted the density profiles along individual arms, as well as the azimuthally averaged profile and the average profile of the relaxed arms, with an isothermal $\beta$-model \citep{cavaliere1978},

\begin{equation}
n_{\rm e}=n_0\left[1+\left(\frac{r}{r_c}\right)^2\right]^{-3\beta/2}.
\end{equation}
We restrict our fits to radii $r>10'$ in order to avoid the cool core. The results are shown in Tab.~\ref{tab:density}. The $\beta$ parameter of the average profile ($\beta=0.71\pm0.05$) is in approximate agreement with
the canonical value for large clusters $\beta\sim2/3$. We obtain a slightly lower value when fitting the average of the relaxed arms ($\beta=0.61\pm0.04$). The best fit azimuthally averaged $\beta$-model profile is shown in
the top right panel of Fig.~\ref{fig:average}.

\begin{table}
\centering
\caption{Results of the analytical modeling of density. The last two rows show the parameters of the azimuthally averaged profile and the average profile of the relaxed arms. The last column contains the indices of the power
law fits to the individual arms beyond $34'$ ($0.7\,\text{Mpc}$) to avoid the influence of the cold front towards the E.}
\begin{tabular}{c|cc|c}
\hline\hline
\multirow{2}{*}{arm}&\multicolumn{2}{c|}{$\beta$-model}&index\\
                    &$r_c\left('\right)$&$\beta$&$r>0.7\,\text{Mpc}$\\
\hline
E &$20.14\pm3.94$&$0.91\pm0.12$&$-1.47\pm0.19$\\
NE&$15.85\pm3.63$&$0.86\pm0.12$&$-1.69\pm0.27$\\
N &$7.45\pm2.48$ &$0.59\pm0.04$&$-1.17\pm0.19$\\
NW&$7.92\pm2.05$ &$0.55\pm0.03$&$-1.64\pm0.23$\\
W &$8.47\pm2.29$ &$0.51\pm0.03$&$-1.56\pm0.15$\\
SW&$18.13\pm2.52$&$0.73\pm0.06$&$-2.13\pm0.16$\\
S &$12.80\pm2.68$&$0.66\pm0.06$&$-1.84\pm0.28$\\
SE&$14.50\pm1.84$&$0.79\pm0.06$&$-1.84\pm0.10$\\
\hline
average&$13.18\pm2.31$&$0.71\pm0.05$&$-1.69\pm0.13$\\
relaxed&$9.53\pm1.94$&$0.61\pm0.04$&$-1.55\pm0.16$\\
\hline\hline
\end{tabular}
\label{tab:density}
\end{table}

Fitting the profiles for the 8 arms independently, we notice significant differences between the best fit parameters. Fixing the core radius $r_c$ to the average profile value of $13.18'$, we measure a spread of $\beta$
values across the eight arms of $0.58<\beta<0.81$. 

Power-law modeling of density profiles of galaxy clusters, $n_{\rm e}\propto r^{-\delta}$, has proved popular due to its simplicity and also due to the fact that the $\beta$-model at radii $r\gg r_c$ behaves like a power-law.
We fit a power-law model separately to the density profile of each arm for $r>0.7\,\text{Mpc}$ ($r>0.4r_{200}$), thus avoiding the influence of the cold front towards the east. The fitted power-law indices are shown in
the last column of Tab.~\ref{tab:density} and show similar trends to the surface brightness slopes in Tab.~\ref{tab:sberr}, with the steepest gradients on the southern side of the cluster. 
We report a relatively flat azimuthally averaged density profile, falling off with radius with an index of $\delta=1.69\pm0.13$ outside $0.7\,\text{Mpc}$ (or $r>0.4r_{200}$), in agreement with the average slope previously reported by \citet{simionescu2011} for the average between only the E 
and NW arms. Fitting a power-law model to the average density profile for $r>0.5r_{200}$ results in a flattening, with $\delta=1.34\pm0.16$.

\subsubsection{Reference Models for the Pressure}

\citet{nagai2007} propose a generalized pressure profile of the form

\begin{equation}
\frac{P(r)}{P_{500}}=\frac{P_0}{\left(c_{500}x\right)^\gamma\left[1+\left(c_{500}x\right)^\alpha\right]^{(\beta-\gamma)/\alpha}},
\label{eqn:pressure}
\end{equation}
where $P_{500}=1.45\times10^{-11}\,\text{erg}\,\text{cm}^{-3}\left(\frac{M_{500}}{10^{15}h^{-1}\,M_{\odot}}\right)^{2/3}E(z)^{8/3}$, $x=r/r_{500}$, $c_{500}$ is the concentration parameter defined at $r_{500}$, and the
indices $\alpha$, $\beta$ and $\gamma$ are the profile slopes in the intermediate, outer and central regions, respectively. $E(z)=\sqrt{\Omega_{\rm m}(1+z)^3+\Omega_\Lambda}$ is the ratio of the Hubble constant at redshift $z$ with
its present value and $M_{500}$ is the total cluster mass enclosed within $r_{500}$.

Using a set of 33~local ($z<0.2$) \xmm\ clusters with data extending to $r<0.6r_{200}$, \citet{arnaud2010} find the best fitting parameters to be
$\left[P_0,c_{500},\alpha,\beta,\gamma\right]_{\text{Arnaud}}=\left[8.403h_{70}^{-3/2},1.177,1.0510,5.4905,0.3081\right]$. Recently, \citet{planck2013} studied the pressure profiles of 62 \emph{Planck} clusters between
$0.02r_{500}<r<3r_{500}$, finding the best fit set of parameters $\left[P_0,c_{500},\alpha,\beta,\gamma\right]_{\text{Planck}}=\left[6.41,1.81,1.33,4.13,0.31\right]$.

Shown in the bottom left panel of Fig.~\ref{fig:average}, we have fitted the average Perseus pressure profile with the generalized pressure model for $r<r_{200}$, leaving $r_{500}$ as the only free parameter (while expressing $M_{500}$ as a function of $r_{500}$ self-consistently) and 
fixing
the other parameters to the
two sets of values mentioned above. The resulting values for $r_{500}$ are in agreement with each other: $r_{500}^{\rm Planck}=59.7'\pm0.4'$ and $r_{500}^{\rm Arnaud}=59.3'\pm0.5'$. Overall, the model is a reasonable fit. 
However, in the cluster outskirts ($0.65-1.0r_{200}$) the \suzaku\ data lie above the model, before dropping below in the last annulus (not shown in Fig.~\ref{fig:average}). 

We fit the \citet{planck2013} pressure profile in the same way to the individual arms and find $r_{500}$ to be $\sim10\%$ larger along the cluster major (east--west) as compared to the minor (north--south) axis.

We note that \citet{simionescu2011} measured $r_{500}=54$~arcmin for the NW arm, which is approximately 10\% lower than the best fit average value for $r_{500}$ using Eqn. \ref{eqn:pressure}. This difference could be attributed to asymmetries intrinsic to the cluster itself (minor axis vs. 
major axis). This uncertainty in measuring $r_{500}$ implies that the value of $r_{200}$ may also be up to 10\% larger than the value adopted throughout this paper, which is based on the mass model of \citet{simionescu2011}.

\subsubsection{Reference Models for the Entropy}

In a cluster formed by gravitational collapse, and in which no additional heating or cooling occurs, the entropy is expected to follow a power-law of the form
\begin{equation}
\frac{K}{K_{500}}=1.47\left(\frac{r}{r_{500}}\right)^{1.1},
\label{eqn:pratt}
\end{equation}
where $K_{500}=106\,\text{keV}\,\text{cm}^2\left(\frac{M_{500}}{10^{14}M_{\odot}}\right)^{2/3}\left(\frac1{f_b}\right)^{2/3}E(z)^{-2/3}$ \citep{voit2005b,pratt2010}. We assumed $f_b=0.15$ and used our best-fit value $r_{500}=59.6'$ in calculating $K_{500}$. 

As shown in Fig.~\ref{fig:deprojection}, in each of the relaxed directions, we observe a flattening of the entropy profile with respect to this expected power-law beyond $\sim1\,\text{Mpc}$ $(\sim0.6r_{200})$.

For the E, SE and NE arms, the characteristic dip in the entropy profiles between radii of 20 and 35~arcmin is jointly caused by increased density and low temperature associated with the eastern cold front. The entropy beyond the cold front ($r>34'$) in these arms increases steadily with 
radius (with the exception of the SE direction, where we observe a drop in entropy beyond $r_{200}$). 

The SW and W arms are affected by sloshing at large radii, which results in an excess density and surface brightness. This leads the entropy to flatten in these arms (forming a dip in the SW case)  at the radii corresponding to the large-scale sloshing; the entropy increases again further out, 
rising steeper than $r^{1.1}$ in both arms.

The bottom right panel of Fig.~\ref{fig:average} shows the average entropy profile for the Perseus Cluster. 
We find an excess of entropy in the cluster center ($r<22'$) with respect to the power-law model in Eqn \ref{eqn:pratt}. This is consistent with the presence of excess ICM heating in the cluster center as favoured by cosmological studies \citep[e.g.][]{voit2005a,cavagnolo2009,mantz2010b}. 
The average profile diverges downwards from a power-law shape at $\sim45'\approx0.55r_{200}$. The ratio of the expected and measured value of the entropy is $\sim1.4$ at the virial radius. 

For comparison, we fitted the normalization of a power-law profile, with its index fixed to 1.1, to the weighed average of the three relaxed arms between $0.2r_{200}<r<0.6r_{200}$, avoiding both the
cool cluster core and the cluster outskirts, as well as the cold front and the large-scale sloshing, present in the other arms. The best-fit normalization was $\sim7\%$ higher than the value calculated using
Eqn.~\ref{eqn:pratt}. Taking the average profile of all the eight arms in the $16.5'<r<82'\,\left(0.2r_{200}<r<1.0r_{200}\right)$ range, we also fitted it with a power-law with both the normalization and the power-law index
as free parameters, obtaining an index of $k=0.81\pm0.06$, significantly flatter than the theoretical value of $k=1.1$.

Using a set of the clusters observed with modest spatial resolution out to the virial radius and beyond, \citet{walker2012c} find, that the entropy profile in the $r>0.2r_{200}$ range can be fitted with an analytical
function

\begin{equation}
K/K\left(0.3r_{200}\right)=A\left(r/r_{200}\right)^{1.1}\exp\left[-\left(r/Br_{200}\right)^2\right],
\label{eqn:entropy1}
\end{equation}
with $\left(A,B\right)=\left(4.4_{-0.1}^{+0.3},1.00_{-0.06}^{+0.03}\right)$. 

We fit the profile from Eqn.~\ref{eqn:entropy1} to the entropy averaged over all 8~arms, and to the subsample of three relaxed arms, with the resulting parameters shown in~Tab.~\ref{tab:entr}. 

\begin{table}
\centering
\setlength{\extrarowheight}{4pt}
\caption{Best fit parameters of the \citet{walker2012c} entropy profile to the averaged entropy distribution for all 8 directions and for the 3 relaxed arms.}
\label{tab:entr}
\begin{tabular}{l|cc}
\hline\hline
                      & A & B               \\
\hline
\citet{walker2012c} &$4.4_{-0.1}^{+0.3}$&$1.0_{-0.06}^{+0.03}$\\
Perseus average&$4.44\pm0.18$&$1.28\pm0.16$\\
relaxed arms&$4.75\pm0.16$&$1.06\pm0.09$\\
\hline\hline		    
\end{tabular}
\end{table}

\subsection{Analysis of the Surface Brightness Fluctuations}

\begin{figure}
\centering
\includegraphics[width=.92\columnwidth]{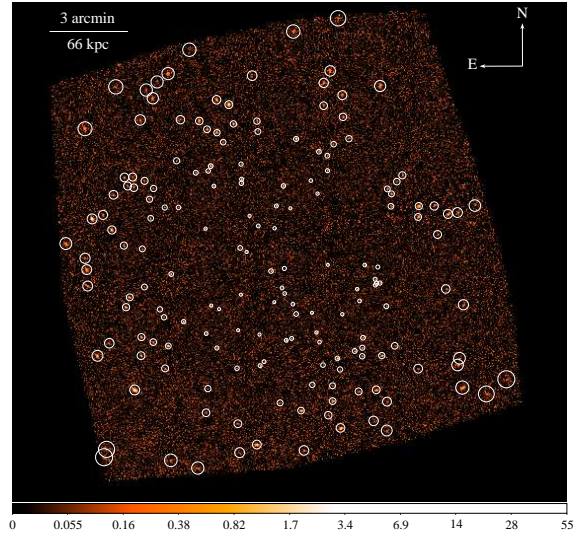}
\caption{Background subtracted and flat-fielded \chandra\ image of the Perseus Cluster outskirts towards the NW. We mark the identified point sources by white circles.}
\label{fig:chandra}
\end{figure}

In order to search for possible bright gas clumps in the cluster outskirts, we have analyzed the surface brightness (SB) fluctuations in the \chandra\ observation targeting the region of the NW arm at $r\sim 0.7r_{200}$, as well as
in the whole \suzaku\ mosaic of the Perseus Cluster. We have measured the power spectrum of fluctuations using the modified $\Delta-$variance method \citep{arevalo2012}, which accounts for gaps in the data due to excluded point sources.

Fig.~\ref{fig:chandra} shows the region of the NW arm at $r\sim0.7r_{200}$ observed with \chandra, and the identified point sources. No obvious extended sources were seen in the image. We determined that the population
of the point sources detected in this region is consistent with that expected from field surveys such as the \chandra\ Deep Field South (see Sect.~\ref{chandraappendix}). The global structure due to the
smooth ICM emission was removed by dividing the image by the best-fit $\beta-$model centered on the core of the Perseus Cluster. 

The power spectrum of the surface brightness fluctuations reveals a slight increase in the 2D power
above the Poisson noise level, which could be either due to the clumpiness of the gas on scales $>20\,\text{kpc}$ or due to unresolved point sources. We estimate the contribution of the faint, unresolved point sources
from the $\log N-\log S$ distribution of resolved point sources \citep[see also][]{churazov2012}, which is reasonably well approximated by the power-law $\displaystyle N(>S)\propto S^{-1.4}$.  The contribution of the faint
point sources to the power spectrum is $\displaystyle P_{\rm unres}\propto \int_0^{S_1} \frac{dN}{dS}S^2dS \propto S_1^{0.6}$, while the contribution of the resolved sources is $\displaystyle P_{\rm resol}\propto \int_{S_1}^{S_2}
\frac{dN}{dS}S^2 dS\propto S_2^{0.6}-S_1^{0.6}$, where $S_1$ and $S_2$ are the smallest and highest detected point source fluxes, respectively. In our case, $S_1\sim 20$ counts and $S_2 \sim 400$ counts. Therefore, we expect
the contribution of the unresolved point source population to our measured power spectrum to be at the level of $\sim20$~per cent relative to the contribution of the resolved bright sources. The power spectrum due to
the unresolved point sources can thus be calculated as $0.2\times(P_{\rm with\, src} -P_{\rm no\,src})$, where $P_{\rm with\, src}$ and $P_{\rm no\, src}$ are the power spectra with and without the resolved sources,
respectively. After the subtraction of the Poisson noise and of the expected power spectrum of the unresolved point sources, the measured 2D power spectrum of the surface brightness fluctuations is consistent with zero. We
therefore conclude, that the slight increase of power above the Poisson noise in the \chandra\ data can be plausibly associated with unresolved point sources. We note, that even if clumping is present in the gas at these
radii ($r\sim0.7r_{200}$), the signal may be difficult to detect because of smearing due to the long line of sight that we probe in the cluster outskirts. The clumps could also simply be small and faint and therefore unresolved by \chandra. 

The study of SB fluctuations with the \suzaku\ satellite is more complicated. Our analysis of the PSF of the \suzaku\ mirrors shows that the power spectrum of SB fluctuations on spatial scales smaller than $\sim120\,\text{kpc}$ is suppressed by a factor of 4 due to the convolution with the 
PSF. To understand the impact of faint point sources, which remain unresolved in our \suzaku\ data, we have compared the SB fluctuations of the \chandra\ observation after excluding only the point sources detected by \suzaku\ with the level of fluctuations after excluding all the \chandra\ 
point sources. We have found that, at a distance of $40-60\,\text{arcmin}$ from the cluster core, corresponding to the location of the deep \chandra\ pointing, the SB fluctuations on scales smaller than $\sim350\,\text{kpc}$ are dominated by the population of point sources unresolved by 
\suzaku. We therefore conclude that \suzaku\ observations of the Perseus Cluster cannot be used to study SB fluctuations on scales smaller than $\sim350\,\text{kpc}$, which is comparable to the field of view of a single \suzaku\ pointing. Structure on larger scales is seen in the form of the 
large-scale sloshing described in \citet{simionescu2012} and throughout this paper.

\section{Discussion}
\label{sect:discus}

The density, entropy, and pressure profiles in the outskirts of the Perseus Cluster show interesting azimuthal variations, as well as intriguing departures from the expected behaviors in the
azimuthally averaged profile shapes.

We report a relatively flat azimuthally averaged density profile, falling off with radius with an index of $\delta=1.69\pm0.13$ outside $0.7\,\text{Mpc}$ (or $r>0.4r_{200}$). There is
currently a large scatter in the values of the density slopes near $r_{200}$ reported in the literature, with values ranging from $\delta=2.53\pm0.25$ for A2142 \citep{akamatsu2011} to
$\delta=1.21\pm0.12$ for the Virgo Cluster \citep{urban2011}, or $\delta=1.24_{-0.56}^{+0.23}$ for A1689 \citep{kawaharada2010}. Measurements which indicate shallow density slopes are
challenging to explain theoretically. Simulations predict relatively steep density profiles in the cluster outskirts with $\delta=2.5$, steepening to $\delta=3.4$ at around $1.3r_{200}$
\citep{roncarelli2006}.

Compared to the expected power-law entropy profile given by Eqn. \ref{eqn:pratt}, we measure an excess in the central $\sim20'$ ($0.3r_{500}$); beyond $\sim40'$ ($\sim0.7r_{500}$), the
profile lies systematically below the expectation. 
Using a combination of SZ and X-ray data for 6~cool core clusters, \citet{eckert2013} have recently argued that the entropy profiles outside $r_{500}$ were in agreement with Eqn.
\ref{eqn:pratt}, which is clearly in tension with our current measurements. \citet{eckert2013} point out that the entropy excess in cluster cores may have caused the normalization of the
$K\propto r^{1.1}$ model to be overestimated in previous publications, where this normalization was allowed as a free parameter in the fit, rather than being fixed based on
Eqn.~\ref{eqn:pratt}. While this is indeed possible, we show that, even when fixing both the normalization and index of the power-law model for the expected entropy behavior, we still find an
entropy deficit at large radii in the azimuthally averaged profile. This is consistent with the conclusion of \citet{walker2013}, who combined the entropy profiles obtained from X-ray
spectroscopy for 13 clusters. 

In addition, the azimuthally averaged pressure profile shows an excess between $0.6r_{200}<r<r_{200}$ with respect to the best-fit model describing the SZ measurements for a sample of
clusters observed with Planck \citep{planck2013}.

In the case of X-ray observations, the quantities that are measured directly are the gas density and temperature. In order to determine which of these quantities contribute primarily to the
deviations of the pressure and entropy from the expected trends at large radii, we may use the self-similar profiles for pressure and entropy (Eqn.~\ref{eqn:pressure} and
Eqn.~\ref{eqn:pratt}, respectively) to solve for the expected density and temperature profiles:
\begin{align}
kT^{\rm expected}(r)&=P(r)^{2/5}K(r)^{3/5} \label{eqn:exptemp}\\
n_{\rm e}^{\rm expected}(r)&=P(r)^{3/5}K(r)^{-3/5}.\label{eqn:expdens}
\end{align}

\begin{figure*}
\begin{minipage}{0.95\columnwidth}
\includegraphics[width=\textwidth]{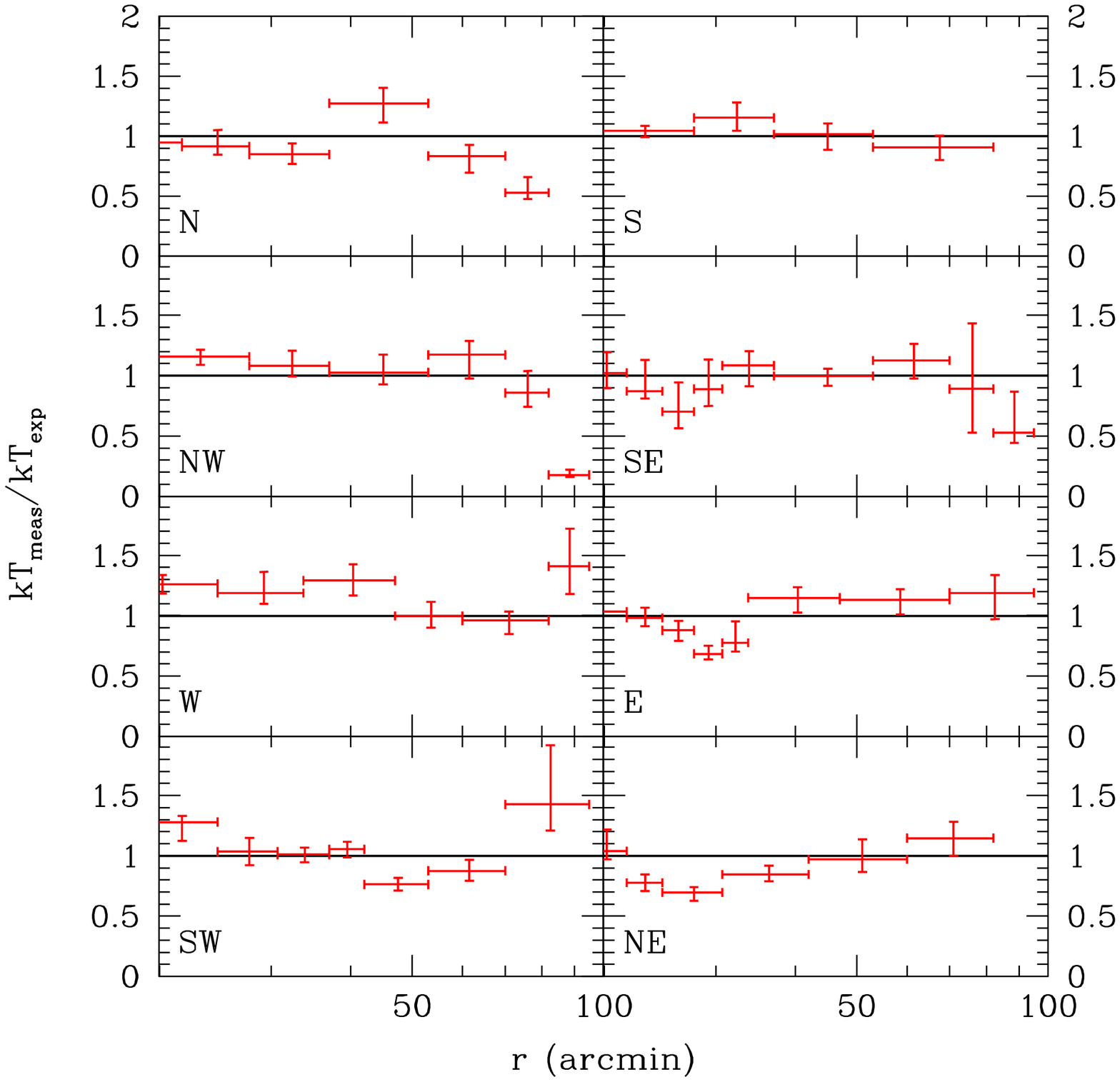}
\end{minipage}
\begin{minipage}{0.95\columnwidth}
\includegraphics[width=\textwidth]{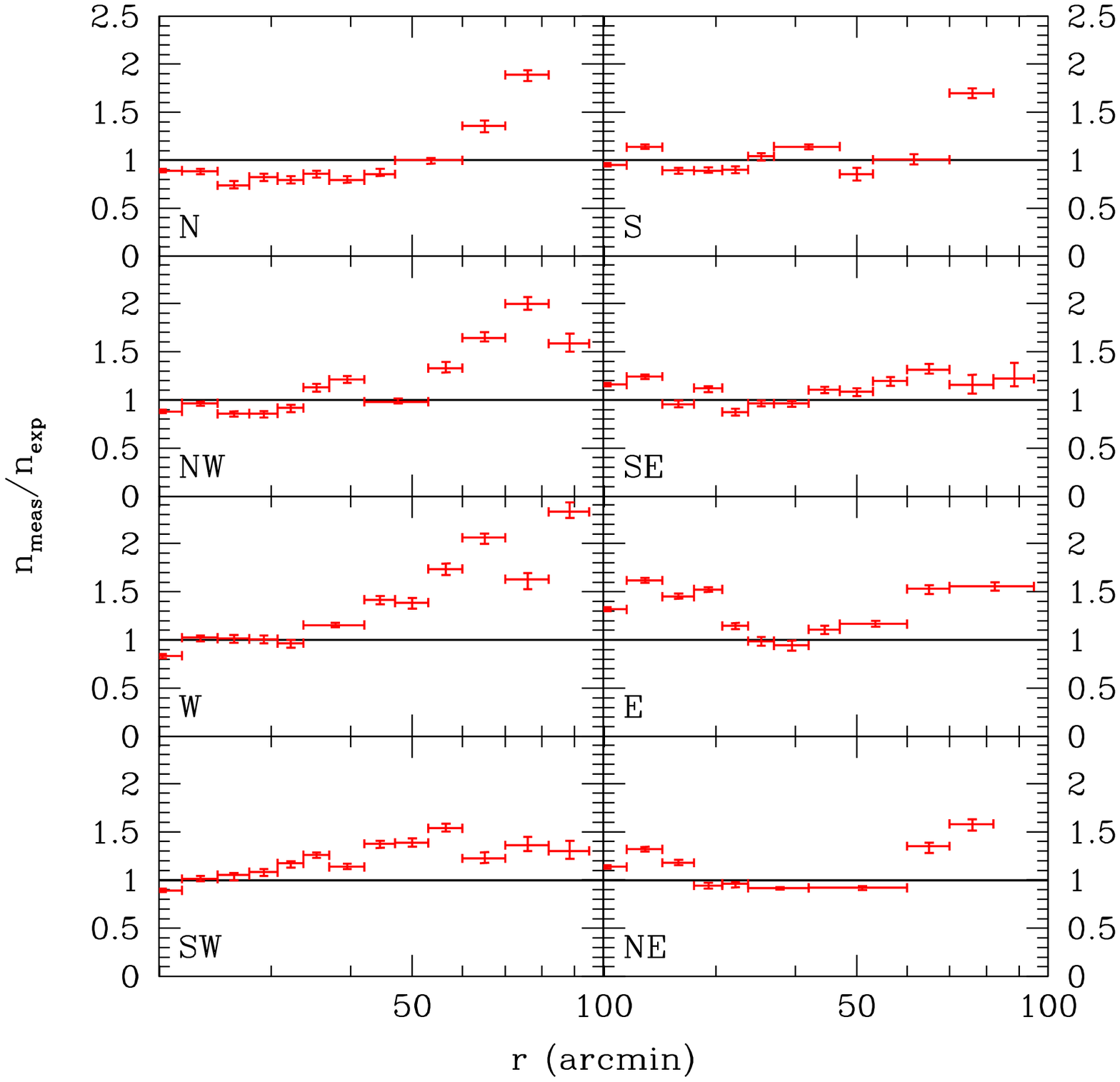}
\end{minipage}
\caption{Ratios between measured ICM temperature (\emph{left}) and density (\emph{right}), and their expectations obtained from Eqns.~\ref{eqn:exptemp} and \ref{eqn:expdens}, respectively.}
\label{fig:expectation}
\end{figure*}

Along each arm, we have determined the ratios between the measured temperatures and densities, and the expected values predicted from the equations given above. These ratios are shown in Fig.~\ref{fig:expectation}. This
allows us to look in more depth at the influence of morphological features in the individual arms on the average density, entropy and pressure profiles.

The eastern cold front is clearly visible as a dip in the entropy profiles along each of the affected arms (NE, E, SE), which coincides both with a temperature decrement and with a density
increase with respect to the expected profiles at $r<30'$. Signatures of the large scale sloshing are present in the SW and W arms, where we see excess density in the $46-60\,\text{arcmin}$
range, causing an apparent flattening of the entropy profiles along these directions.

The density in the cluster outskirts is higher than the expected value for all of the arms. At the virial radius ($82'$), the biggest excess is seen along the relaxed arms (N, NW, S). This
coincides with the cluster's minor (north-south) axis, where the entropy was observed to flatten most significantly with respect to the expected power-law model. In the case of temperature,
the measured and expected values in the outskirts are consistent within the 2~$\sigma$ confidence level, with the only exception of the outermost points of the N and NW arms.
We conclude therefore that the inconsistency between the expected and measured entropy and pressure profiles can be explained primarily by an overestimation of the density due to gas clumping
in the outskirts. 

\begin{figure}
\includegraphics[width=0.95\columnwidth]{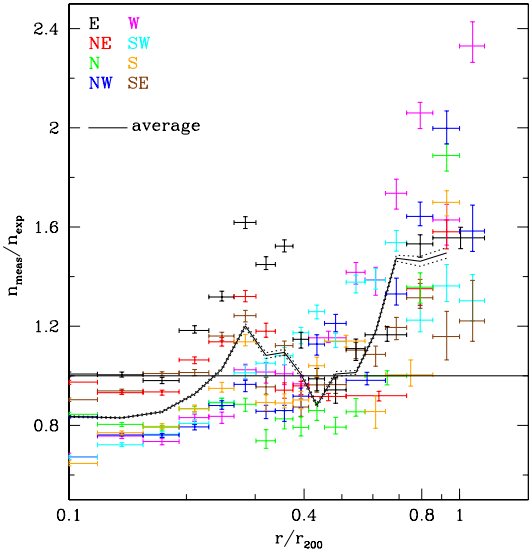}
\caption{Profiles of the ratios of the measured over expected gas densities for the eight different arms of the Perseus Cluster (\emph{points}) and the azimuthally averaged profile (\emph{lines}).}
\label{fig:overdensity}
\end{figure}

In Fig.~\ref{fig:overdensity}, we compare the ratios of the measured over expected gas densities for the eight different arms of the Perseus Cluster by overplotting the individual panels of
the right-hand side of Fig.~\ref{fig:expectation}, as well as the azimuthal average. If the density in the cluster outskirts is indeed overestimated primarily due to gas clumping, then the
square of this plotted ratio is essentially equivalent to the gas clumping factor defined as $C=\frac{\left\langle n_{\rm gas}^2\right\rangle}{\left\langle n_{\rm gas}\right\rangle^2}$.

The azimuthally averaged gas clumping exhibits a peak at $0.2-0.4r_{200}$, which is caused by the presence of the eastern cold front. Beyond $0.4r_{200}$, the level of gas clumping in
the averaged profile increases steadily with radius.

The values reported here are lower than the gas clumping factors initially presented in \cite{simionescu2011}, who find a $\sqrt{C}$ of 2.5--4 in the range from $0.8-1.0\,r_{200}$ for the NW
arm. This is partly due to the different baseline model for the entropy (here, the normalization was fixed to the predictions from numerical simulations, while \citealt{simionescu2011} used
the best-fit normalization for the observed profile) and partly due to the possible mass modeling bias associated with the instrument calibration, and which affected the clumping estimation
based on the gas mass fraction excess with respect to the cosmic mean.

Simulations by \citet{nagai2011} show that gas clumping is more pronounced in dynamically active systems. From this, one might naively expect this effect to be most significant along the cluster's major axis, where the gas accretion predominantly takes place. We observe the opposite 
trend: the highest clumping factor inside $r_{200}$ is seen along the cluster's minor axis. This could be explained if the clumps were more easily destroyed in more dynamically active regions, which may be beyond the gas physics and spatial resolution of current state-of-the-art 
simulations.
Although our \chandra\ analysis found no direct detection of gas clumps at $r\sim0.7r_{200}$ along the NW arm, we cannot rule out the presence of clumping because, even if this effect is important, the signal may be
difficult to detect due to smearing caused by the long line of sight that we probe in the cluster outskirts. It is also possible that the individual clumps are small and faint, and therefore unresolved by \chandra\ - analogous to a fine mist, where the individual drops are unresolved to our eyes. 

Alternative explanations for the flattening of the entropy profiles near the virial radius, such as weakening of accretion shocks proposed by \citet{lapi2010} and \citet{cavaliere2011}  or electron ion non-equilibrium \citep{hoshino2010,akamatsu2011}, would cause the observed 
temperatures to be lower than the expected profile given by Eqn.~\ref{eqn:exptemp}, but would not produce an excess in the observed density and pressure. These effects may be partly responsible for the temperature biases seen in the outermost annuli towards the N and NW, but appear 
not to be important elsewhere within the Perseus Cluster.

The discontinuities in the inferred deprojected temperature and density of the ICM along the NW arm, implying a sharp downturn in entropy and pressure beyond $r_{200}$, suggest -- at face value -- the presence of a
shock front. The apparent density and temperature jumps at $r_{200}$ along the NW arm, however, imply inconsistent Mach numbers. After correcting for the underlying density profile using our best-fit $\beta$-model, we find a density jump of $n_{\rm down}/n_{\rm up}=1.61$ at this 
discontinuity,
implying a Mach number of $\mathcal{M}_{\rm n}=1.43$. After correcting for the underlying temperature profile, using the \citet{planck2013} pressure profile combined with the best-fit $\beta$-model, we find a temperature
jump of $T_{\rm down}/T_{\rm up}=4$ at the $r_{200}$, implying a Mach number $\mathcal{M}_{\rm T}=3.17$.

These $\mathcal{M}$ values are within the range determined by \citet{akamatsu2013}, from a systematic X-ray analysis of six giant radio relics near the outskirts of four clusters of galaxies observed with \suzaku.
\citet{akamatsu2013} report significant temperature jumps across the relics for CIZA 2242.8-5301, Abell 3376, Abell 3667NW, and Abell 3667SE, with the Mach numbers estimated from the X-ray temperature or pressure profiles
being in the range from 1.5--3. In their sample, the shocks are also more clearly seen in the temperature than in the surface brightness or density profiles.

However, the discontinuities in the temperature and density profiles near the virial radius are difficult to interpret in light of the results shown in Fig. \ref{fig:expectation}. If the observed discontinuity were a shock, the post-shock temperature immediately inside the virial radius would be 
higher than expected based on the self similar profiles, which is not seen. Instead, the temperature discontinuity is created by the fact that the temperature immediately outside $r_{200}$ appears to be biased low, under the assumption that the self-similar profiles for the entropy and 
pressure hold beyond $r_{200}$.  We therefore cannot draw any robust conclusions regarding the presence of a shock at the virial radius along this direction, although the thermodynamic properties of the gas beyond $r_{200}$ along the NW arm certainly suggest the onset of intriguing 
physical processes which go beyond self-similarity. 

\section{Conclusions}
\label{sect:concl}

We have presented the results of the analysis of \suzaku\ Key Project observations of the Perseus Cluster from its center out to $\sim2.6\,\text{Mpc}$ along eight azimuthal directions. We detected the ICM emission out to the
virial radius ($82'$) in all directions. In five arms, we were able to study the ICM beyond the virial radius. Our main conclusions are:

\begin{itemize}

\item The azimuthally averaged density profile for $r>0.4r_{200}$ is relatively flat, with a best-fit power-law index $\delta=1.69\pm0.13$ significantly smaller than expected from numerical simulations. The entropy profile
in the outskirts lies systematically below the power-law behavior expected from large-scale structure formation models which include only the heating associated with gravitational collapse. The pressure profile beyond $\sim0.6r_{200}$
shows an excess with respect to the best-fit model describing the SZ measurements for a sample of clusters observed with Planck \citep{planck2013}.

\item The inconsistency between the expected and measured density, entropy, and pressure profiles can be explained primarily by an overestimation of the density due to gas clumping in the outskirts. There is no evidence for
a bias in the temperature measurements, with the exception of the outermost two data points towards the N and NW.

\item We found significant differences in thermodynamic properties of the ICM at large radii along the different arms. Along the cluster minor (north--south) axis, we find a flattening of the entropy profiles outside
$\sim0.6r_{200}$, while along the major (east--west) axis, the entropy rises all the way to the outskirts. Correspondingly, the inferred gas clumping factor is typically larger along the minor than along the major axis.

\item We find no bright gas clumps or surface brightness fluctuations associated with gas clumping in a deep \chandra\ observation of the Perseus Cluster outskirts ($0.7r_{200}$) along the NW direction. The clumping signal
may, however, be difficult to detect due to smearing caused by the long line of sight that we probe in the cluster outskirts.

\end{itemize}

\section*{Acknowledgments}

We acknowledge the support by \suzaku\ grants NNX09AV64G and NNX10AR48G, \chandra\ grant GO2-13144X, NASA ADAP grant NNX12AE05G and \xmm\ grant NNX12AB64G. This work was supported in part by the U.S. Department of Energy
under contract number DE-AC02-76SF00515. AM acknowledges the support by the~NSF grant AST-0838187.

\bibliographystyle{aa}
\bibliography{clusters}

\appendix

\section{Projected Results}\label{appa}


In Table~\ref{tab:projtemp} and \ref{tab:projnorm}, we show the projected values for the ICM temperature and XSPEC normalization, respectively. Throughout the sections of this Appendix, the XSPEC normalization is defined as $\int n_{\rm e}n_{\rm H} {\rm d}V\times\frac{10^{-14}}{4\pi\left[D_{\rm A}(1+z) \right]^2}\frac{1}{20^2\pi}$~cm$^{-5}$~arcmin$^{-2}$.

\begin{table*}
\setlength{\extrarowheight}{4pt}
\centering
\caption{Projected temperatures along the individual arms. The properties of the annuli are given in arcminutes, the radius is the distance from the cluster center to the inner edge of the given annulus.}
\begin{tabular}{cc|cccccccc}
\hline\hline
\multicolumn{2}{c|}{Annulus}&\multicolumn{8}{c}{Temperature (keV)}\\
Radius&Width&N&NW&W&SW&S&SE&E&NE\\
\hline
6.7  & 3   &$5.95^{+0.20}_{-0.18}$&$6.69^{+0.10}_{-0.17}$&$6.98^{+0.45}_{-0.35}$&$6.41^{+0.16}_{-0.26}$&$6.55^{+0.24}_{-0.38}$&$5.76^{+0.15}_{-0.10}$&$6.02^{+0.21}_{-0.13}$&$5.67^{+0.15}_{-0.18}$\\
9.7  & 3   &$5.83^{+0.17}_{-0.20}$&$7.22^{+0.16}_{-0.24}$&$7.73^{+0.30}_{-0.40}$&$6.65^{+0.12}_{-0.24}$&$7.39^{+0.26}_{-0.28}$&$5.85^{+0.14}_{-0.10}$&$5.68^{+0.16}_{-0.10}$&$5.78^{+0.13}_{-0.16}$\\
12.7 & 3   &$5.55^{+0.17}_{-0.19}$&$6.73^{+0.20}_{-0.16}$&$7.88^{+0.30}_{-0.40}$&$6.77^{+0.18}_{-0.20}$&$6.57^{+0.16}_{-0.22}$&$5.92^{+0.13}_{-0.12}$&$5.77^{+0.14}_{-0.14}$&$5.30^{+0.10}_{-0.12}$\\
15.7 & 3   &$5.55^{+0.20}_{-0.18}$&$6.73^{+0.24}_{-0.20}$&$7.23^{+0.30}_{-0.40}$&$6.77^{+0.26}_{-0.20}$&$6.23^{+0.18}_{-0.22}$&$6.09^{+0.15}_{-0.15}$&$5.87^{+0.15}_{-0.16}$&$5.45^{+0.19}_{-0.10}$\\
18.7 & 3   &$5.27^{+0.26}_{-0.20}$&$6.07^{+0.30}_{-0.18}$&$6.43^{+0.42}_{-0.28}$&$6.43^{+0.20}_{-0.24}$&$5.97^{+0.24}_{-0.30}$&$5.85^{+0.19}_{-0.21}$&$5.47^{+0.16}_{-0.20}$&$5.05^{+0.18}_{-0.09}$\\
21.7 & 3   &$5.23^{+0.55}_{-0.25}$&$6.43^{+0.30}_{-0.50}$&$6.13^{+0.30}_{-0.45}$&$5.55^{+0.36}_{-0.22}$&$6.53^{+0.50}_{-0.45}$&$5.31^{+0.34}_{-0.26}$&$5.43^{+0.22}_{-0.26}$&$4.47^{+0.17}_{-0.17}$\\
24.7 & 3   &$4.98^{+0.70}_{-0.45}$&$7.18^{+0.65}_{-0.55}$&$7.15^{+1.30}_{-0.60}$&$5.08^{+0.50}_{-0.25}$&$5.73^{+0.45}_{-0.45}$&$5.11^{+0.26}_{-0.30}$&$4.71^{+0.17}_{-0.20}$&$4.31^{+0.26}_{-0.26}$\\
27.7 & 3   &$4.83^{+0.35}_{-0.55}$&$5.58^{+0.40}_{-0.40}$&$6.18^{+0.45}_{-0.60}$&$6.13^{+0.45}_{-0.45}$&$5.21^{+0.54}_{-0.22}$&$5.03^{+0.24}_{-0.34}$&$4.56^{+0.10}_{-0.27}$&$3.99^{+0.24}_{-0.22}$\\
30.7 & 3   &$4.68^{+0.45}_{-0.30}$&$5.23^{+0.30}_{-0.35}$&$6.28^{+0.85}_{-0.35}$&$5.17^{+0.38}_{-0.20}$&$6.63^{+0.55}_{-0.55}$&$5.28^{+0.35}_{-0.35}$&$5.09^{+0.34}_{-0.28}$&$4.99^{+0.22}_{-0.48}$\\
33.7 & 3.3 &$5.13^{+0.35}_{-0.55}$&$5.43^{+0.50}_{-0.30}$&$6.08^{+0.80}_{-0.35}$&$4.63^{+0.16}_{-0.28}$&$4.98^{+0.40}_{-0.40}$&$5.38^{+0.40}_{-0.30}$&$5.78^{+0.55}_{-0.40}$&$5.08^{+0.50}_{-0.40}$\\
37   & 5   &$4.78^{+0.40}_{-0.35}$&$4.58^{+0.45}_{-0.35}$&$5.68^{+0.45}_{-0.45}$&$4.21^{+0.24}_{-0.16}$&$5.78^{+0.45}_{-0.85}$&$4.95^{+0.44}_{-0.24}$&$4.88^{+0.90}_{-0.45}$&$4.05^{+0.32}_{-0.36}$\\
42   & 5   &$5.03^{+0.45}_{-0.55}$&$4.68^{+0.50}_{-0.35}$&$4.73^{+0.40}_{-0.35}$&$4.21^{+0.26}_{-0.20}$&$4.05^{+0.90}_{-0.70}$&$4.75^{+0.26}_{-0.44}$&$5.85^{+0.60}_{-0.80}$&$5.65^{+0.80}_{-0.80}$\\
47   & 6   &$3.48^{+0.40}_{-0.20}$&$5.43^{+0.60}_{-0.60}$&$4.59^{+0.24}_{-0.40}$&$3.43^{+0.30}_{-0.18}$&$3.33^{+0.35}_{-0.60}$&$5.08^{+0.40}_{-0.50}$&$5.18^{+0.85}_{-0.5}$&$4.33^{+0.65}_{-0.40}$\\
53   & 7   &$3.53^{+0.50}_{-0.40}$&$4.03^{+0.50}_{-0.30}$&$4.48^{+0.45}_{-0.30}$&$4.23^{+0.40}_{-0.35}$&\multirow{2}{*}{$3.78^{+0.95}_{-0.40}$}&$5.03^{+0.55}_{-0.65}$&$4.68^{+0.60}_{-0.70}$&$4.98^{+0.65}_{-0.60}$\\
60   & 10  &$2.58^{+0.35}_{-0.30}$&$3.59^{+0.34}_{-0.22}$&$4.18^{+0.40}_{-0.30}$&\multirow{3}{*}{$5.11^{+0.67}_{-0.54}$}&     &$4.08_{-0.35}^{+0.60}$&$4.23^{+0.60}_{-0.30}$&\multirow{2}{*}{$4.63^{+0.65}_{-0.45}$}\\
70   & 12  &$2.53^{+0.45}_{-0.40}$&$2.47^{+0.40}_{-0.24}$&$4.03^{+0.30}_{-0.60}$&		       &$4.45^{+0.90}_{-0.90}$&$3.45^{+0.90}_{-0.60}$&\multirow{2}{*}{$5.30^{+1.60}_{-1.20}$}&     \\
82   & 13  &--                    &$0.97^{+0.32}_{-0.34}$&$5.75^{+3.00}_{-1.50}$&		       &--		      &$2.75^{+4.00}_{-1.50}$&			     &--	           \\		  
\hline\hline
\end{tabular}
\label{tab:projtemp}
\end{table*}

\begin{table*}
\setlength{\extrarowheight}{4pt}
\centering
\caption{Projected XSPEC spectrum normalizations, in units of $10^{-3}$ times the standard definition from Appendix \ref{appa}.}
\begin{tabular}{cc|cccccccc}
\hline\hline
Radius&Width&N&NW&W&SW&S&SE&E&NE\\
\hline
6.7  & 3   &$715.50^{+11.00}_{-13.00}$&$722.25^{+7.00}_{-6.50}$&$599.50^{+17.00}_{-9.00}$&$589.50^{+8.00}_{-11.00}$&$604.50^{+11.00}_{-13.00}$&$966.50^{+10.00}_{-10.00}$&$1171.50^{+17.00}_{-14.00}$&$1047.50^{+15.00}_{-10.00}$\\
9.7  & 3   &$421.75^{+4.50}_{-7.00}$  &$394.30^{+4.20}_{-3.20}$&$402.75^{+8.00}_{-7.00}$ &$404.75^{+5.50}_{-4.00}$ &$427.75^{+5.50}_{-5.50}$  &$610.75^{+6.00}_{-5.50}$  &$710.25^{+9.00}_{-6.50}$   &$596.25^{+5.00}_{-8.50}$	 \\
12.7 & 3   &$269.75^{+4.00}_{-3.50}$  &$262.70^{+3.00}_{-2.60}$&$269.25^{+5.50}_{-4.00}$ &$282.25^{+3.00}_{-3.50}$ &$299.25^{+4.50}_{-3.50}$  &$422.25^{+5.00}_{-3.00}$  &$512.75^{+6.50}_{-5.00}$   &$405.25^{+5.00}_{-4.00}$	 \\
15.7 & 3   &$184.30^{+3.40}_{-2.60}$  &$188.90^{+2.00}_{-3.00}$&$209.75^{+4.00}_{-3.50}$ &$217.70^{+3.00}_{-2.60}$ &$224.70^{+3.40}_{-3.40}$  &$292.90^{+4.40}_{-2.20}$  &$423.25^{+5.00}_{-5.00}$   &$308.25^{+3.50}_{-4.00}$	 \\
18.7 & 3   &$125.90^{+2.40}_{-3.20}$  &$148.10^{+2.20}_{-3.20}$&$163.25^{+4.00}_{-3.00}$ &$171.10^{+1.80}_{-3.80}$ &$172.10^{+4.60}_{-2.60}$  &$218.30^{+3.60}_{-3.40}$  &$330.25^{+5.50}_{-5.00}$   &$228.25^{+3.50}_{-3.50}$	 \\
21.7 & 3   &$83.90^{+3.00}_{-3.00}$   &$105.50^{+4.20}_{-2.40}$&$140.70^{+3.80}_{-3.60}$ &$137.75^{+3.50}_{-3.50}$ &$121.75^{+3.50}_{-4.50}$  &$148.25^{+4.00}_{-3.50}$  &$240.25^{+6.00}_{-5.50}$   &$157.25^{+4.00}_{-3.50}$	 \\
24.7 & 3   &$53.30^{+3.60}_{-2.60}$   &$75.30^{+1.80}_{-3.00}$ &$99.25^{+3.50}_{-5.00}$  &$102.25^{+4.00}_{-4.50}$ &$67.50^{+3.80}_{-2.60}$   &$84.90^{+2.40}_{-2.60}$	 &$149.25^{+3.00}_{-4.00}$   &$88.25^{+3.50}_{-4.00}$	 \\
27.7 & 3   &$42.10^{+2.00}_{-2.00}$   &$61.30^{+1.80}_{-2.00}$ &$74.30^{+3.00}_{-3.00}$  &$85.10^{+2.40}_{-2.00}$  &$55.50^{+1.80}_{-2.00}$   &$62.90^{+2.20}_{-2.40}$	 &$97.10^{+2.40}_{-2.60}$    &$52.50^{+2.00}_{-1.80}$	 \\
30.7 & 3   &$31.85^{+1.40}_{-1.30}$   &$53.50^{+1.40}_{-1.80}$ &$61.10^{+1.60}_{-2.40}$  &$66.50^{+1.80}_{-2.00}$  &$41.25^{+1.60}_{-1.30}$   &$42.25^{+1.40}_{-1.50}$	 &$51.35^{+1.80}_{-1.40}$    &$37.05^{+1.00}_{-1.40}$	 \\
33.7 & 3.3 &$25.85^{+1.20}_{-1.10}$   &$44.85^{+1.40}_{-1.80}$ &$50.70^{+1.60}_{-2.20}$  &$52.95^{+1.50}_{-1.40}$  &$37.05^{+1.50}_{-1.60}$   &$33.85^{+0.90}_{-1.30}$	 &$35.15^{+1.20}_{-1.50}$    &$25.65^{+1.30}_{-0.90}$	 \\
37   & 5   &$17.62^{+1.05}_{-0.70}$   &$33.05^{+1.40}_{-1.70}$ &$49.25^{+1.50}_{-2.00}$  &$38.05^{+1.20}_{-0.90}$  &$27.95^{+2.00}_{-1.30}$   &$24.62^{+0.85}_{-0.95}$	 &$25.10^{+1.60}_{-2.00}$    &$23.05^{+0.90}_{-1.00}$	 \\
42   & 5   &$14.22^{+0.75}_{-0.55}$   &$21.35^{+1.00}_{-1.00}$ &$34.25^{+1.20}_{-2.00}$  &$27.45^{+0.80}_{-1.20}$  &$12.95^{+1.00}_{-1.10}$   &$19.02^{+0.60}_{-0.80}$	 &$19.45^{+0.80}_{-1.30}$    &$14.22^{+0.75}_{-0.50}$	 \\
47   & 6   &$11.83^{+0.80}_{-0.75}$   &$13.78^{+0.65}_{-0.50}$ &$25.85^{+1.10}_{-0.80}$  &$19.55^{+0.90}_{-1.10}$  &$9.85^{+1.20}_{-0.90}$    &$12.57^{+0.45}_{-0.50}$	 &$14.13^{+0.55}_{-0.45}$    &$9.97^{+0.50}_{-0.85}$	 \\
53   & 7   &$7.97^{+0.50}_{-0.50}$    &$15.12^{+0.95}_{-0.60}$ &$18.95^{+1.10}_{-1.00}$  &$11.32^{+0.65}_{-0.60}$  &$7.83^{+0.90}_{-0.75}$    &$8.47^{+0.55}_{-0.55}$	 &$12.43^{+0.85}_{-0.75}$    &$8.47^{+0.50}_{-0.50}$     \\											
																
60   & 10  &$7.62^{+0.85}_{-0.50}$    &$10.12^{+0.45}_{-0.60}$ &$11.03^{+0.75}_{-0.50}$  &$4.85^{+0.30}_{-0.40}$   &$3.62^{+0.30}_{-0.45}$    &$4.51^{+0.50}_{-0.26}$	 &$7.03^{+0.30}_{-0.55}$     &$5.98^{+0.35}_{-0.45}$ 	\\
70   & 12  &$3.62^{+0.35}_{-0.35}$    &$5.42^{+0.65}_{-0.40}$  &$4.75^{+0.46}_{-0.22}$	 &$2.59^{+0.22}_{-0.22}$   &$3.25^{+0.22}_{-0.34}$    &$1.81^{+0.24}_{-0.18}$	 &$2.63^{+0.24}_{-0.20}$     &$2.43^{+0.22}_{-0.22}$     \\
82   & 13  &--                        &$1.22^{+0.65}_{-0.35}$  &$2.73^{+0.18}_{-0.24}$	 &$0.86^{+0.14}_{-0.13}$   &--			      &$0.62^{+0.21}_{-0.11}$	 &$1.49^{+0.19}_{-0.18}$     &--                         \\
\hline\hline
\end{tabular}
\label{tab:projnorm}
\end{table*}

\section{Background-Related Systematic Uncertainties}
\label{sect:systematics}

\subsection{CXFB Variations}
\label{subs:cxfbvar}

To test the influence of possible CXFB variations over the area of our mosaic, we have examined spectra extracted from $10'$~wide annular regions at  $r>95'$. In total, we examined 23~regions (three from each
of~seven\footnote{The N~arm was not used due to the influence of the background cluster as discussed in Sect.~\ref{subs:cxfb}} arms plus one from the W and one from the S arms that extend further out). Starting from our
default background model, we simultaneously refitted the normalizations of all CXFB components in each region, with the exception of the LHB (whose variation does not have a significant influence on the results). From the
measured values, we estimate the dispersion of the individual CXFB components across the mosaic. For the power-law component, we used the complete set of 23 normalizations. In the case of the anisotropic CXFB components, we
took the standard deviation of the normalization of regions selected from the given and the two adjacent arms (in the vicinity of the N arm, we used only two arms, since no data from this arm has been used in the CXFB
analysis). Tab.~\ref{tab:variations} shows the average values and their variations for the anisotropic components and the power-law component. 

The standard deviations were used as the bracketing values to upscale and downscale, one at a time, the normalizations of the individual CXFB components, in each arm refitting in each case
the ICM model.

\begin{table}
\centering
\caption{XSPEC normalizations and their standard deviations for the CXFB model components along the individual arms. The values are normalized to $10^{-3}$ times the standard definition from Appendix \ref{appa}.}
\begin{tabular}{c|cc}
\hline\hline
\multirow{2}{*}{arm}&\multicolumn{2}{c}{component}\\
&GH&0.6 keV\\
\hline
E &$3.28\pm2.02$&$0.400\pm0.168$\\
NE&$5.18\pm2.28$&$0.298\pm0.122$\\
N &$5.08\pm1.93$&$0.231\pm0.068$\\
NW&$5.02\pm1.42$&$0.194\pm0.077$\\
W &$3.01\pm1.61$&$0.547\pm0.104$\\
SW&$3.25\pm1.44$&$0.350\pm0.091$\\
S &$2.97\pm1.13$&$0.282\pm0.188$\\
SE&$3.25\pm1.36$&$0.594\pm0.175$\\
\hline
&\multicolumn{2}{c}{power-law}\\
\hline
all&\multicolumn{2}{c}{$1.28\pm0.15$}\\
\hline\hline
\end{tabular}
\label{tab:variations}
\end{table}

\begin{figure*}
\centering
\includegraphics[width=.47\textwidth]{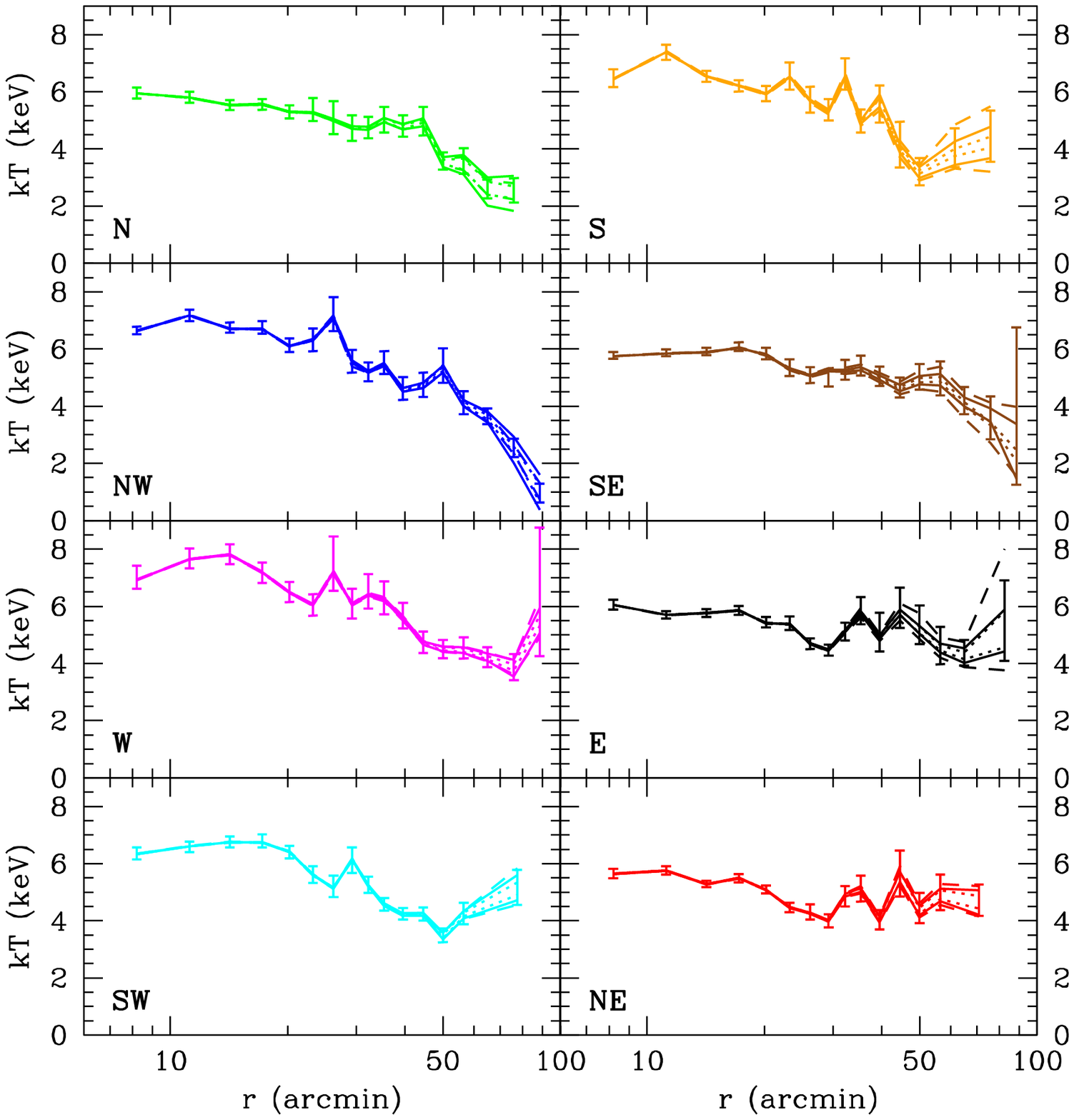}
\includegraphics[width=.47\textwidth]{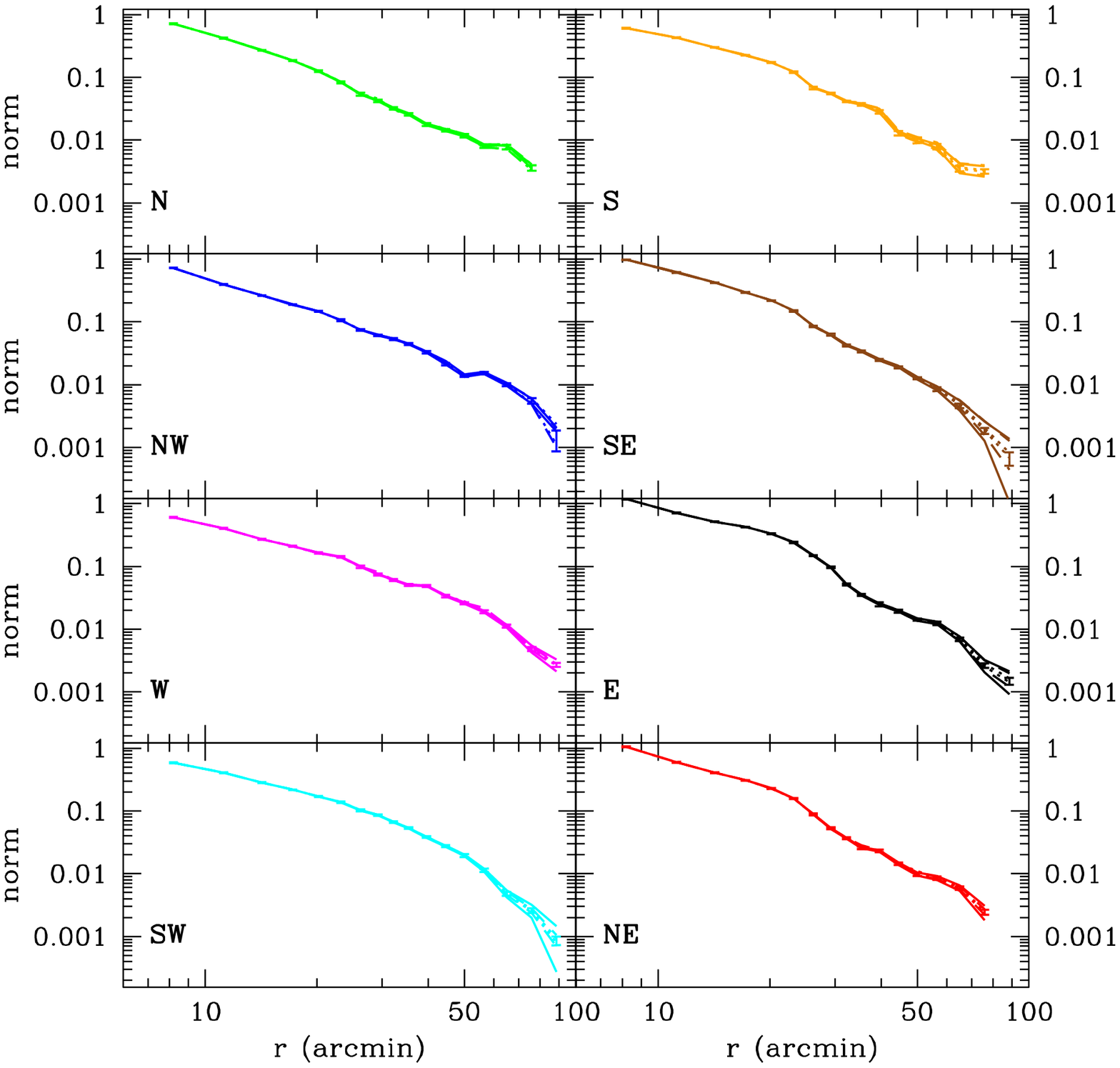}
\caption{Projected temperature and normalization profiles and the systematic errors determined from up- and downscaling the CXFB model parameters by the values shown in Table~\ref{tab:variations}. The individual lines - full (power-law), dotted (GH), and dashed (hot component) - 
show the bracketing values for the measurements resulting from potential background variations. }
\label{fig:systemp}
\end{figure*}

The uncertainties in the ICM temperature and normalization profiles obtained in this way are plotted over the original results in Fig.~\ref{fig:systemp}. The systematic errors, with the exception of a few data
points beyond the virial radius, lie within the statistical error bars of the individual values, and the trends remain robust.

\begin{figure*}
\centering
\includegraphics[width=.95\textwidth]{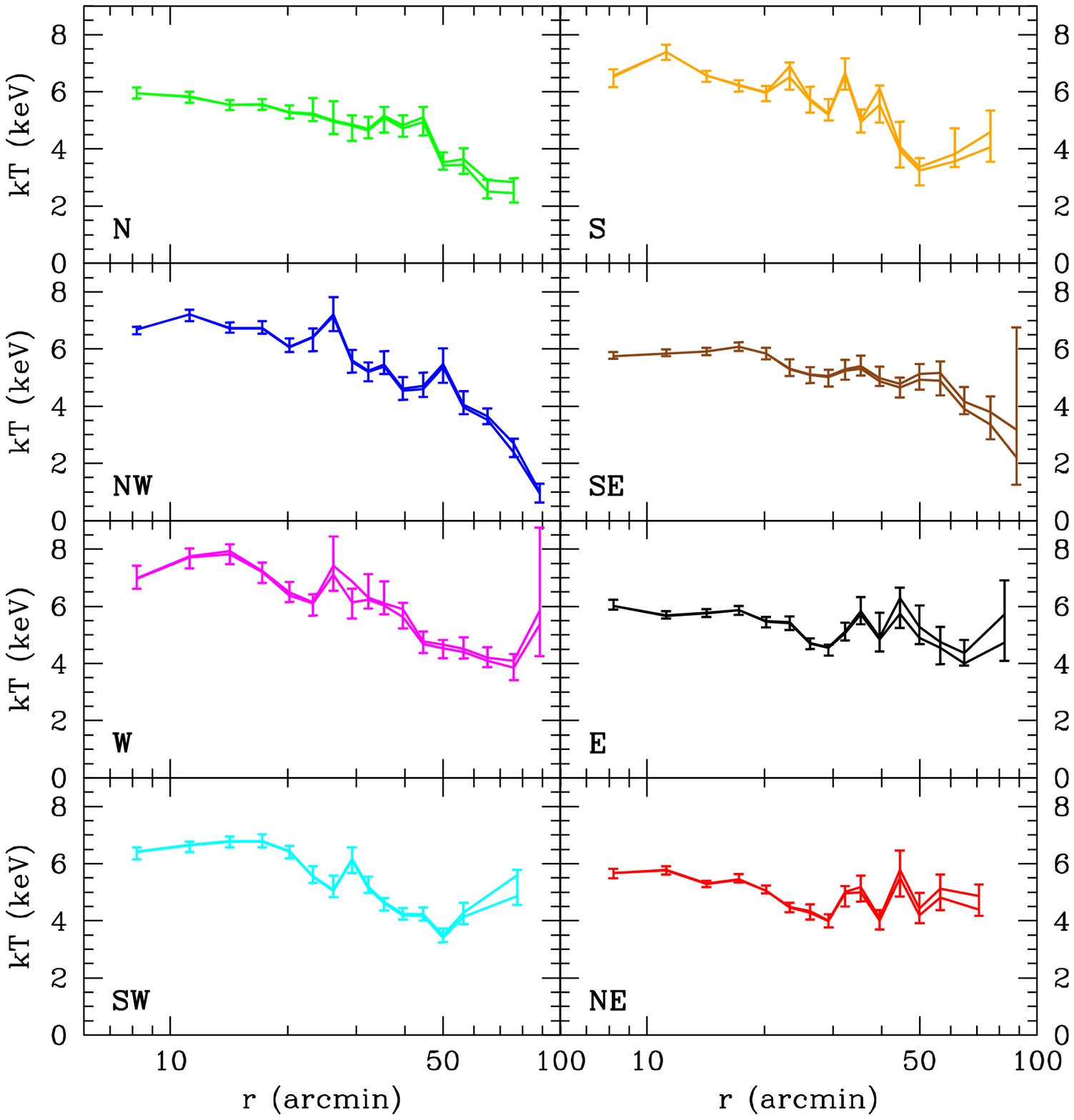}
\caption{Statistical errors of the projected temperature profiles (\emph{points}) and the systematic errors estimated using bootstrap analysis (\emph{lines}).}
\label{fig:boottemp}
\end{figure*}

\begin{figure*}
\centering
\includegraphics[width=.95\textwidth]{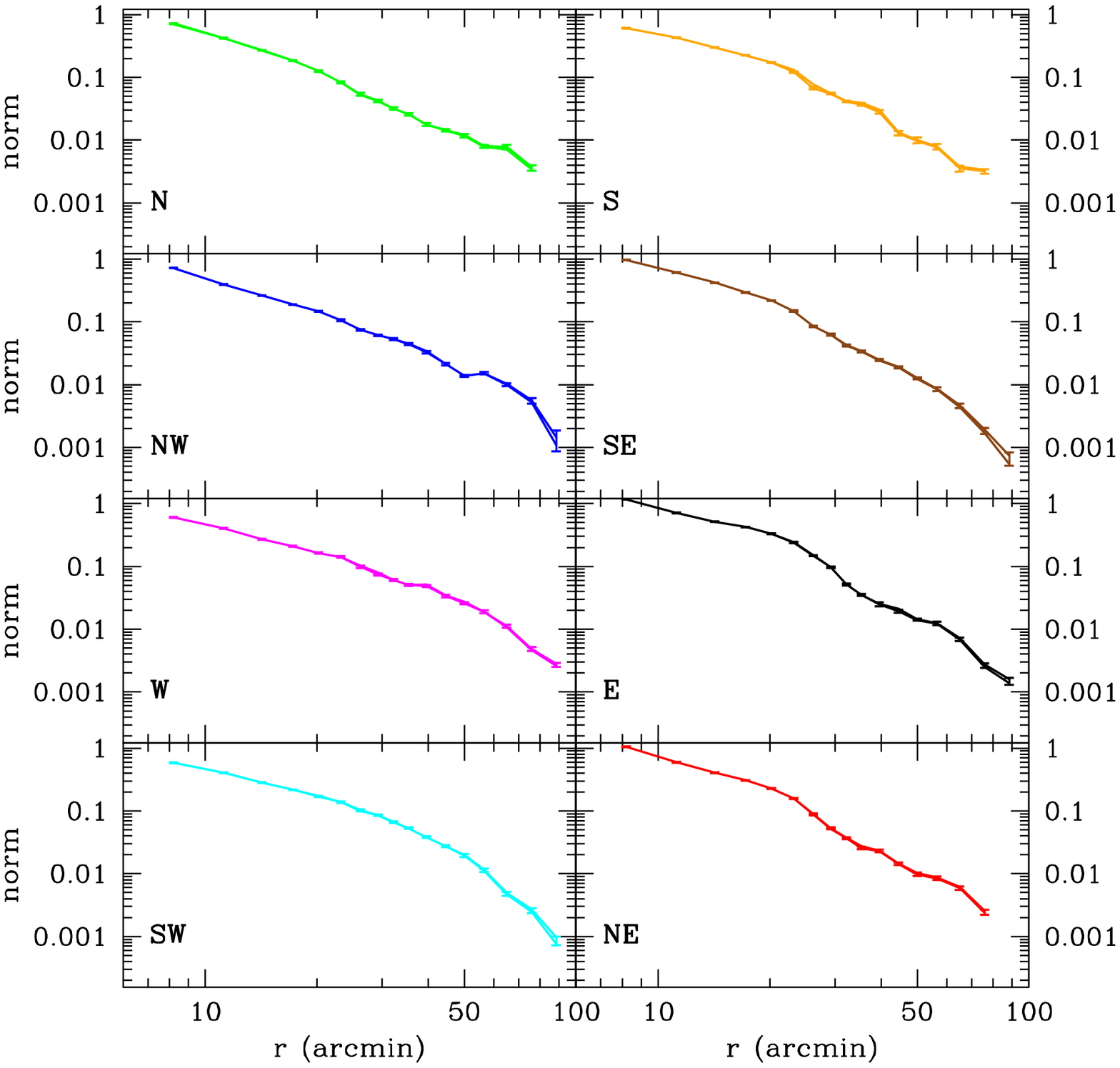}
\caption{Same as Fig.~\ref{fig:boottemp} but for normalizations.}
\label{fig:bootnorm}
\end{figure*}

To further investigate the influence of the CXFB variations on our results, we followed a bootstrap approach, as described in \citet{simionescu2013}. In order to obtain a set of spectra with approximately the same number of counts, as well as to extract a statistically meaningful number of 
spectra from each arm, we divided each of the background pointings into quadrants and divided the event files from each detector into parts with exposure times of approximately 5~ks. In total, we extracted 272~spectra per detector from all the arms in this way, with at least 24~spectra per 
detector per arm. 

From each arm we randomly selected $N$ out of the $N$ spectra that had been extracted from that arm, allowing for repetition, and simultaneously used the sets from all the (seven) arms to re-determine the best-fit CXFB model in the same way as the original model, described in the main 
text. We use this new CXFB model to re-fit the ICM emission.

We repeated this procedure 1000~times, each time with a different randomly selected set of background spectra, and determined the distributions of the best-fit temperatures and normalizations and their 68\% confidence intervals. These are shown in Fig.~\ref{fig:boottemp} (temperature) 
and Fig.~\ref{fig:bootnorm} (normalization), respectively. We find that the uncertainties due to background fluctuations are smaller than the 1-$\sigma$ statistical error bars for all of our measurements.

\subsection{Uncertainty of the Non-X-ray Background}
\label{subs:nxbsyst}

\begin{figure*}
\centering
\includegraphics[width=.47\textwidth]{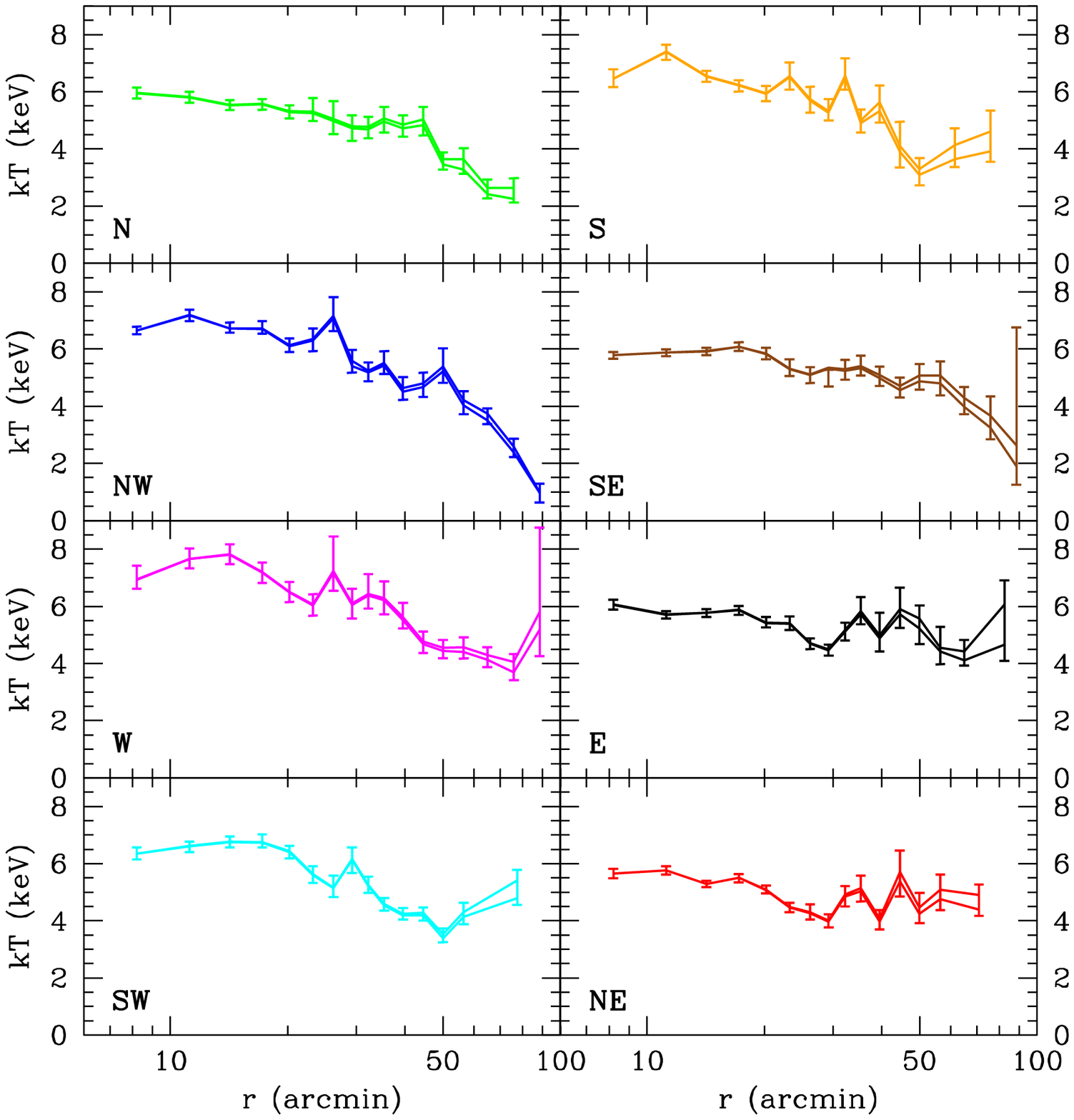}
\includegraphics[width=.47\textwidth]{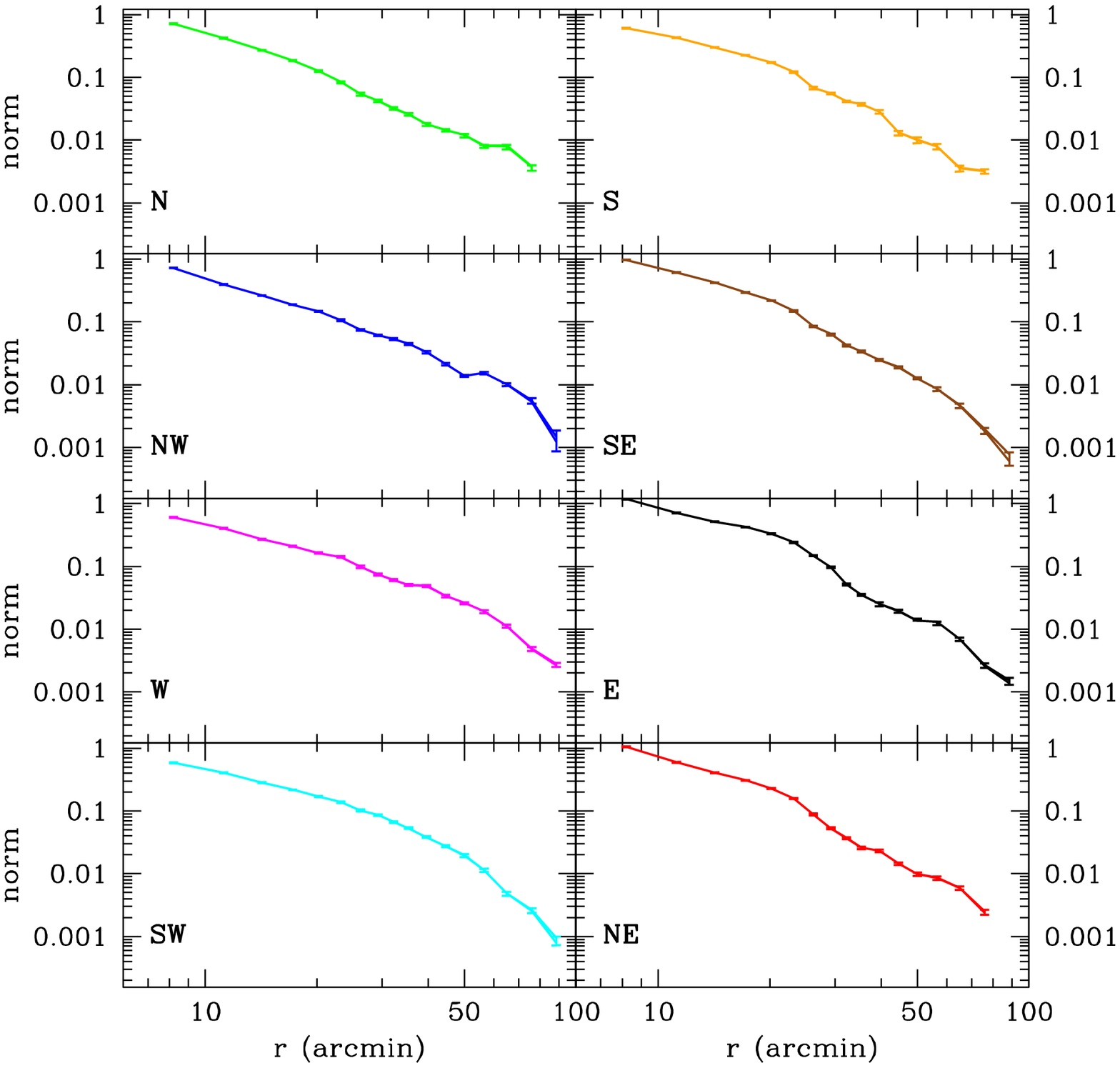}
\caption{Systematic errors on the projected temperature
(\emph{left panel}) and XSPEC normalization (\emph{right panel}) profiles
determined by conservatively up- and downscaling the NXB by 3\% for all detectors simultaneously. Points show the statistical errors of
the best-fit values, lines show the bracketing values for the measurements resulting from potential NXB variations.}
\label{fig:nxbsyst}
\end{figure*}

The \suzaku\ non-X-ray background (NXB) is known to be remarkably stable, with an uncertainty smaller than $3\%$ \citep{ishisaki2007}. 
In order to examine the sensitivity of our results to potential uncertainties in the NXB, we simultaneously up- or downscaled the NXB by 3\% for all detectors and
refitted the spectra using our default model. The bracketing values for the
projected temperatures and XSPEC normalizations are shown in
Fig.~\ref{fig:nxbsyst}. In all cases the systematic uncertainties are smaller than the statistical errors. We note that this is a pessimistic estimate of the background fluctuations, as the NXB level is unlikely to be either increased or decreased by the same (maximal) amount in all detectors simultaneously.

\subsection{Point Source Identification in {\it Suzaku}}
\label{subs:ps}

Correctly identifying and excluding point sources when studying the ICM, especially with the modest spatial resolution of \suzaku, provides another source of uncertainty. To test the influence of the point
source identification method, we re-analysed the data for the SW arm excluding two sets of point sources obtained with different (extreme) techniques. In the ``minimal'' selection we chose, by
visual inspection, only the most obvious sources, identifying only three sources within $r<95'$ and one at $r>95'$. In the ``maximal'' selection, we obtained a set of 22~point sources in the $r<95'$ region and 12 sources
with $r>95'$ by running the CIAO tool \texttt{wavdetect} on the Perseus mosaic with a more aggressive significance threshold of $10^{-4}$. The three visually identified sources were part of the maximal set. To account for
the different sensitivity to point source detection, we refitted the original CXFB model using the data extracted from $r>95'$ with the power-law normalization as the only free parameter for both point source sets, and used
these updated CXFB models in the analysis.

\begin{figure}
\centering
\includegraphics[width=.92\columnwidth]{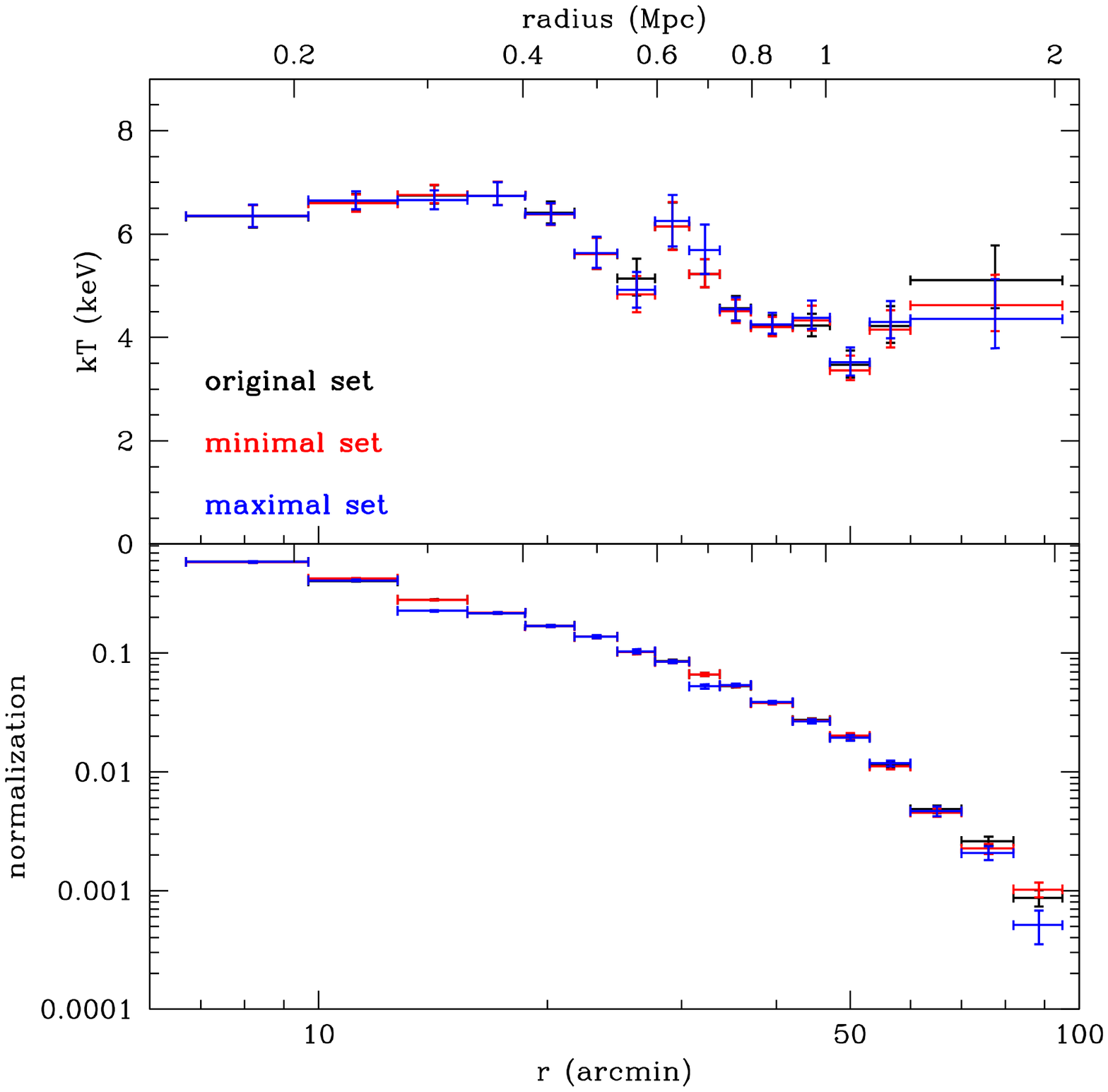}
\caption{Projected temperature (\emph{top}) and XSPEC normalization (\emph{bottom}) profiles for three different sets of point sources excluded in the SW arm.}
\label{fig:pstest}
\end{figure}

Fig.~\ref{fig:pstest} shows the projected temperature and normalization profiles for the original, minimal and maximal sets of point sources. We see a good agreement between the measured ICM properties independent of the
excluded point sources, with the exception of the normalization beyond $r_{200}$, where the measured value is about $60\%$ of the original when excluding the maximal set of point sources. We carried out an
analogous test in all the arms where we performed deprojection beyond the virial radius (NW, W, SE, E). Some influence of the point source exclusion on the spectral fitting results is found in the E and W arms, where the
normalizations in the outermost annulus (outside $r_{200}$) decrease by $\sim30\%$ and $\sim15\%$, respectively, when the maximal point source set is used, compared to the original set. The main conclusions of our analysis, however,
remain robust.

\subsection{Point Sources in the Chandra Data}
\label{chandraappendix}

The point source analysis in the deep \cha \ observation centered around 0.7$r_{200}$ along the NW arm has been performed using the procedure discussed in \cite{ehlert2013}. In short, this procedure begins with an aggressive search for candidate sources using the {\it CIAO} routine 
\wav \ \citep{freeman2002}, and
refines the source catalog using the \ae \ point-source analysis package\footnote{The {\em ACIS Extract} software package and User's Guide are available at \texttt{http://www.astro.psu.edu/xray/acis/acis\_analysis.html.}}
\citep{broos2010,broos2012}. We use \ae \ to simulate the \cha \ point spread function\footnote{We define the PSF region as the region that approximates the 90\% Enclosed Energy Fraction (EEF) of the local PSF, averaged
across five energies (0.277, 1.497 4.510, 6.400, and 8.600 \keV) } (PSF) at the positions of each candidate point source in each of the four observations of this field. \ae \ also computes the background counts within
background regions surrounding the sources, accounting for contributions from neighboring sources. 

In order to refine the catalog and remove spurious sources, we then utilize the no-source binomial probability $P_{\rm b}$ to determine the likelihood that source counts are due to fluctuations in the background \citep[see Appendix A of][]{weisskopf2007}. 
For this study, we present the results for all sources that satisfy \thresh \ in the full band ($0.5-8.0 \keV$), although the results are consistent with those measured in the soft ($0.5-2.0 \keV$) and hard ($2.0-8.0 \keV$)
bands. The final set of identified point sources is shown in Fig.~\ref{fig:chandra}.

We used the same procedure discussed in \citet{ehlert2013} to determine, for each position in the field of view, the minimum point source flux to which our detections are sensitive (i.e. the sensitivity map). The full band point source flux limit for the combined 150~ks of \cha\ observations is 
$\sim 1 \times 10^{-15} \ergpcmsqps$. 

The cumulative number density of point sources above a given flux ($S$) is calculated as 

\begin{equation}
N (> S) = \sum_{S_{\rm i} > S} \frac{1}{\Omega_{\rm i}}
\end{equation}

\noindent where $\Omega_{\rm i}$ is the total survey area sensitive to the $i^{th}$ source flux $S_{\rm i}$. The dominant uncertainty is the Poisson noise in the total number of sources. The expected Poisson fluctuations for
a sample of size $n$ are estimated using the 1-$\sigma$ asymmetric confidence limits of \cite{gehrels1986}. 

\begin{figure}
\includegraphics[width=0.92\columnwidth,angle=270]{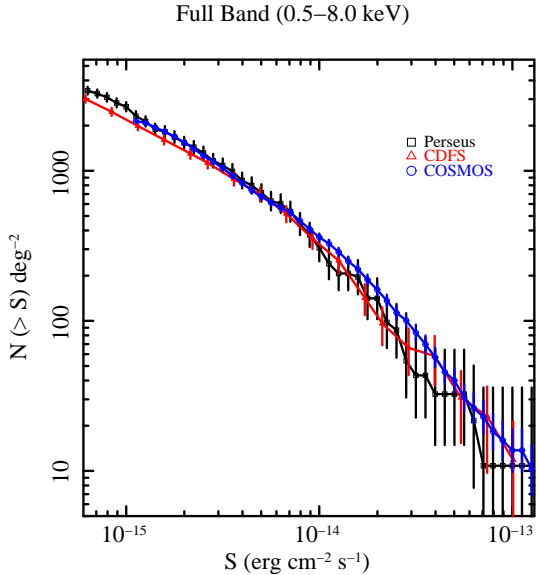}
\caption{Cumulative {\it Chandra} point source number counts (log$N$-log$S$) in the full band ($0.5-8.0 \keV$) in black. In red we show the cumulative number counts for the \cdfs \ in the same energy band
\citep{lehmer2012} and in blue the number counts from the COSMOS survey \citep{ehlert2013}.} 
\label{fig:logNlogS}
\end{figure}

The $\log{N}-\log{S}$ cumulative number counts for sources in this region of the Perseus Cluster are shown in Fig.~\ref{fig:logNlogS}, together with the \cdfs\ results in the same energy band
\citep[][]{ehlert2013,lehmer2012} and the results from the COSMOS survey \citep{ehlert2013}. All fluxes have been corrected for Galactic absorption.

Based on the $\log{N}-\log{S}$ curve of the Perseus field, there is no significant evidence for additional point sources above the levels expected from deep and medium-deep X-ray surveys in the field, down to a full band
flux limit of $\sim 1 \times 10^{-15} \ergpcmsqps$. 

We therefore do not detect X-ray bright, unresolved clumps, suggesting that if gas clumping is important here, the individual clumps must be spatially extended with respect to the \chandra\ PSF and/or below the flux
detection threshold.

\subsection{Point Sources Unresolved by \suzaku\ -- Comparison With \xmm\ Data}
\label{subs:xmm}

An observation of the E arm outskirts at a radius $r=93'$ ($1.1r_{200}$) was performed with \xmm\ on 2011~August~4 with an exposure time of 17~ks for EPIC/MOS and 13~ks for EPIC/pn. 

The data were reduced using the \xmm\ Science Analysis System (SAS) version~12.0.1. To remove the soft proton flares in the EPIC/MOS and EPIC/pn observations, we created hard band ($10-12\,\text{keV}$ for the MOS and
$10-14\,\text{keV}$ for pn) count rate histograms. After fitting them with a Gaussian, we removed all periods with count rates exceeding the mean by more than $2\sigma$. We then repeated the process in the soft band
($0.3-1.0\,\text{keV}$). The exposure times after cleaning were 15~ks and 11.5~ks for MOS and pn detectors, respectively.

We stacked the images from all three detectors in the $0.7-7.0\,\text{keV}$ range and used the result as an input to the task \texttt{ewavelet} to identify point sources down to $5\sigma$ significance. Using only the central
$12'$ of the pointing (to avoid systematic uncertainties associated with measurements at large off-axis angles), we found 43~candidate point sources. We extracted spectra for 34 of these sources, which had not been detected
with \suzaku. We used neighbouring regions to extract background spectra. We created the response matrices and ancillary response files using the SAS tasks \texttt{rmfgen} and \texttt{arfgen}, respectively.

The spectra of each individual point source were fitted with an absorbed power-law model with a fixed spectral index of 1.52, corresponding to the CXFB power-law component index. Our analysis indicates that the \xmm\ data were able to resolve point sources down to a flux limit of $
\sim1\times10^{-14}\,\text{erg}\,\text{cm}^{-2}\,\text{s}^{-1}$ in the $0.5-8.0\,\text{keV}$ band. The flux of the brightest excluded source not detected by \suzaku\
was $9.57\times10^{-14}\,\text{erg}\,\text{cm}^{-2}\,\text{s}^{-1}$. The total surface brightness of these sources of $2.09\times10^{-15}\,\text{erg}\,\text{cm}^{-2}\,\text{s}^{-1}\,\text{arcmin}^{-2}$ accounted for
$\sim31\%$ of the power-law CXFB component measured by \suzaku\ in the $0.5-8.0\,\text{keV}$ range. We compared this to a surface brightness value of
$1.76\times10^{-15}\,\text{erg}\,\text{cm}^{-2}\,\text{s}^{-1}\,\text{arcmin}^{-2}$ calculated from $\log N-\log S$ \chandra\ measurements in the
$10^{-14}-10^{-13}\,\text{erg}\,\text{cm}^{-2}\,\text{s}^{-1}$ flux range. The difference between the power-law flux expected from the $\log N-\log S$ and that measured with \xmm\ is $3.3\times10^{-16}\,\text{erg}\,\text{cm}^{-2}\,\text{s}^{-1}\,\text{arcmin}^{-2}$, and can be due to 
variability of the CXFB power-law component that occurs between the two observations, and to possible calibration offsets.
From Tab.~\ref{tab:cxfb}, this difference amounts to $\sim6\%$ of the total CXFB power-law flux. The assumed value for the cosmic variance of the power-law component used in Sect.~\ref{subs:cxfbvar} of $\sim11\%$ therefore appears conservative.

All the point sources detected with \suzaku\ have \xmm\ counterparts, in the area of the sky that is covered by both observatories.

In conclusion, our analysis of the \xmm\ data confirms the robustness of the \suzaku\ results with respect to our treatment of unresolved point sources.

\section{Stray Light Contamination}
\label{sect:stray}

Even though reduced by the pre-collimator mounted on top of the X-ray Telescope (XRT) optics, stray light radiation, caused by light reflection along routes other than the standard light paths, is an important source of systematic error in the \suzaku\ analysis
\citep{mori2005,serlemitsos2007}. The analysis of faint diffuse sources, such as galaxy cluster outskirts, can in principle be influenced by the presence of a nearby bright source, e.g. a cool cluster core, which needs to be accounted for.
Fortunately the design of \suzaku's XRT allows for observational strategies that leave parts of the detector shaded from this stray light \citep{takei2012}. 

\begin{figure}
\centering
\includegraphics[width=.92\columnwidth]{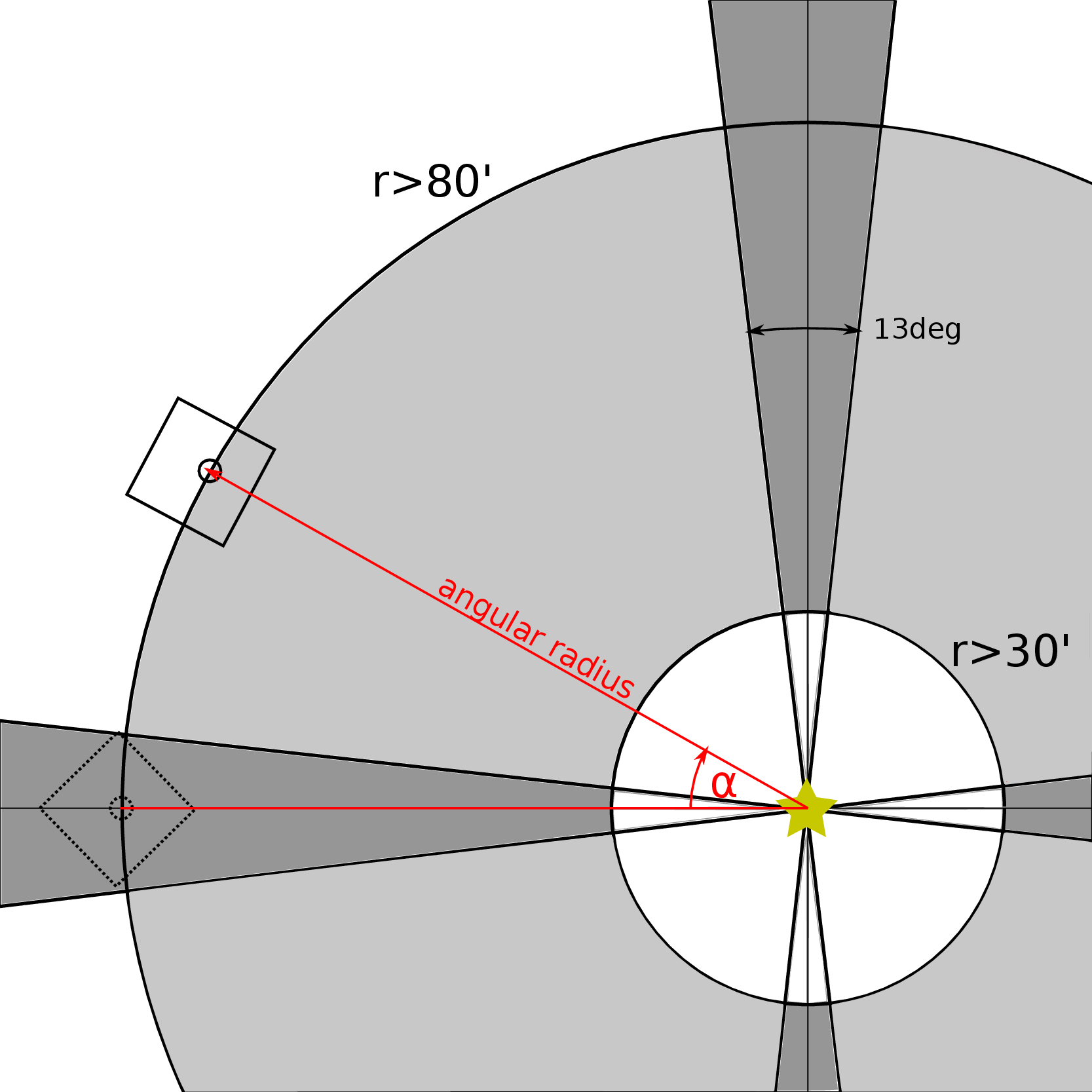}
\caption{Schematic of the stray light shaded configurations of the XRT w.r.t. the bright source. For angular radii $r<30'$ stray light influences the whole detector (inner white circle). For $r>30'$ four
wedge-shaped areas of angular width 13~degrees are shaded from stray light (dark grey), whose position depends on the detector's roll angle. For $r>80'$ the interior of a circle centered on the bright source with radius $r$ passing through the boresight
(light grey) is shaded from stray light.}
\label{fig:strayfree}
\end{figure}

Through communication with the \suzaku\ team we were able to identify regions shaded from stray light in our observations (see also Fig.~\ref{fig:strayfree}):

\begin{itemize}
  \item for angular radii, $r$, smaller than 30~arcmin, the whole detector is affected to some degree by stray light. However, the high intrinsic brightness of the Perseus Cluster at these radii makes stray light unimportant to the present analysis.
  \item for $r>30'$, the stray light will be shaded within a wedge-shaped area of angular width 13~degrees, centered on the bright source (dark grey areas in Fig.~\ref{fig:strayfree}). In the ideal ``diamond configuration''
  (cf. Sect.~\ref{sect:obsanal}), the symmetry axis passes through the boresight. Deviating from the diamond configuration by an angle $\alpha$ moves the shade on the detector so that the angle between the wedge axis and the flowline between the boresight and the bright
  source is also $\alpha$.\footnote{Clearly then
  there are four such wedge shaped areas -- rotating the detector by $90^\circ$ restores the diamond configuration.}
  \item for $r>80'$, the stray light shaded area is the \emph{interior} of a circle centered on the bright source passing through the boresight (light grey areas in Fig.~\ref{fig:strayfree}).  
\end{itemize}

To test for the systematic  errors due to stray light in the diamond configuration pointings, we extracted spectra from stray light contaminated and stray light shaded regions of pointings containing a significant area of each, namely SW3, SW4, N3, SE3 and NE3.5. For each
individual pointing, we independently fitted the spectra from both regions using a single temperature plasma model with the CXFB described above. Doing the fitting twice, we first kept the model metallicity fixed to 0.3~Solar
and then set it free. The results are shown in Tab.~\ref{tab:stray}.

\begin{table*}
\setlength{\extrarowheight}{4pt}
\begin{center}
\caption{Results of the stray light influence testing.}
\label{tab:stray}
\begin{tabular}{c|ccc|ccc}
\hline\hline
\multirow{2}{*}{pointing}&\multicolumn{3}{c|}{contaminated area}&\multicolumn{3}{c}{shaded area}\\
&kT (keV)& metallicity (Solar)&flux $\left(\times10^{-11}\,\text{erg}\,\text{cm}^{-2}\,\text{s}^{-1}\right)$&kT (keV)& metallicity (Solar)&flux $\left(\times10^{-11}\,\text{erg}\,\text{cm}^{-2}\,\text{s}^{-1}\right)$\\
\hline
\multirow{2}{*}{SW3}  &$4.072_{-0.473}^{+0.610}$&$0.3^\dagger$            &$2.580_{-0.067}^{+0.057}$&$4.215_{-0.276}^{+0.305}$&$0.3^\dagger$            &$2.555_{-0.038}^{+0.035}$\\
                      &$4.084_{-0.465}^{+0.595}$&$0.396_{-0.223}^{+0.281}$&$2.585_{-0.071}^{+0.065}$&$4.256_{-0.276}^{+0.296}$&$0.406_{-0.132}^{+0.151}$&$2.567_{-0.040}^{+0.034}$\\                     
\multirow{2}{*}{SW4}  &$4.361_{-0.728}^{+1.071}$&$0.3^\dagger$            &$1.198_{-0.020}^{+0.016}$&$4.362_{-0.477}^{+0.922}$&$0.3^\dagger$            &$1.350_{-0.027}^{+0.020}$\\
                      &$4.352_{-0.752}^{+1.100}$&$0.179_{-0.179}^{+0.350}$&$1.196_{-0.023}^{+0.018}$&$4.492_{-0.467}^{+0.800}$&$0.978_{-0.416}^{+0.552}$&$1.368_{-0.029}^{+0.023}$\\                      
\multirow{2}{*}{SE3}  &$4.354_{-0.395}^{+0.592}$&$0.3^\dagger$            &$2.358_{-0.049}^{+0.041}$&$5.077_{-0.439}^{+0.484}$&$0.3^\dagger$            &$2.319_{-0.036}^{+0.036}$\\
                      &$4.372_{-0.380}^{+0.565}$&$0.576_{-0.229}^{+0.280}$&$2.377_{-0.048}^{+0.040}$&$5.068_{-0.434}^{+0.476}$&$0.358_{-0.151}^{+0.172}$&$2.322_{-0.037}^{+0.038}$\\                     
\multirow{2}{*}{NE3.5}&$3.360_{-0.377}^{+0.588}$&$0.3^\dagger$            &$1.406_{-0.028}^{+0.027}$&$6.235_{-1.068}^{+1.380}$&$0.3^\dagger$            &$1.588_{-0.031}^{+0.026}$\\
                      &$3.358_{-0.386}^{+0.593}$&$0.267_{-0.204}^{+0.249}$&$1.406_{-0.024}^{+0.023}$&$6.214_{-1.079}^{+1.387}$&$0.243_{-0.243}^{+0.339}$&$1.585_{-0.032}^{+0.026}$\\                     
\multirow{2}{*}{N3}   &$3.895_{-0.500}^{+0.568}$&$0.3^\dagger$            &$1.835_{-0.041}^{+0.036}$&$4.325_{-0.342}^{+0.481}$&$0.3^\dagger$            &$1.901_{-0.032}^{+0.028}$\\
                      &$3.840_{-0.523}^{+0.581}$&$0.174_{-0.174}^{+0.237}$&$1.826_{-0.043}^{+0.037}$&$4.413_{-0.338}^{+0.451}$&$0.731_{-0.209}^{+0.247}$&$1.925_{-0.030}^{+0.026}$\\                      		      		      
\hline\hline
\multicolumn{5}{l}{$^\dagger$ fixed value}
\end{tabular}
\end{center}
\end{table*}

It is noteworthy that all the pointings used in the stray light contamination test are indexed 3 and higher, meaning the spectra come from regions with $r>0.5r_{200}$ (as seen in Fig.~\ref{fig:octopus}; the virial
radius lies in pointings indexed 5). These regions have relatively low signal-to noise when compared to the inner regions, which limits the constraining power of the fits. 

The temperature differences between the shaded and unshaded areas of the detectors lie within $1\sigma$ of each other, with the exception of pointing NE3.5, which shows a large discrepancy in the measured temperature. However, the X-ray flux measured in the potentially 
contaminated area of NE3.5 does not show an excess compared to the shaded region, which we would expect if stray light were responsible for this temperature discrepancy. In 3 out of 5 cases, the metallicities remain
consistent between the contaminated and the stray-light free areas, as well as with the original fixed value. The other two cases show lower metallicity in the contaminated areas.

AO~4 and AO~5 pointings (square, as opposed to the diamond configuration) in which we measured the CXFB (indices 6 and higher; a total of 10 pointings), were used to check for the stray light influence outside $80'$. In these cases, we fitted the spectra with the
appropriate CXFB model in a given arm, allowing for an additional thermal plasma component in the stray light influenced regions. Having accounted for possible ICM emission by checking for the presence of significant thermal plasma emission in the stray light free region, we find clear 
stray light contamination in just one out of 10 pointings (E6). 

Based on these fits we conclude that the influence of the stray light contamination is not significant for our results.

\section{Comparison With \rosat\ Observations}

\begin{figure*}
\centering
\includegraphics[width=.47\textwidth]{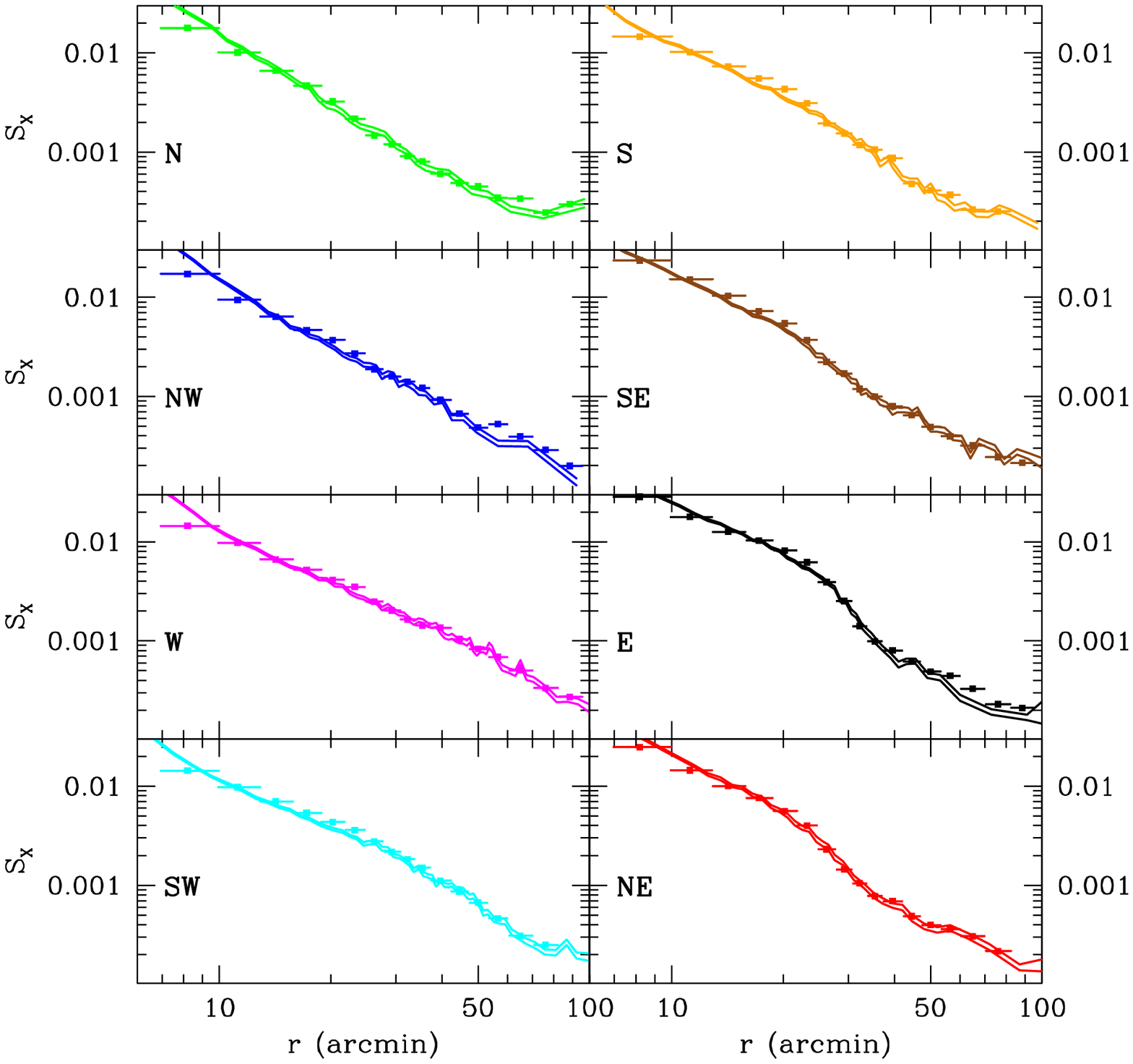}
\includegraphics[width=.50\textwidth]{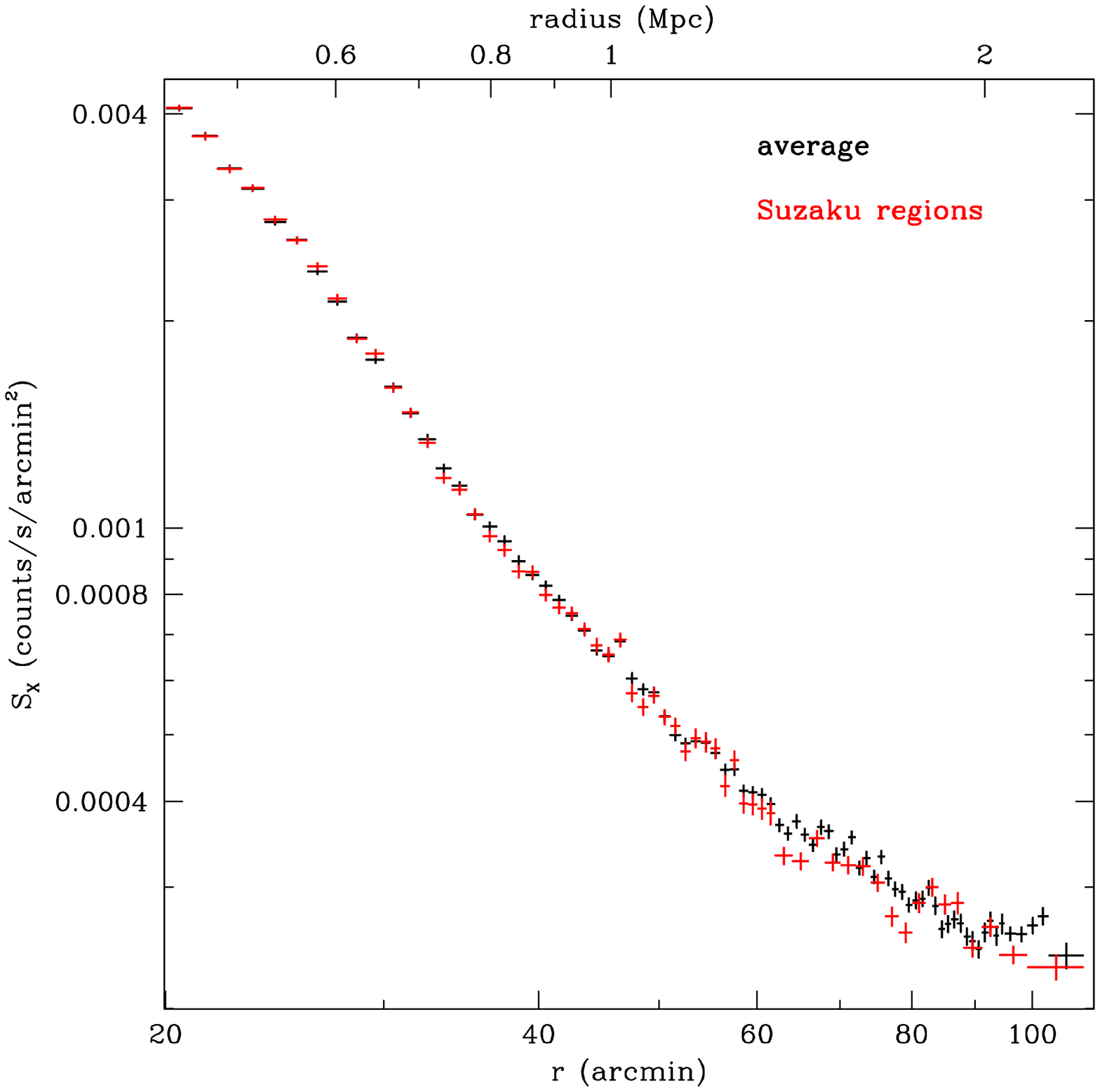}
\caption{{\it Left panel: }Comparison of the ICM surface brightness in the $0.7-2.0\,\text{keV}$ range, as obtained by \rosat\ and \suzaku. \rosat\ confidence intervals are delimited by lines; the surface brightness derived
from \suzaku\ observations is plotted with squares (the error bars are smaller than the plotting symbols). The $y$-axis units are given in $\text{counts}/\text{s}/\text{arcmin}^2$. {\it Right
panel:} Average \rosat\ surface brightness profiles of the Perseus Cluster, using the full azimuthal coverage (\emph{black}) and using only the regions observed by \suzaku\ (\emph{red}).}
\label{fig:rosatsb}
\end{figure*}

We used the \rosat\ mosaic observation of the Perseus Cluster, for which the data reduction is described in \citet{simionescu2012}, to obtain the surface brightness profiles in the
$0.7-2.0\,\text{keV}$ range in the regions corresponding to the \suzaku\ coverage. In the left panel of Fig.~\ref{fig:rosatsb}, we compare these profiles to the number of ROSAT counts
predicted from the best-fit \suzaku\ models in each annulus using the ROSAT response. We measure a very good overall agreement between the profiles, with some small deviations between 53--60
arcmin along the NW and 60--70 arcmin along the E arm.

The fraction of the surface area of the Perseus Cluster covered by the \suzaku\ mosaic decreases with radius, and amounts to only $\sim30\%$ in the outskirts. To check whether this coverage offers a representative sample of the ICM properties, we measured the X-ray surface brightness 
using the \rosat\ mosaic, obtaining first an average profile of the full azimuthal angle and, second, a profile where we
consider only the regions observed by \suzaku. The right panel of Fig.~\ref{fig:rosatsb} shows that both profiles are consistent across the entire observed region. 

\section{Temperature Bias at Low Signal-to-Background Ratio}

\citet{leccardi2007,leccardi2008} argue that commonly used maximum likelihood estimators may introduce a downward bias in the best-fit temperatures for data with low statistics and
relatively low signal-to-background ratios. This is because, even if the CXFB and NXB models are perfectly understood (for an ``infinite'' background exposure time and ideal calibration), if the background parameters are
fixed throughout the fit, the ICM model will adjust to compensate for the statistical fluctuations in the background data due to Poisson noise.

To test for a possible bias this effect may introduce, we fitted the data in the two outermost annuli ($70'<r<82'$ and $82'<r<95'$, or annuli~16 and 17, respectively) in the regions, where we
were not tying the temperatures between the neighbouring annuli. In total, we used five regions from annulus~16 -- N, NW, W, S and SE arms -- and three regions from annulus~17 --  NW, W and
SE arms -- with different values for the best-fit temperatures and different ICM-to-CXFB flux ratios. For each individual region we simulated 250~realizations of its best-fit model (cluster
plus CXFB). To each of these simulated spectra, we added the expected contribution from the NXB, computed by taking the NXB spectrum and scaling the number of counts and corresponding Poisson
noise to the duration of the on-source observation. Finally, we fitted each of the simulated spectra, using the original NXB spectrum.

\begin{figure*}
\centering
\begin{minipage}{0.95\columnwidth}
\includegraphics[width=.9\columnwidth]{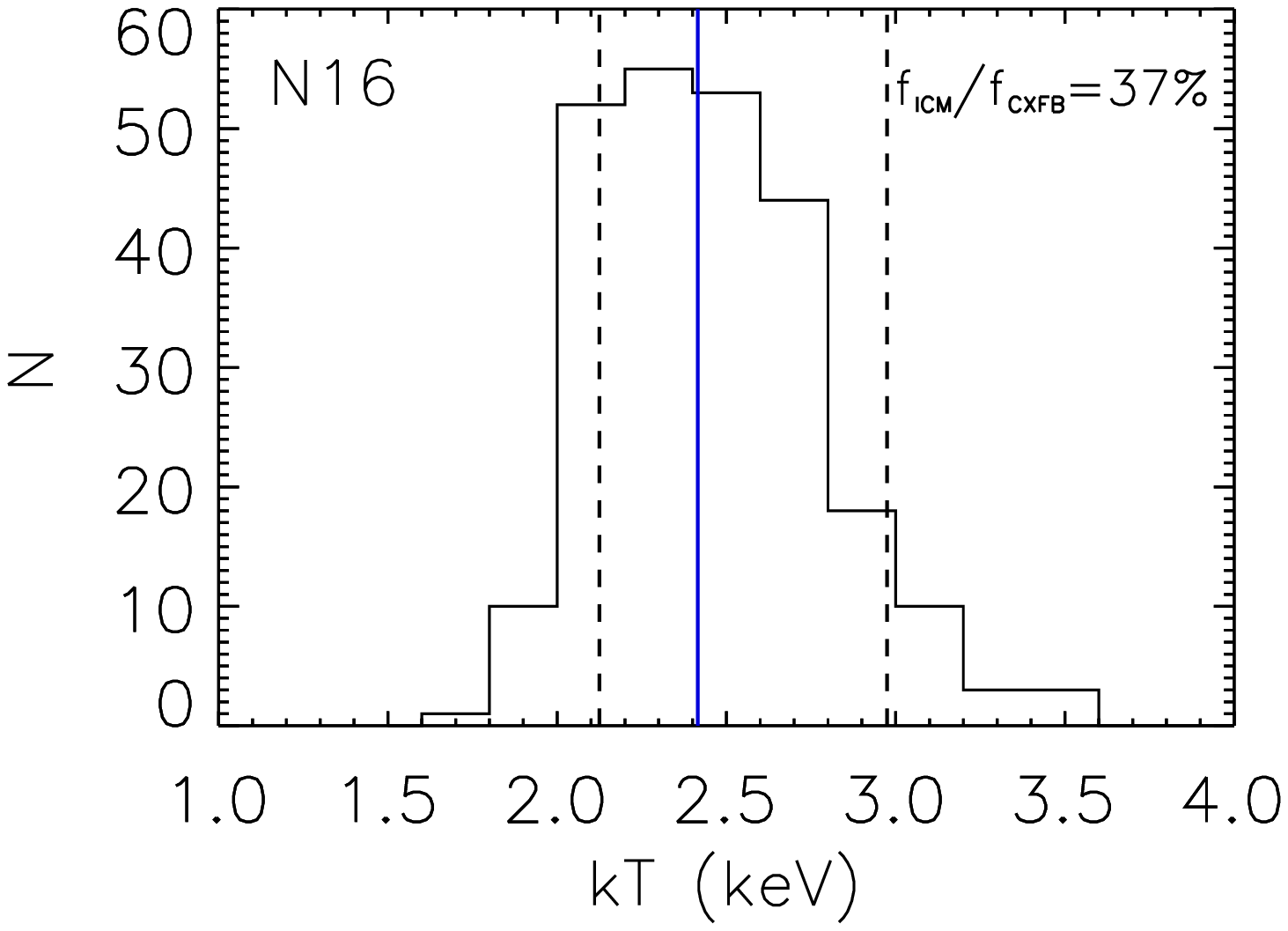}
\end{minipage}
\begin{minipage}{0.95\columnwidth}
\includegraphics[width=.9\columnwidth]{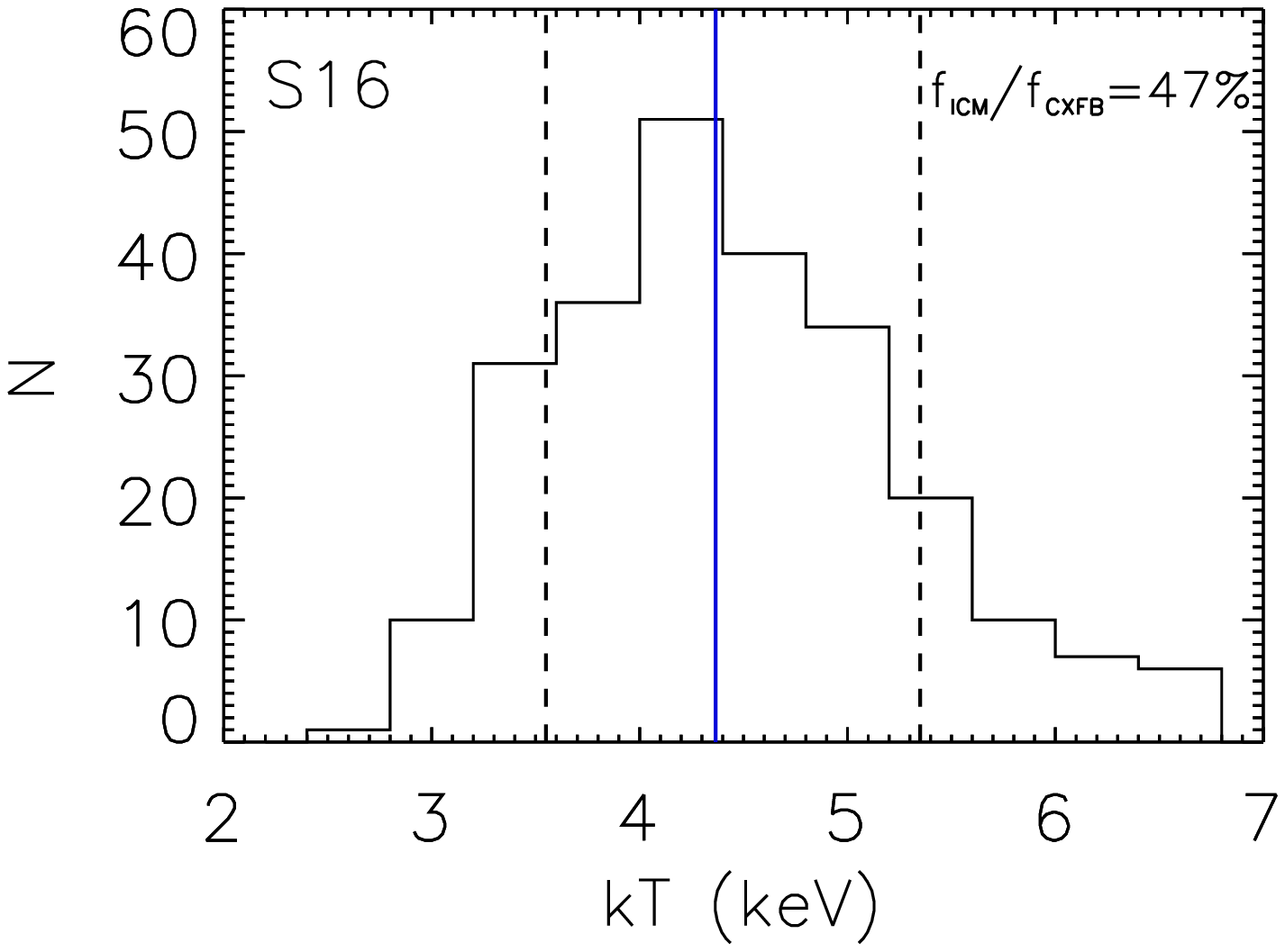}
\end{minipage}
\begin{minipage}{0.95\columnwidth}
\includegraphics[width=.9\columnwidth]{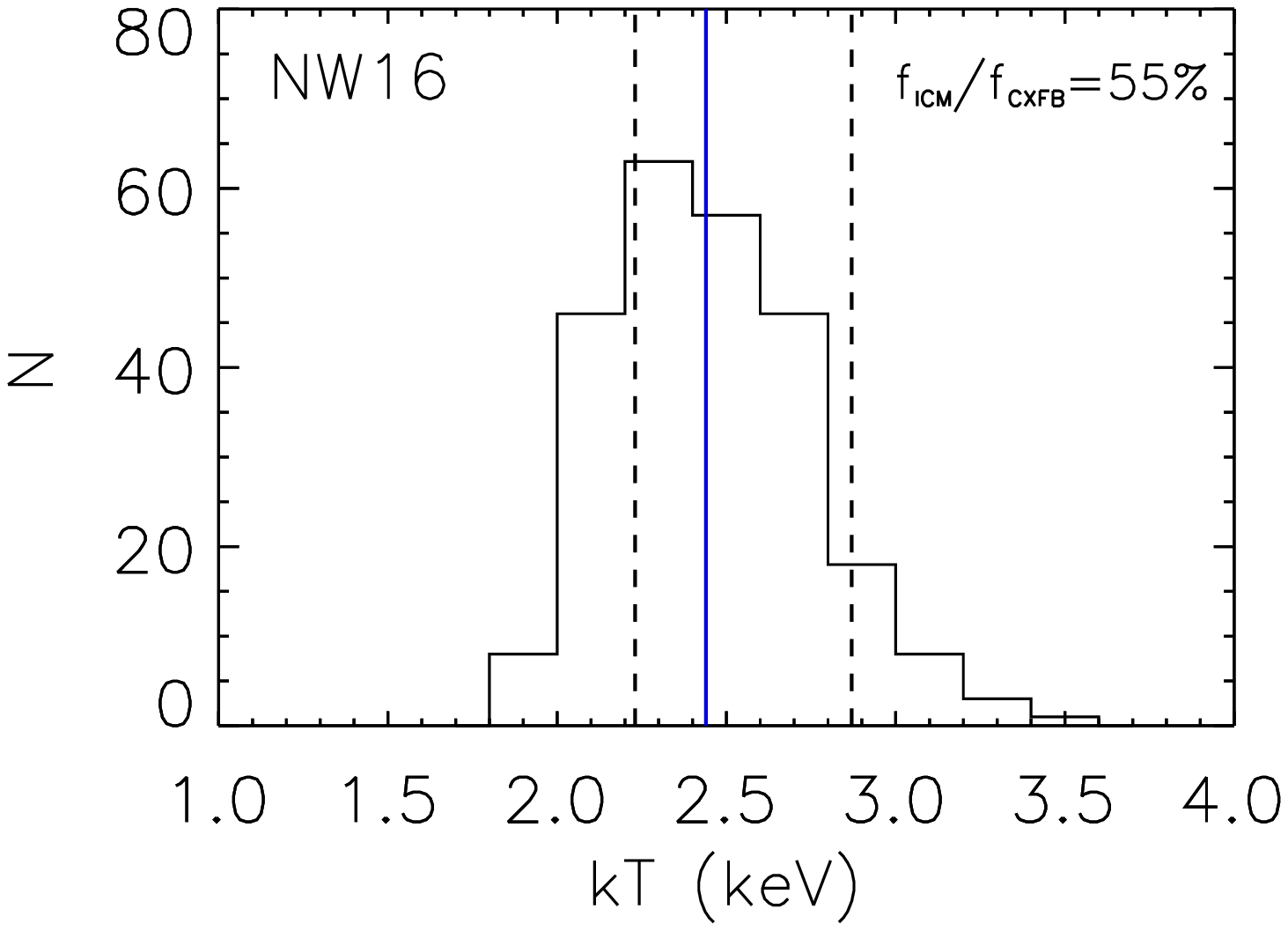}
\end{minipage}
\begin{minipage}{0.95\columnwidth}
\includegraphics[width=.9\columnwidth]{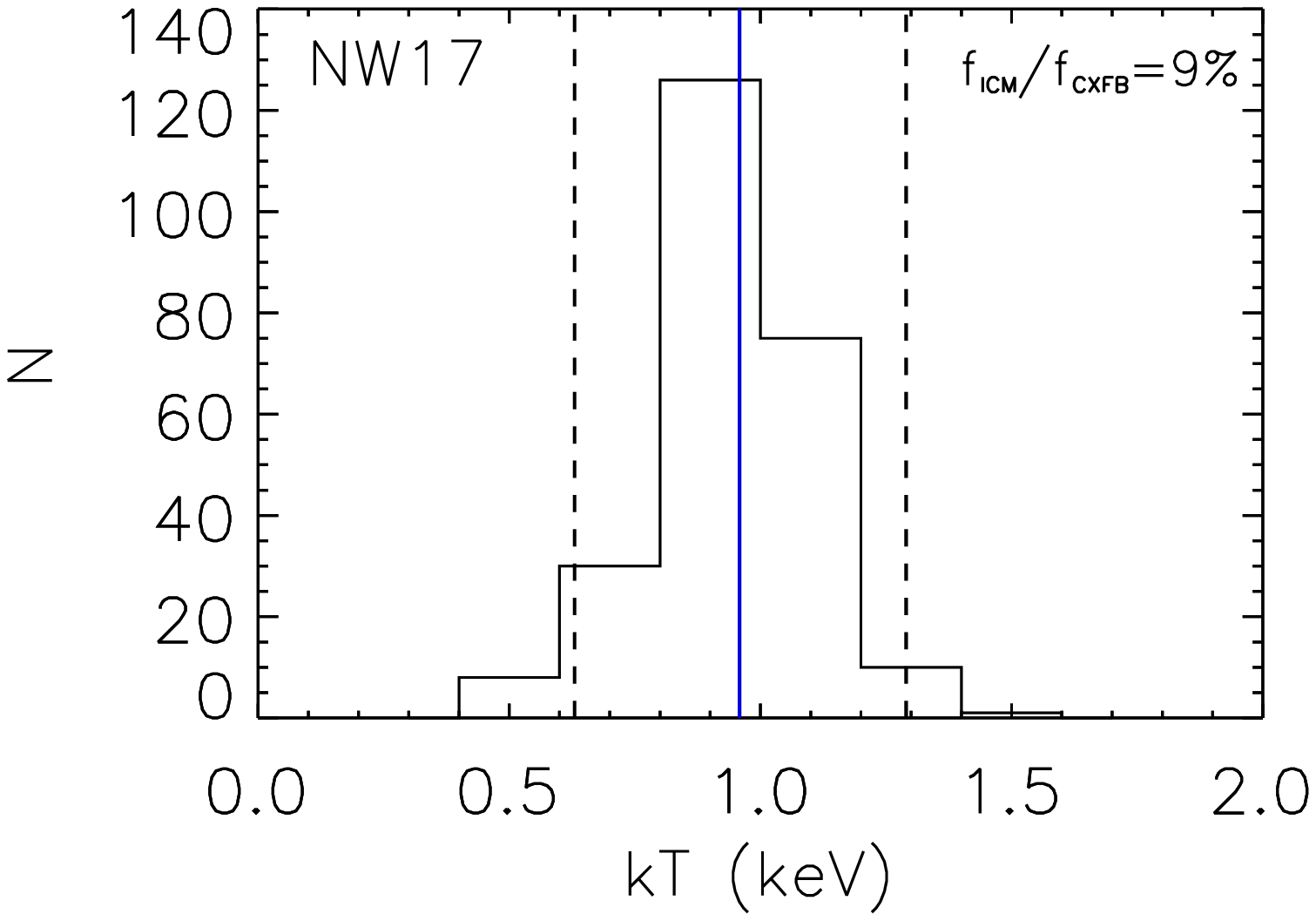}
\end{minipage}
\begin{minipage}{0.95\columnwidth}
\includegraphics[width=.9\columnwidth]{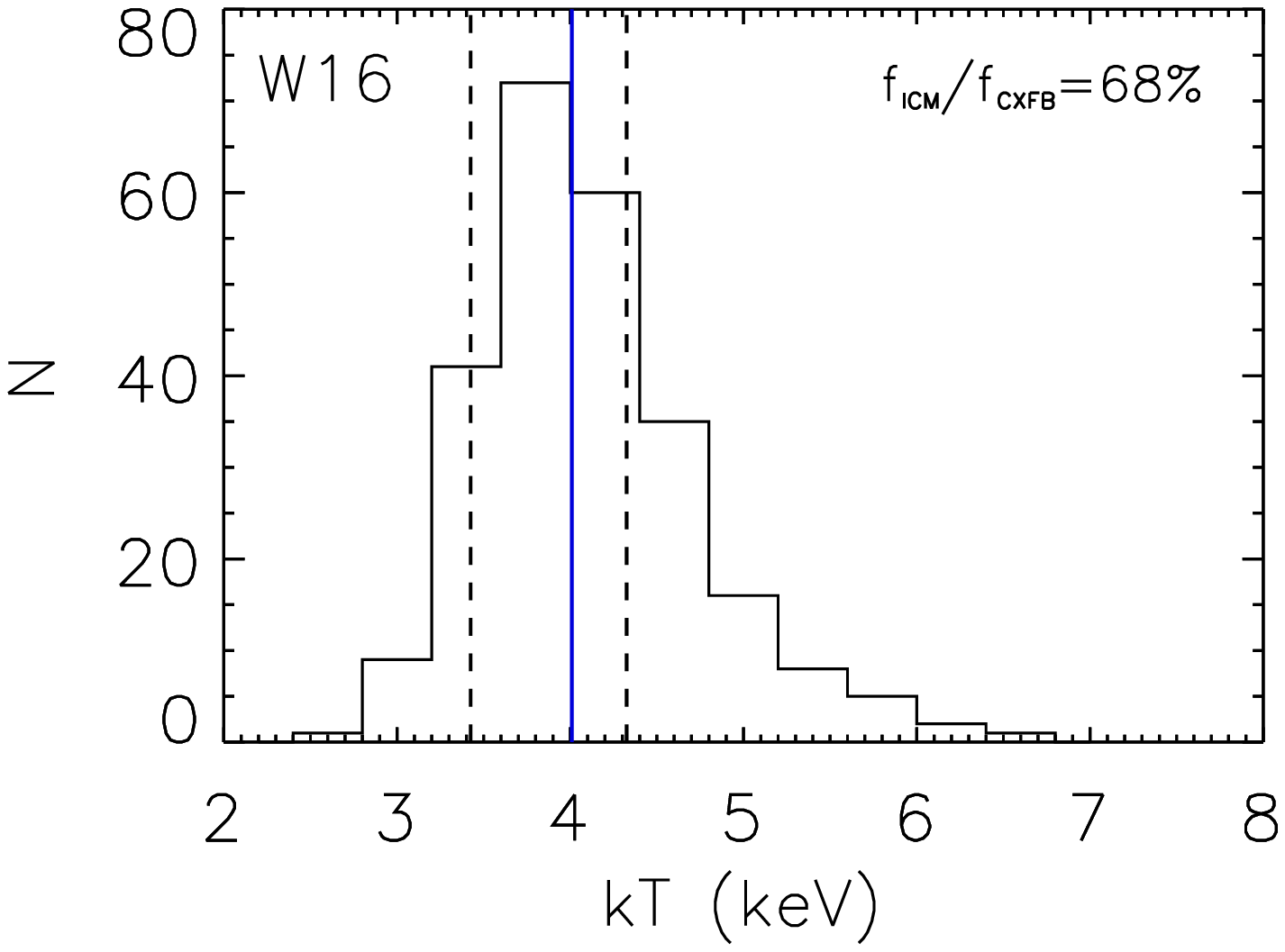}
\end{minipage}
\begin{minipage}{0.95\columnwidth}
\includegraphics[width=.9\columnwidth]{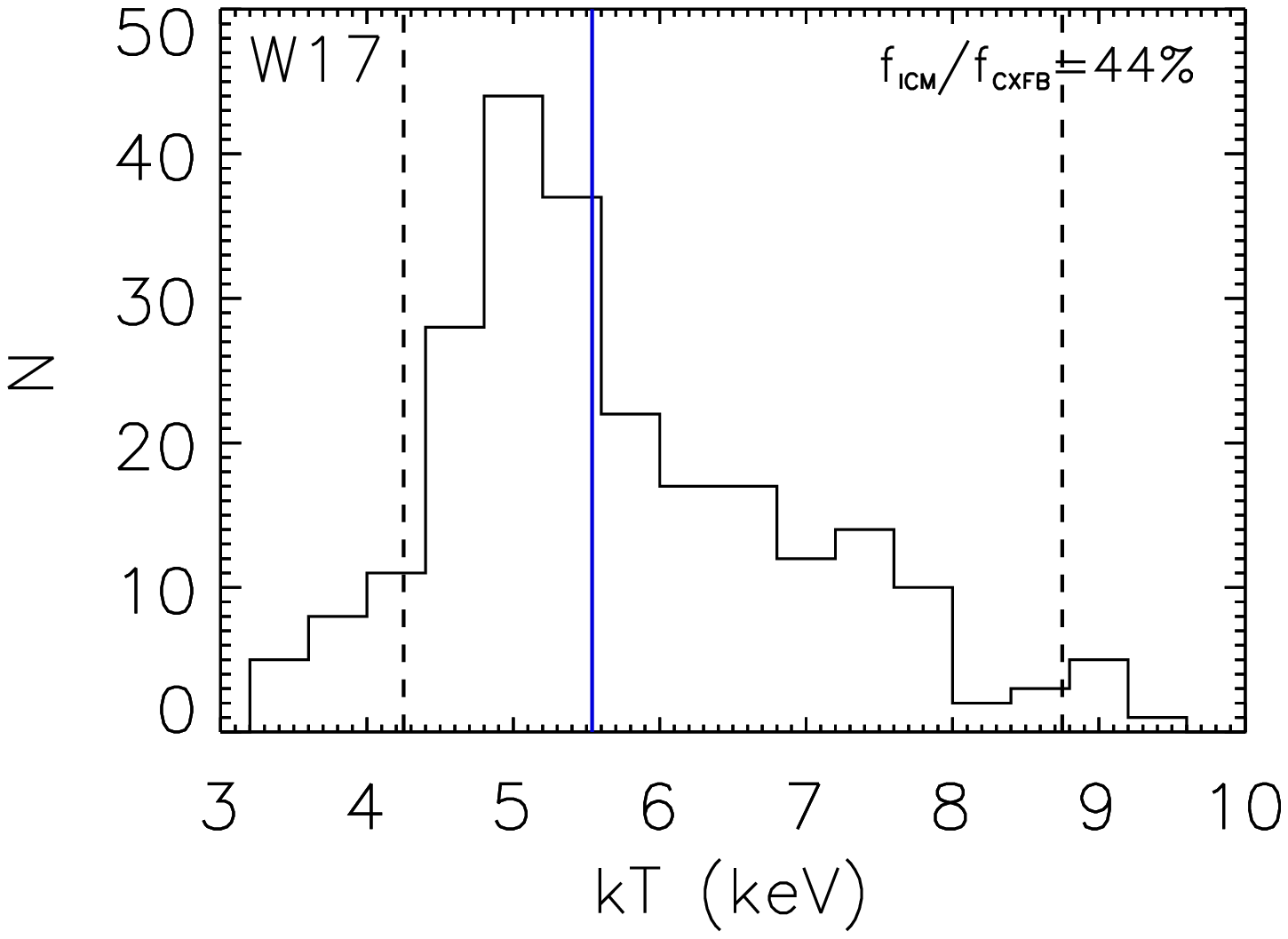}
\end{minipage}
\begin{minipage}{0.95\columnwidth}
\includegraphics[width=.9\columnwidth]{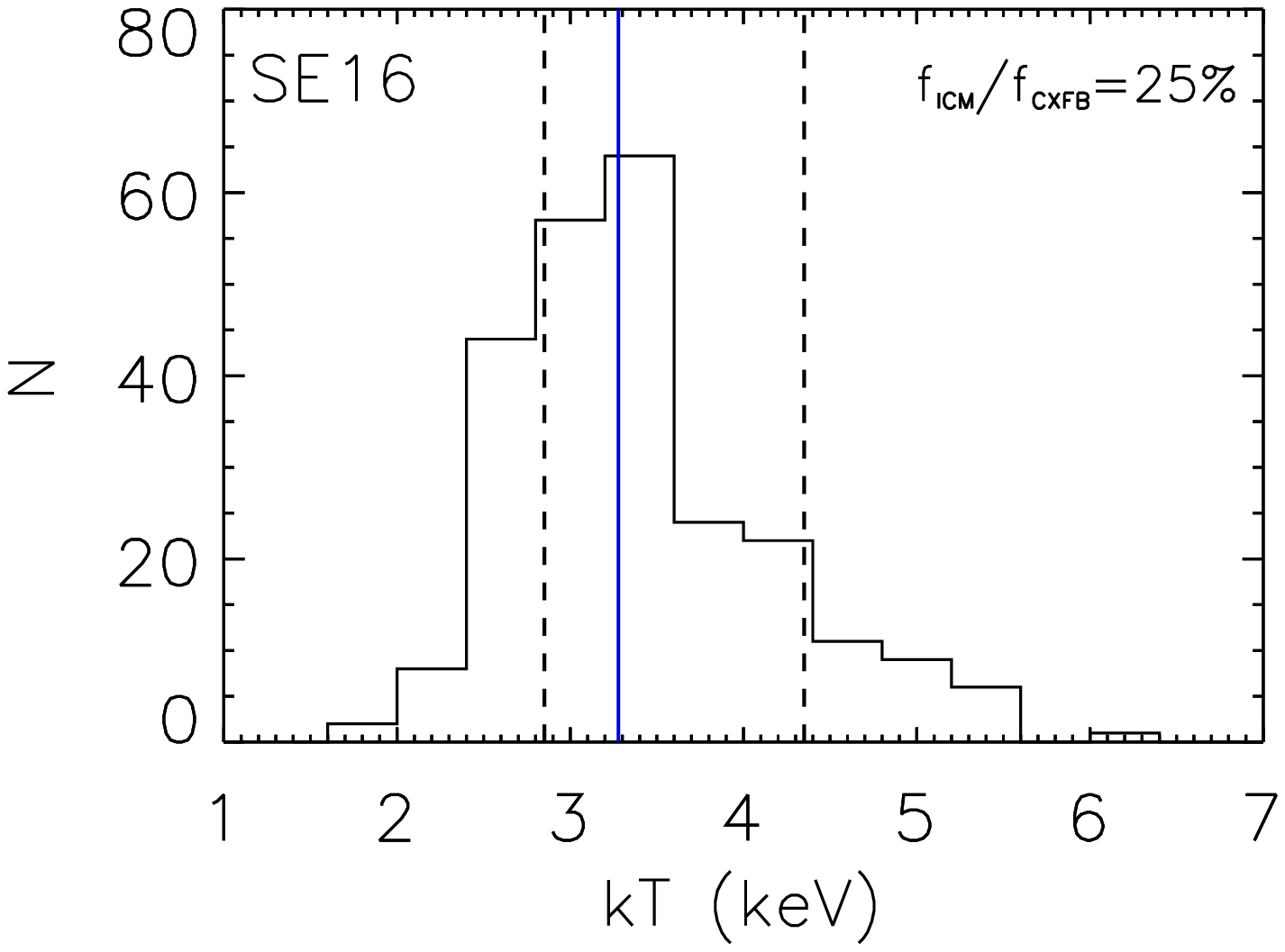}
\end{minipage}
\begin{minipage}{0.95\columnwidth}
\includegraphics[width=.9\columnwidth]{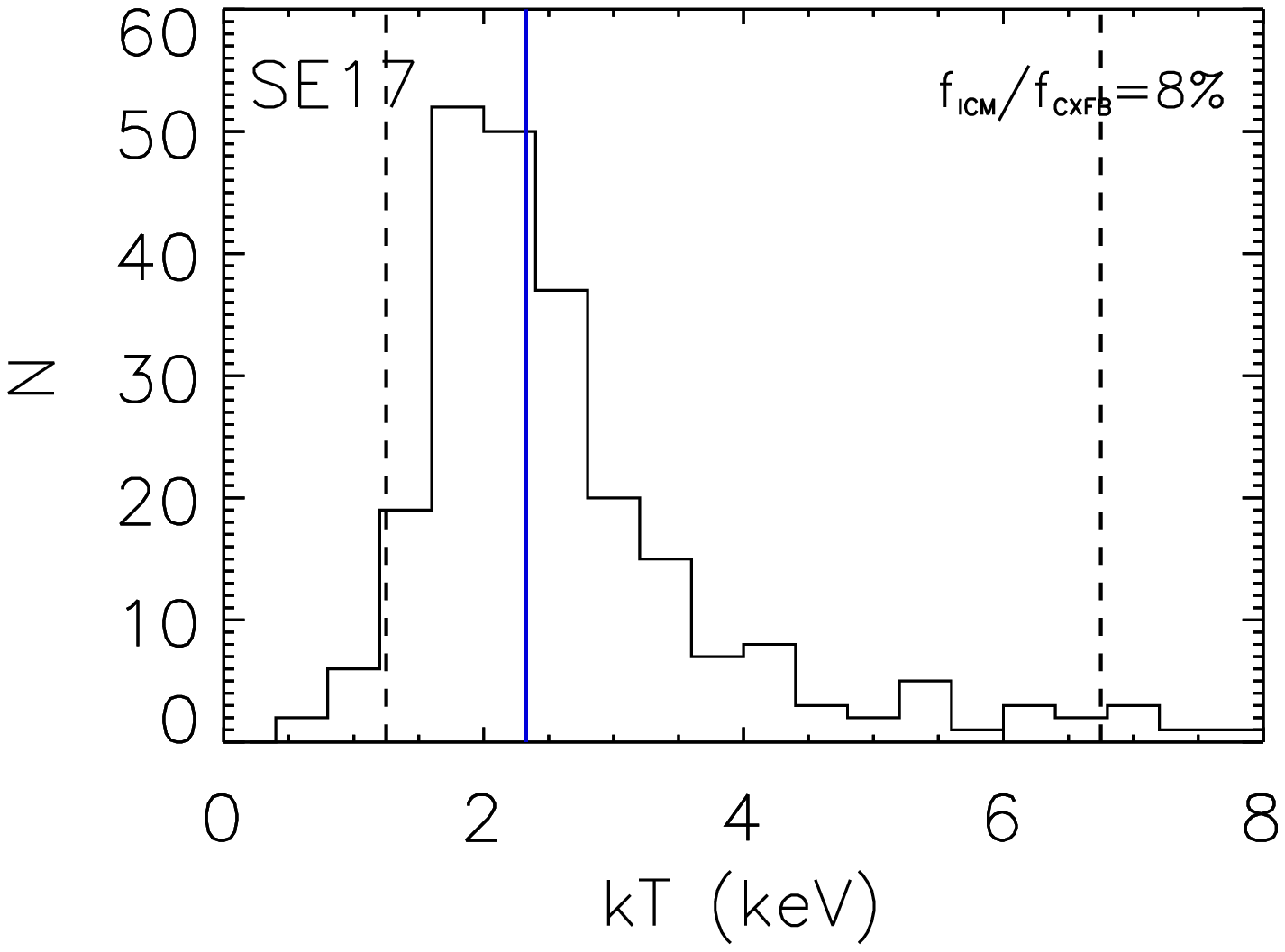}
\end{minipage}
\caption{Histograms of the best-fit temperatures obtained from 250 simulated realizations of the spectra in eight regions of the cluster outskirts. The ICM-to-CXFB flux ratios in the $0.7-7.0\,\text{keV}$ range are labeled.
The thick dashed lines show the $1\sigma$ confidence intervals of the best-fit temperature in the given region. Median of each of the distributions is shown in solid blue line.}
\label{fig:histogram}
\end{figure*}

We show the histograms of the best-fit temperature values in Fig.~\ref{fig:histogram}. The distributions are generally slightly skewed towards lower temperatures with the mode being equal or lower than the median of the
distribution. However, in all cases, both the mode and the median lie inside the $1\sigma$ confidence interval of the original best-fit temperature.

\section{An Example of the Spectrum}

\begin{figure*}
\includegraphics[angle=270,width=0.9\textwidth]{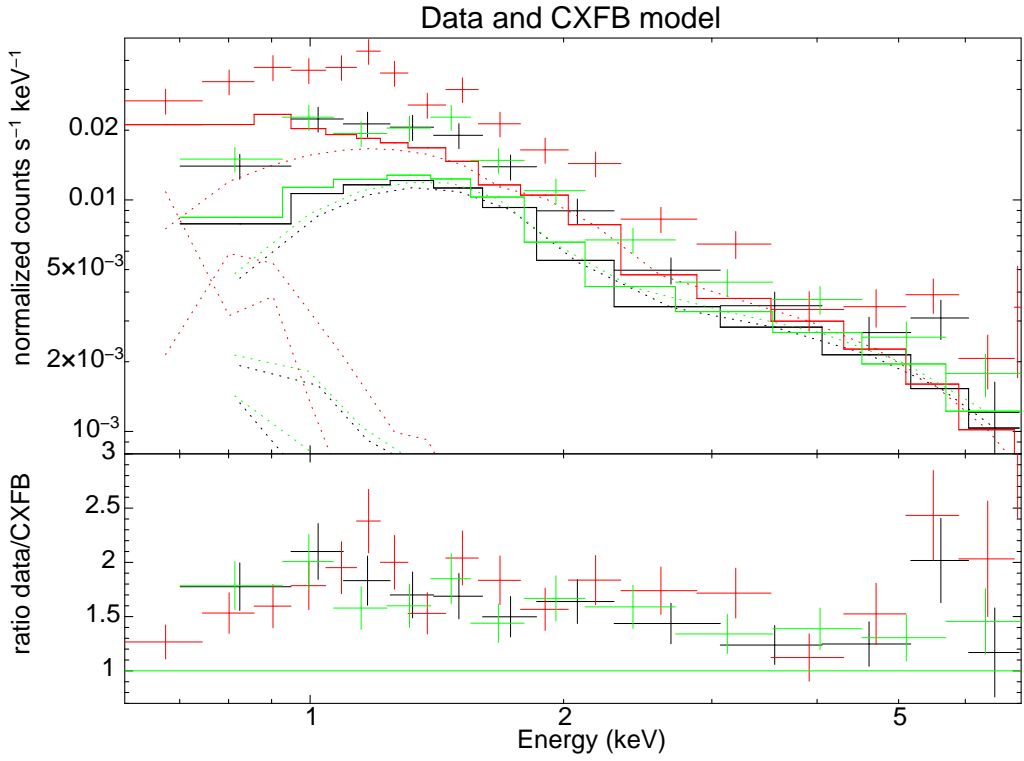}
\caption{Example of spectra of the last annulus within $r_{200}$ ($70'<r<82'$) in the NW~arm. Data from all three detectors are shown -- XIS0 (black), XIS1 (red) and XIS3 (green). The CXFB model is overplotted with thick lines.}
\label{fig:spec_cxfb}
\end{figure*}

\begin{figure*}
\includegraphics[angle=270,width=0.9\textwidth]{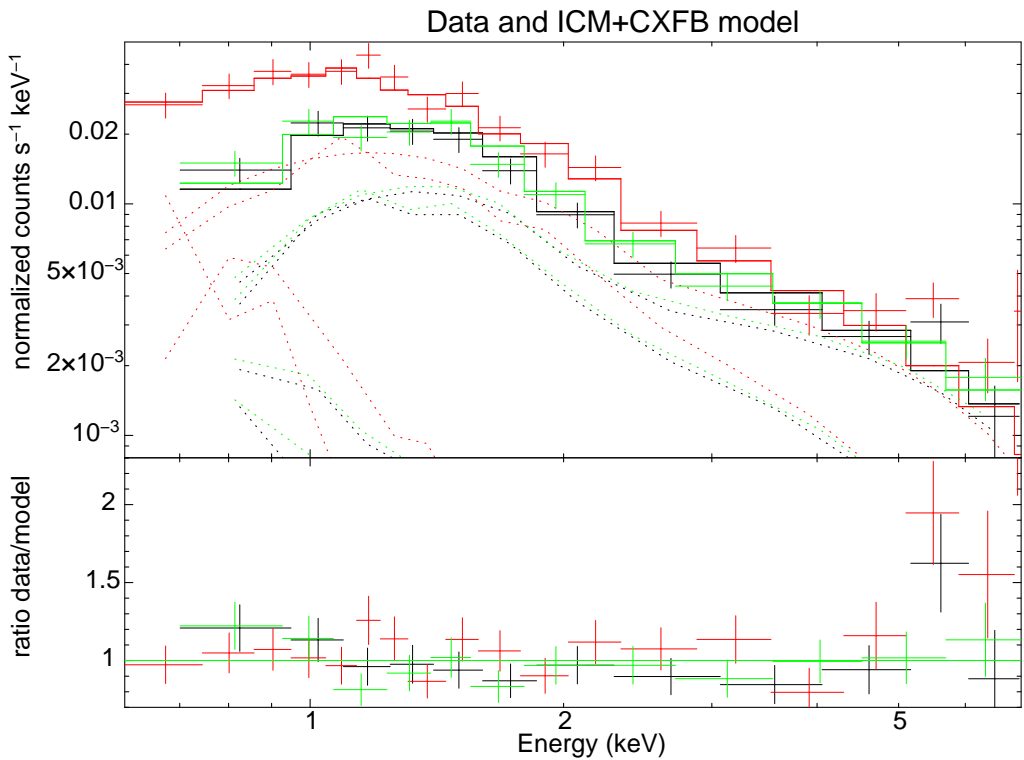}
\caption{Same as Fig.~\ref{fig:spec_cxfb}, the thick lines now show the best-fit model, including the ICM emission.}
\label{fig:spec_all}
\end{figure*}

In Fig.~\ref{fig:spec_cxfb} and Fig.~\ref{fig:spec_all}, we show the spectrum extracted from the last annulus inside $r_{200}$ ($70'<r<82'$ or $1.5\,\text{Mpc}<r<1.8\,\text{Mpc}$) along the NW arm as an example of data with
relatively low ICM/CXFB flux ratio. In this case, $\frac{f_{\rm ICM}}{f_{\rm CXFB}}\sim55\%$ in the $0.7-7.0\,\text{keV}$ range.

In the bottom panel of Fig.~\ref{fig:spec_cxfb}, we plot the ratio between the X-ray and CXFB fluxes. We see a clear excess, representing the strength of the ICM signal. Fig.~\ref{fig:spec_all} shows the data with the
over-plotted model including the ICM emission, for comparison.


\end{document}